\documentclass[11pt,a4paper]{article}
\usepackage[margin=1in,marginparwidth=2.2cm,marginparsep=1.5mm]{geometry}
\usepackage{amsmath,amssymb,amsthm,mathtools}
\usepackage{braket}
\usepackage[nointegrals]{wasysym}
\usepackage{booktabs}
\usepackage{longtable}
\usepackage{array}
\usepackage{multirow}
\newcolumntype{L}[1]{>{\raggedright\arraybackslash}p{#1}}
\usepackage{url}
\usepackage[table]{xcolor}
\usepackage{float}
\usepackage{graphicx}
\usepackage[affil-it]{authblk}

\usepackage{xr-hyper}%
\usepackage[hidelinks,pdfauthor={J. Senellart, B. Stott, S. Wein, E. Cuenca Gomez, P. Senellart, S. Mansfield, P.-E. Emeriau}]{hyperref}
\usepackage{tablefootnote}
\DeclareMathOperator{\per}{Per}

\newcommand{\C}{\mathbb C}
\newcommand{\E}{\mathbb E}

\newcommand{\hollow}{$\circ$\,}

\usepackage[textsize=tiny]{todonotes}

\usepackage{pemod}
\usepackage{listings}
\newcommand{\miPic}[2][1.0]{\begin{tikzpicture}[scale=#1, baseline={([yshift=-.5ex]current bounding box.center)}]#2\end{tikzpicture}}
\newcommand{\miFock}{\miPic{\peSource{.45,0}{fock}\draw[pe/wire] (.47,0) -- (1.1,0); \peTrain{.4,0}{.17}{3}{fock}}}
\newcommand{\miSup}{\miPic{\peSource{.45,0}{fock}\draw[pe/wire] (.47,0) -- (1.1,0); \peTrain{.4,0}{.17}{3}{sup}}}
\newcommand{\miSq}{\miPic{\peSqueezer{.45,0}\draw[pe/wire] (.47,0) -- (1.1,0); \peTrain{.4,0}{.17}{3}{sq}}}
\newcommand{\miEm}{\miPic{\peSource{.45,0}{spin}\draw[pe/wire] (.47,0) -- (1.2,0); \peTrain{.45,0}{.17}{4}{ent}}}
\newcommand{\miUshot}{\miPic[1.0]{%
  \foreach \y in {.1,.35,.6} {\draw[pe/wire] (0,\y) -- (.25,\y); \draw[pe/wire] (.75,\y) -- (.95,\y); \peDet{.95,\y}}
  \peU{.25,0}{.75,.7}{\scriptsize$U$}}}
\newcommand{\miUtick}{\miPic[1.0]{%
  \foreach \y in {.1,.35,.6} {\draw[pe/wire] (0,\y) -- (.25,\y); \draw[pe/wire] (.75,\y) -- (.95,\y); \peDet{.95,\y}}
  \peU{.25,0}{.75,.7}{\scriptsize$U_t$}
  \foreach \i in {1,2,3} \draw[pePh, line width=.45pt] (1.03+\i*.11,.29) -- ++(0,.12);
  \draw[-{Stealth[length=2pt]}, black!45, line width=.4pt] (1.1,.15) -- ++(.42,0);}}
\newcommand{\miLoops}{\miPic[1.0]{%
  \draw[pe/wire] (0,0) -- (1.55,0);
  \draw[black!70, line width=.5pt, rounded corners=2.5pt] (.3,.07) -- (.3,.35) -- (.6,.35) -- (.6,0); \peBS{.3,0}
  \draw[black!70, line width=.5pt, rounded corners=3pt] (.8,.07) -- (.8,.55) -- (1.3,.55) -- (1.3,0); \peBS{.8,0}
  \peDet{1.55,0}}}
\newcommand{\miFF}{\miPic[1.0]{%
  \foreach \y in {.1,.35,.6} \draw[pe/wire] (0,\y) -- (.2,\y);
  \peU{.2,0}{.55,.7}{\tiny$U_0$}
  \draw[pe/wire] (.55,.1) -- (.65,.1); \peDet{.65,.1}
  \foreach \y in {.35,.6} {\draw[pe/wire] (.55,\y) -- (.85,\y); \draw[pe/wire] (1.15,\y) -- (1.3,\y); \peDet{1.3,\y}}
  \draw[pe/cl] (.78,.1) -- (1.0,.1) -- (1.0,.25);
  \peU{.85,.25}{1.15,.7}{\tiny$V_p$}}}
\newcommand{\miNL}{\miPic[1.0]{%
  \foreach \y in {.1,.35,.6} {\draw[pe/wire] (0,\y) -- (.9,\y); \peDet{.9,\y}}
  \peNL{.45,.35}}}

\makeatletter
\def\pe@stripsp#1{\zap@space#1 \@empty}
\def\pe@recordcite#1{\@for\pe@k:=#1\do{\edef\pe@k{\expandafter\pe@stripsp\expandafter{\pe@k}}%
  \expandafter\gdef\csname pe@cited@\pe@k\endcsname{}}}
\AddToHook{begindocument/before}{\let\citation\pe@recordcite}
\AddToHook{begindocument/end}{\let\citation\@gobble}
\newif\ifpe@skipbib
\let\pe@origbibitem\bibitem
\def\bibitem#1{\ifpe@skipbib\egroup\global\pe@skipbibfalse\fi
  \@ifundefined{pe@cited@#1}{\global\pe@skipbibtrue\setbox\z@\vbox\bgroup}{\pe@origbibitem{#1}}}
\AtBeginDocument{\let\pe@origendbib\endthebibliography
  \def\endthebibliography{\ifpe@skipbib\egroup\global\pe@skipbibfalse\fi\pe@origendbib}}
\makeatother

\theoremstyle{definition}

\theoremstyle{remark}

\author[1]{Jean Senellart}
\author[1]{Benjamin Stott}
\author[1]{Stephen Wein}
\author[2]{Esperanza Cuenca Gomez}
\author[1,3]{Pascale Senellart}
\author[1]{Shane Mansfield}
\author[1]{Pierre-Emmanuel Emeriau}
\affil[1]{Quandela, Massy, France}
\affil[2]{NVIDIA}
\affil[3]{Centre for Nanosciences and Nanotechnology (C2N), CNRS, Universit\'e Paris-Saclay, Palaiseau, France}
\date{\small \today}

\hypersetup{pdftitle={The simulability frontiers of photonic quantum computing}}
\makeatletter
\let\pe@maketitle\maketitle
\let\pe@@maketitle\@maketitle
\let\pe@and\and
\AtBeginDocument{\let\pe@author\@author}
\makeatother

\title{The simulability frontiers of photonic quantum computing}

\begin{document}
\sloppy
\maketitle

\begin{abstract}
Estimates of when photonic devices outgrow classical simulation are hard to compare: even for
single-photon boson sampling, quoted thresholds range from 50 to 98 photons, because they rest on
different tasks and assumptions. We propose as a common unit the time a classical computer needs to draw
one sample from the device's exact output distribution, up to floating-point rounding. Exactness is the only criterion that applies
equally to every kind of light, since the cost of approximate sampling depends on the accuracy
required. The simulability frontier is the device size at which our fastest exact sampler produces one
sample per second on a fixed machine; it is a lower bound on classical capability, not a hardness
threshold. In a universal interferometer the time per sample grows exponentially with the number of
detected photons, so doubling the machine's speed moves the frontier by only about one photon, whereas
a better algorithm changes the growth rate. We measure the frontier with \texttt{photonic-engines}, an
open-source GPU library of exact samplers for single photons, entangled photons from quantum
emitters, vacuum--single-photon superpositions and squeezed light. Each sampler is validated against a
simulator that shares no code with it.
On one NVIDIA H100 GPU, with Haar-random interferometers, twice as many modes as photonic inputs and
photon-number-resolving detectors, the frontier lies at $38.3$ single photons and at $31.0$ entangled
photons from one emitter, whose entanglement makes each sample more expensive. For squeezed light, a
timing model fitted to measurements places it at $41.5$ squeezed photonic inputs of mean photon number $0.5$. For our
samplers, coherence alone makes simulation harder: sources that emit a photon with probability one half
reach $64.2$ when vacuum and photon are mixed incoherently, as for lossy single photons, but only $50.0$
when they are in coherent superposition. Feed-forward from mid-circuit measurement of up to three modes
lowers the single-photon frontier only from $38.3$ to $38.1$.
\end{abstract}

\section{Introduction}\label{sec:intro}\label{sec:thesis}

The prospect of a computational advantage has driven the development of photonic quantum computing
over the past decade. Boson sampling~\cite{AA2013} became its first milestone, and experiments with
squeezed light~\cite{Madsen2022} and with single photons from quantum emitters~\cite{Maring2024} now
reach regimes where classical simulation becomes costly. For each such device, a practical question
is the size at which it stops being classically simulable at the rate it runs. That size decides
where a demonstration is worth attempting, which imperfection to reduce first, and which classical
machine to compare against.

The literature gives no answer that can be compared across devices. For
single-photon boson sampling alone, published thresholds range from 50 to 98
photons~\cite{CC2018,Neville2017,Dalzell2020}, because they rest on different tasks, sampling
guarantees and hardware assumptions.
Other devices are characterised by other quantities: the memory of a simulator for time-bin
loops~\cite{Novak2025}, transmission thresholds for lossy
devices~\cite{GarciaPatron2019,OszmaniecBrod2018,Qi2020}, or the time a supercomputer needs for a
single experiment~\cite{Madsen2022,Bulmer2022}. Complexity results~\cite{CC2018,CC2020} give no
wall-clock time. None of these statements converts into another, and none depends on the rate
at which samples are produced (Supplementary Sec.~\ref{sec:thresholds} and
Table~\ref{tab:T2}).

Part of the difficulty is that classical simulation names several tasks with very different costs
(Table~\ref{tab:tasks}, Sec.~\ref{sec:frontier}). Strong simulation returns every output probability
and is limited by memory. Estimating a probability or an expectation value to additive error costs
polynomial time~\cite{Gurvits2005,AaronsonHance2014,LimOh2025,Estimator,Seron2026}, but yields no
samples. Exact sampling draws outcomes from exactly the device's output distribution, whereas
approximate sampling draws them from a nearby one. In this work, we focus on exact sampling. It is
what a device does, and it remains hard outside the observables that efficient estimators cover.
It is also the only task whose cost compares across kinds of light, since the cost of an
approximate sampler depends on the error it is allowed and on the test that checks it.

We therefore introduce the \textbf{simulability frontier}, the device size at which the fastest
exact sampler we implement produces a given number of samples per second on a fixed classical
machine. The frontier is a lower bound on classical capability rather than a hardness threshold,
since better algorithms or more computing power could move it. Because the cost of exact sampling grows exponentially with
the number of detected photons, faster hardware moves the frontier only logarithmically, whereas a
better algorithm changes its growth rate. To make comparisons between inputs meaningful, we time
every input under the same protocol on the same machine, and an audit records the algorithm, the
code and the run behind each number (Sec.~\ref{sec:frontier}).

To measure the frontier, we developed \texttt{photonic-engines}, a GPU library of exact samplers
released as open source (MIT). It covers single photons, vacuum--single-photon
superpositions, squeezed light and the entangled photon trains of quantum emitters, combined with
feed-forward, single-mode non-linear phases, time-bin loops, loss and partial distinguishability
(Table~\ref{tab:modalities}, Sec.~\ref{sec:algorithms}). Several samplers implement exact sampling rules
derived in a companion paper~\cite{Methods}, and the library also includes the estimator of a second
companion paper~\cite{Estimator}. Each sampler is validated against a separate simulator that shares
no code with it (Supplementary Sec.~\ref{supp:validation}). To our knowledge, the library is the
fastest public exact sampling code for the inputs it covers on generic, full-depth interferometers
(Supplementary Table~\ref{tab:coverage}): the only accelerated exact sampler for single photons on
record, a four-chip reconfigurable engine, takes about $80\,$s per sample for $40$ photons in $60$ modes~\cite{Morse2024},
against $1.0\,$s for ours on the same instance size. Other GPU codes compute
permanents~\cite{SUperman} or sample squeezed light~\cite{GBSKernels}; the latter, run as shipped on our squeezed instances,
reaches about $5$ photonic inputs at one sample per second, against $41.5$ for ours (Supplementary Sec.~\ref{sec:nogpu}).

\begin{figure}[t]\centering
\includegraphics[width=0.78\textwidth]{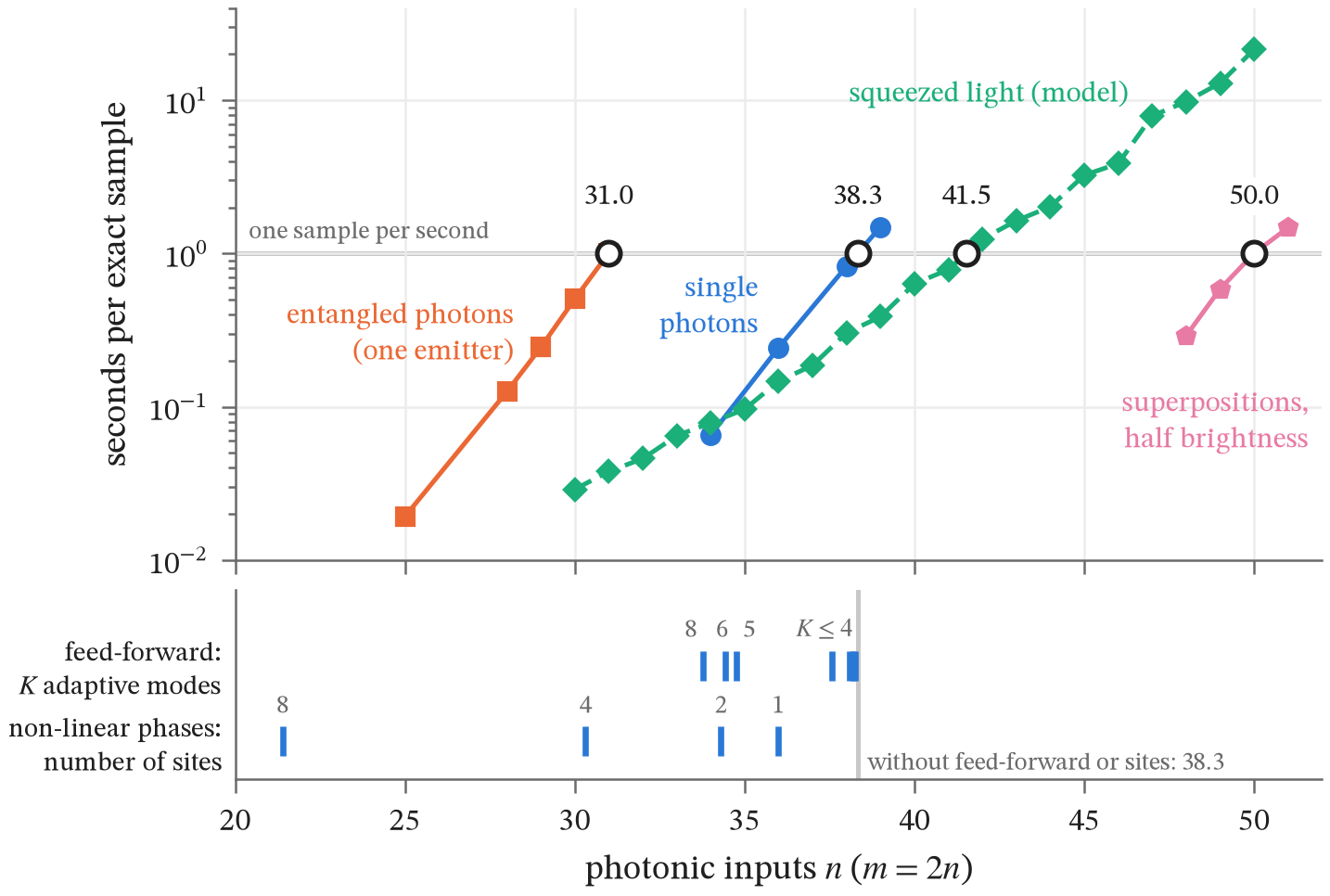}
\caption{\textbf{The one-hertz simulability frontier, demultiplexed in space.} Seconds per exact
sample against the number $n$ of photonic inputs, with $m = 2n$ modes, on one H100. Open circles mark
the one-second crossings $n^*$, with their values. In the
strip below, each mark is the one-hertz frontier $n^*$ of single photons with $K$ adaptive modes or
one, two, four or eight non-linear sites; the grey line is $n^*$ without either option. Superpositions
are at half brightness: each source emits a photon with probability one half. For comparison, approximate
squeezed-light samplers report $10^7$ samples in $115\,$min on eight H100~\cite{Dodd2025} and in
about $62\,$min on up to $288$ A100~\cite{Oh2024MPS}, on their own instances, which are not
placed at any $n$ here.}\label{fig:money}
\end{figure}

On a single NVIDIA H100 GPU, the frontier at one sample per second lies between about thirty and fifty
photonic inputs, depending on the input (Fig.~\ref{fig:money}).
For the samplers we evaluate, coherence between the vacuum and single-photon components of a source
makes simulation markedly harder for sources that emit a photon with the same probability, whereas feed-forward on a few modes barely
changes the cost. When a device is used tick by tick, the cost is set by the state carried from one
tick to the next rather than by the length of the photon train. In particular, a device can outpace
one GPU without any of its samples being hard: reproducing in real time the output stream of a fast,
lossy device can require tens of GPUs even when its typical sample lies well inside the frontier.

The paper is organised as follows. Section~\ref{sec:frontier} defines the frontier and its
measurement, and Sec.~\ref{sec:algorithms} presents the library and its samplers.
Sections~\ref{sec:demux} and~\ref{sec:pertick} give the frontier for devices whose photon trains are
sent into space and for devices used tick by tick, and Sec.~\ref{sec:imperfections} adds loss and
distinguishability. Section~\ref{sec:companions} gives the memory and precision frontiers,
Sec.~\ref{sec:hardware} compares the frontier with existing devices and with energy, and
Sec.~\ref{sec:discussion} discusses the limits of our approach.

\section{The simulability frontier and its measurement}\label{sec:frontier}

The frontier is the device size at which our fastest exact sampler takes one second per sample, on a
fixed machine and for a specified family of interferometers. Comparing the frontiers of two inputs
requires timing both the same way, with equally optimised code. This section makes both statements
precise.

\begin{table}[t]\centering\small
\renewcommand{\arraystretch}{1.35}
\caption{\textbf{What ``classical simulation'' can mean.} Each task returns a different object at a
different cost. Costs are for $n$ single photons in $m$ modes, with $\beta \approx 1.8$ the base of the
exponential cost, Eq.~\eqref{eq:beta}; the last column gives the largest
device one H100 handles at $m = 2n$. The shaded row is the task this paper measures.}\label{tab:tasks}
\begin{tabular}{@{}L{2.7cm}L{3.9cm}L{4.1cm}L{3.6cm}@{}}
\toprule
\textbf{task} & \textbf{returns} & \textbf{cost} & \textbf{reach on one H100}\\
\midrule
strong simulation & every output probability, stored as one table & $\binom{m+n-1}{n}$ amplitudes in memory & $13$ photons, set by memory (Sec.~\ref{sec:memory})\\
exact probability & one output probability, $|\per(U_S)|^2$ & one permanent, $O(n2^n)$; \#P-hard~\cite{AA2013} & the inner step of every exact sampler\\
estimated probability & a probability or expectation value to $\pm\varepsilon$ & polynomial, $O(n^2/\varepsilon^2)$~\cite{Gurvits2005,LimOh2025,Estimator,Seron2026} & no samples; blind to exponentially small probabilities\\
\rowcolor{black!8} exact sampling & samples from the device's distribution & a few permanents per sample, $O(n\beta^n)$~\cite{CC2018,CC2020,Methods} & $38.3$ photons at one sample per second\\
approximate sampling & samples from a nearby distribution & set by the error allowed~\cite{Oh2024MPS,Dodd2025} & depends on the test that validates it\\
\bottomrule
\end{tabular}
\end{table}

Let $t_A(\theta)$ be the time, in seconds, that an exact sampling algorithm $A$ needs per sample for
device parameters $\theta$, on a fixed machine and for a specified family of random interferometers. We take the
smallest of these times over the algorithms we implement (Supplementary Table~\ref{tab:T1}),
\begin{align}
  t(\theta) \coloneqq \min_A t_A(\theta).\label{eq:envelope}
\end{align}
The minimum runs over tested algorithms only, so $t(\theta)$ bounds the cost of the best exact method from above. Exact methods we have not implemented, such as the treewidth sampler for low-depth circuits \cite{OhTreewidth2022}, are discussed where they could lower the frontier. Because the time per sample varies from shot to
shot with the number of photons detected, $t_A(\theta)$ is an average over shots. We measure it by
$t_S$, the mean over every detected photon number except the rarest, which together occur in at most
a fraction $1/S$ of shots (Eq.~\eqref{eq:tS}).

The \textbf{simulability frontier at rate $R$} is the set of devices simulated at exactly that rate,
\begin{align}
  F_R \coloneqq \{\theta : t(\theta) = 1/R\}.\label{eq:frontier}
\end{align}
Unless stated otherwise, $R = 1\,$Hz on one H100 PCIe GPU.
The rate is a throughput, not the latency of a single sample. The samples are independent, and
the GPU draws them in batches, of at most $256$ shots for the headline frontiers. The rate is the number of samples drawn
divided by the wall-clock time of the timed batches, with kernel compilation excluded
(Supplementary Sec.~\ref{supp:repro}). One sample per second thus means a sustained stream of
independent samples at that rate, not one sample returned a second after it is requested.

The axis is the number $n$ of \textbf{photonic inputs}: the modes of the interferometer that receive
light, whether a single photon, one entangled photon of an emitter's train, a vacuum--single-photon
superposition or a single-mode squeezed state. The interferometer has $m = cn$ modes in all, detection
resolves photon number, and no outcome is post-selected. Losses are uniform: every photon reaches the
detectors with the same probability $\eta$, the transmission (Sec.~\ref{sec:imperfections}). For devices that process their photons one time step at a time
(Sec.~\ref{sec:pertick}), the size is instead set by what the simulator must carry from one step to the
next: the delay $L$ of a loop or the number $k$ of emitters. The \textbf{interpolated frontier} $n^*$
is where the measured time per sample crosses one second: we time the two integer sizes on either side
of the crossing and interpolate between them in log rate. The tables quote $n^*$ to one decimal.
Timing noise gives it an uncertainty of at most $0.03$ for the lossless single-photon, emitter and
superposed results, timed with at least $256$ shots per size, and about $0.2$ at $\eta = 0.5$; scans
timed with fewer shots, such as those of Fig.~\ref{fig:adaptive}, can be off by a few tenths. Changing how we measure, for
instance the batch size, moves $n^*$ by up to $0.2$, so we only compare values measured the same way.
The \textbf{integer frontier} $\lfloor n^* \rfloor$ is, up to timing noise, the largest $n$ measured
at one sample per second or faster.

\paragraph{Why exact sampling.} Table~\ref{tab:tasks} places exact sampling, the task measured here,
among the tasks that ``classical simulation'' can mean. The samplers are \emph{algorithmically
exact}: in exact arithmetic each draws from the target distribution. In practice they run in floating-point
arithmetic, and we time them only where the rounding error at each step of a sample stays below
$10^{-3}$ (Sec.~\ref{sec:precision}).
Exactness makes inputs comparable, because the cost of an approximate sampler depends on how much
error it is allowed and on the test used to check it, and neither has a common meaning for single
photons and squeezed light. A tolerance in total-variation distance would not help either, since at the sizes of interest it
cannot be checked, whereas the error of an exact sampler is rounding alone, which we measure and bound
(Sec.~\ref{sec:precision}). All exact samplers in this paper are chain rules: they draw a sample one photon or one mode at
a time, each from its probability conditioned on the outcomes already drawn, so a sample needs a few
marginal probabilities rather than the whole output distribution (Sec.~\ref{sec:algorithms}).

\paragraph{Algorithm, engineering and hardware.}\label{sec:decomposition} The speed of a sampler depends on
three factors: the \emph{algorithm}, which sets the work per sample; the \emph{engineering}, which sets
how fast a given machine executes that work through memory layout, batching, kernel fusion and
arithmetic format; and the \emph{hardware}, meaning the GPU model and the number of GPUs. Only the
algorithm sets how the work grows with $n$. For every sampler in this paper, the cost per sample
grows at large $n$ as
\begin{align}
  t \propto \mathrm{poly}(n)\,\beta^n,\label{eq:beta}
\end{align}
with a base $\beta > 1$ that depends on the algorithm and on the mode ratio $c$, for instance
$\beta = 1.82$ for single photons at $m = 2n$ (Clifford and Clifford write it
$\rho(c)$~\cite{CC2020}). Engineering
and hardware change the speed by a constant factor $X$, which moves the frontier by
\begin{align}
  \Delta n^* = \frac{\ln X}{\ln\beta}.\label{eq:deltan}
\end{align}\label{sec:deltan}%
With $\beta = 1.82$ for single photons at $m = 2n$, code twice as fast moves $n^*$ by $+1.2$, eight GPUs
by $+3.5$ (verified by direct timing), $16{,}384$ GPUs of an exascale supercomputer by $+16.2$, and reaching
$n^* = 100$ by hardware alone would take $\approx 10^{16}$ GPUs. Lowering $\beta$ takes a better algorithm
(Sec.~\ref{sec:fockalg}).
To separate the three factors, we change one at a time for single photons at $m = 2n$. Replacing
Clifford and Clifford's 2018 algorithm by their 2020 one, with the same Triton kernels on one H100,
raises $n^*$ from $32.5$ to $35.7$. Rewriting the 2020 algorithm as an optimised CUDA kernel raises it
to $38.3$. Running that kernel on eight H100 instead of one adds another $3.5$. When we compare our code with another
library, the two usually differ in algorithm and engineering at once, so we report only the overall
speed ratio and do not say which factor causes it (Supplementary Sec.~\ref{sec:nogpu}).

\paragraph{Which devices the frontier covers.}\label{sec:classes}\label{sec:axes}\label{sec:lowdepth}
A device can use its stream of photons in two ways. It can route them into separate spatial modes,
so that each shot sends all $n$ photonic inputs through the interferometer at once; we call this
\emph{demultiplexed}. Or it can process them \emph{tick by tick}, one time step after another, as a
loop or an emitter train does; the simulator then carries a state from one tick to the next, the
\emph{register}, and the cost grows with the size of that register rather than with $n$.
Independently, the interferometer is universal and full-depth, a network of loops acting at each tick,
or low-depth with only local couplings. For low-depth circuits, the treewidth sampler of Oh \emph{et al.} has an exact cost that grows as
$2^{O(D/c)}\,\mathrm{poly}(n)$ at depth $D$ and $c = m/n$~\cite{OhTreewidth2022}, which is polynomial
in $n$ at fixed depth, so there is no frontier in $n$. Calibrated to our measured speed, it
beats our chain-rule samplers below about $12$ to $17$ layers at $n = 38$ and $m = 76$, which is $15$
to $25\,\%$ of the depth of a universal interferometer (Supplementary Sec.~\ref{supp:depth}). We
measure the first two kinds of interferometer and label every result with its kind.

\paragraph{The statistic.}\label{sec:statistic} The cost per sample is a truncated expectation,
\begin{align}
  t_S \coloneqq \sum_{N \le N_S} P(N)\,\E[\mathrm{cost} \mid N],\label{eq:tS}
\end{align}
with $P(N)$ the exact law of the detected photon number $N$ and $N_S$ the smallest $N$ with
$P(N > N_S) \le 1/S$. The conditional mean $\E[\mathrm{cost} \mid N]$ is measured wherever at least
five shots have $N$ detected photons, and taken from a fit $\log t = a + bN$ at larger $N$; for every
reported value we record what fraction of $t_S$ comes from the fit. Photon numbers above $N_S$ are left
out, yet a run of $S$ samples contains at least one of them with probability up to about $1 - 1/e
\approx 63\,\%$, so $t_S$ is not the average cost of a real run of $S$ samples. For single-photon and emitter inputs, $t_S$ agrees with the plain
mean to within $0.16$ in $n^*$ at every $S$ we tried.
For squeezed inputs, whose photon number has no cap, the result depends on $S$: $n^* = 41.5$,
$37.6$ and $35.7$ at $S = 10^4$, $10^6$ and $10^8$ ($m = 2n$, calibrated timing model). We quote
$S = 10^4$; the full mean is dominated by shots about four times brighter than a typical one. The
squeezed value of $41.5$ is therefore a modelled value of this truncated statistic, and refers to
$t_{10^4}$ throughout the paper.

\paragraph{Parity across rows.}\label{sec:parity} Comparing the frontiers of two inputs, such as
single photons and emitter photons, is meant to show how much harder one kind of light is to
simulate. If one sampler were less optimised than the other, part of the gap would come from the code
rather than from the physics. Every quoted row therefore satisfies four conditions:
\begin{enumerate}
  \item it runs the fastest algorithm we have for its input;
  \item it runs on GPU kernels optimised to the same level as those of the other rows
  (Sec.~\ref{sec:primitive});
  \item it is timed the same way; and
  \item it is measured with the same statistic, $t_S$.
\end{enumerate}
A result that fails any of these conditions is said to be \emph{outside the parity audit} and is
excluded from every comparison between inputs. It is marked with a hollow symbol (\hollow) in figures
and supplementary tables, and with a footnote in Table~\ref{tab:modalities}. The audit (Supplementary Sec.~\ref{supp:kernels}) records, row by row, the
algorithm, the code path, whether the number is measured or modelled, the arithmetic and the run
stamp, so that a reader can check which part of a difference between two results comes from the algorithm. Scans within
one input, such as loss or coherence, run in a single sampler and need no such check. Every sampler
is also validated against a separate reference simulator that shares no code with it, agreeing
within sampling noise, or to $10^{-13}$--$10^{-16}$ where probabilities are compared directly (Supplementary Sec.~\ref{supp:validation}); the full
protocol, written as a proposed shared standard, is Supplementary Sec.~\ref{supp:protocol}.\label{sec:validation}

\paragraph{Conservative and provisional results.}\label{sec:provisional} The frontier is a minimum
over the algorithms we \emph{evaluated}, so a low $n^*$ can come from the physics or from a gap in
our algorithms, and the measurements alone cannot tell which. We call a result \emph{conservative}
when the frontier changes smoothly and in one direction as a physical parameter varies, and
\emph{provisional} when it jumps or reverses, which points to the algorithm rather than the physics. Even a
smooth frontier could be lowered by a better algorithm. Coherence is an example of the first kind (Sec.~\ref{sec:coherence}), and the phase of a
non-linear site of the second (Sec.~\ref{sec:nonlinear}).

\section{The framework and its samplers}\label{sec:algorithms}

All the samplers measured in this paper live in one library and follow one idea, the chain rule
of Sec.~\ref{sec:frontier}: they draw a sample step by step from conditional probabilities. They
differ only in how they compute those probabilities for each kind of light, and that is where their
costs differ. This section introduces the library and the devices it covers, then each sampler in
turn. The derivations are in the companion paper~\cite{Methods}, and the implementation details in
Supplementary Sec.~\ref{supp:samplers}.

We introduce \texttt{photonic-engines}, a Python library built on PyTorch, with GPU kernels written
in CUDA and in Triton, a language for writing GPU kernels. All its engines share one description of input states, output spaces and
imperfections, so uniform loss and partial distinguishability are modelled the same way for every
input. Setting the transmission to $\eta = 1$ gives exactly the lossless result for every engine,
which our tests check. Besides the samplers, the library contains strong simulators, which compute
the full output distribution for single-photon, superposed, emitter and squeezed inputs
(Supplementary Sec.~\ref{supp:slos}), and the estimator of a companion paper for emitter
inputs~\cite{Estimator}. Each engine is checked against a separate simulator that shares no code with
it (Supplementary Sec.~\ref{supp:validation}). Every frontier value is produced by one timing script
from a fixed random seed, and the record of that run is listed for every number in the audit of
Supplementary Sec.~\ref{supp:kernels} (see also Supplementary Sec.~\ref{supp:software}). Where each
number is quoted, we say whether it is measured, comes from a calibrated model, or is a bound.

A photonic device is specified by three choices (Table~\ref{tab:modalities}). The \emph{source}
emits single photons, vacuum--single-photon superpositions, squeezed light, or the entangled photon
train of a quantum emitter. The \emph{interferometer} is universal and used once per shot, universal
and used tick by tick, or a network of time-bin loops. \emph{Additional elements} between sources and
detectors add feed-forward from mid-circuit detection or single-mode non-linear phases. Every cell marked \CIRCLE\ in
Table~\ref{tab:modalities} has a sampler in \texttt{photonic-engines} that is validated and
timed here; Supplementary Fig.~\ref{fig:modalities} shows the full grid with its circuits.

\begin{table}[t]\centering\small
\setlength{\tabcolsep}{4pt}
\caption{\textbf{Modalities covered.} Rows: the light source that feeds each photonic input.
Columns: the interferometer and the additional elements. \CIRCLE: a sampler is implemented, validated
against a separate reference simulator and timed; the number is its frontier at one sample per second
on one H100. For demultiplexed devices this is $n^*$ in photonic inputs at $m = 2n$; for devices used
tick by tick it is the number $k$ of emitters, or the number $N^*$ of photons that pass through a loop
of delay $L$. \LEFTcircle: no separate sampler is needed, because the feed-forward and non-linear
samplers accept any input that the demultiplexed samplers handle; not timed separately. ---: nothing
is carried from one tick to the next, so each tick is one shot of the first column. Out of scope:
threshold detection, multimode squeezers, fusion- and measurement-based optics, and non-linear
resources beyond single-mode phases.}\label{tab:modalities}
\renewcommand{\arraystretch}{1.25}
\newcommand{\stk}[1]{\renewcommand{\arraystretch}{1}\begin{tabular}[b]{@{}c@{}}#1\end{tabular}}
\begin{tabular}{@{}llccccc@{}}
\toprule
& & \multicolumn{3}{c}{interferometer} & \multicolumn{2}{c}{additional elements}\\
\cmidrule(lr){3-5}\cmidrule(l){6-7}
& & \miUshot & \miUtick & \miLoops & \miFF & \miNL\\
\multicolumn{2}{@{}l}{source} & \stk{universal,\\per shot} & \stk{universal,\\per tick} & loops$^{e}$ & feed-forward & \stk{non-linear\\phase}\\
\midrule
\miFock & single photons & \CIRCLE\ $38.3$ & --- & \CIRCLE\ $N^* > 7.5{\cdot}10^4$ & \CIRCLE\ $38.3$--$33.8^{d}$ & \CIRCLE\ $36.1$\\
\miSup & superposed & \CIRCLE\ $50.0$ & --- & \CIRCLE\ $N^* = 267.4$ & \LEFTcircle & \LEFTcircle\\
\miSq & squeezed & \CIRCLE\ $41.5^{a}$ & --- & \CIRCLE\ $N^* \ge 4000$ & \LEFTcircle & \LEFTcircle\\
\miEm & emitter & \CIRCLE\ $31.0$ & \CIRCLE\ $k \approx 21$--$22$ & \CIRCLE\ $N^* > 3142^{b}$ & \CIRCLE\ $k \approx 21$--$22^{c}$ & \LEFTcircle\\
\bottomrule
\end{tabular}

\smallskip{\raggedright\footnotesize $^{a}$Modelled crossing of the truncated statistic $t_{10^4}$ (Sec.~\ref{sec:statistic}).
$^{b}$Outside the parity audit: this sampler builds its tables of photon-number sectors on the CPU, not on the GPU.
$^{c}$Emitter trains with feed-forward on the spins or on the routing: once a tick is limited by
computation rather than by launching GPU work ($k \ge 11$), feed-forward adds no time per tick.
$^{d}$$n^*$ from $K = 1$ to $K = 8$ adaptive modes: $38.3$ at $K = 1$, $38.1$ at $K = 3$, $33.8$ at
$K = 8$.
$^{e}$One loop of delay $L$. With single photons a single loop is efficiently simulable at any
delay~\cite{Novak2025,Lubasch2018}, so that cell gives the reach of our sampler, not a hardness
boundary. Values at the shortest delays measured: single photons at every delay measured and one
emitter for $L \le 4$, both lower bounds set by the largest size we searched; superposed and squeezed
light at $L = 1$. Squeezed light falls fastest with $L$ and crosses $41.5$ at $L \approx 6$; emitters
whose photons are held in loops reach $1440$ at $k = 3$, $L = 1$.\par}
\end{table}

\paragraph{The chain rule.} As in Sec.~\ref{sec:frontier}, every sampler draws a sample
$s = (s_1, \dots, s_J)$ one element at a time: either the output mode of each photon, placed one by
one, or the photon count of each mode, read one by one. It draws from
\begin{align}
  P(s) = \prod_{j=1}^{J} P(s_j \mid s_{<j}),\label{eq:chain}
\end{align}
where each conditional is a ratio of two marginal probabilities, $P(s_{\le j})/P(s_{<j})$. The
samplers differ in how they compute these marginals. For single photons, placing the photons in a
random order turns the marginal of the first $j$ photons into a sum of squared $j \times j$ permanents.
For the other inputs it becomes a sum over the components of the input state, a permanent whose
entries are matrices, or a loop hafnian, the generalisation of the permanent that describes Gaussian
states. The last steps of a sample dominate its cost, which is where the factor $\beta^n$ of
Eq.~\eqref{eq:beta} comes from. The derivations are in the companion paper~\cite{Methods}, except for
emitter trains, which are treated here.

\paragraph{Single photons.}\label{sec:fockalg} We use Clifford and Clifford's 2020
algorithm~\cite{CC2020}. It places the photons one at a time, obtains all the permanents a step needs
from one shared sum, and saves work when several photons land in the same output mode. Our CUDA kernel
runs these sums for many samples at once. To compare algorithms fairly, we reimplemented the other
published exact samplers for single photons with the same Triton kernels and timed them on the same
random interferometers. At $m = 2n$, Clifford and Clifford's 2018 algorithm reaches $n^* = 32.5$, two
other published variants $31.7$ and $30.9$, and their 2020 algorithm $35.7$; our CUDA kernel raises it
to $38.3$, and to $42.6$ at $m = n$. Computing the full output distribution first and then sampling
from it is limited by memory to $n \le 11$--$13$ at $m = 2n$ (Supplementary Fig.~\ref{fig:envelope}).
The operation counts we measure along sampled chains match Clifford and Clifford's theorem to
$0.5\,\%$, and the measured $\beta$ is $1.82$ at $m = 2n$ and $1.78$ at $m = n$.

\paragraph{Photons from quantum emitters.}\label{sec:emitteralg} A train of photons emitted by $k$
quantum emitters and sent through a universal interferometer is a matrix-product state with bond
dimension at most $2^k$~\cite{Schon2005}: the emitters' spins carry the entanglement between photons.
The sampler places the photons one at a time while keeping track of the spins, so each step needs
permanents whose entries are $2^k \times 2^k$ matrices rather than numbers. Its cost is the
single-photon cost multiplied by a number of terms that grows like $n/2$.

\paragraph{Vacuum--single-photon superpositions.} Given how many photons are detected, the input is an
equal superposition of every choice of the sources that emitted them, a Dicke state. The sampler places
the photons following a random order of the sources and keeps the saving from repeated output modes of
the single-photon sampler. It remains exact under loss.

\paragraph{Squeezed light.}\label{sec:gbsalg} Here the sampler reads the photon count of one mode at a
time. Tracing out the modes not yet read leaves a Gaussian state on the modes already read, so the
probability of their counts is a loop hafnian of a matrix built from its covariance; the probability of
the next count, given the earlier ones, is a ratio of two such loop
hafnians~\cite{QuesadaArrazola2020,Quesada2022,Bulmer2022}. We evaluate each loop hafnian with the
algorithm of Bj\"orklund, Gupt and Quesada~\cite{BGQ2019}, which sums over $2^{N/2}$ subsets of photon
pairs at $O(N^3)$ each for $N$ detected photons, and we group repeated output modes as Bulmer
\emph{et al.} do~\cite{Bulmer2022}; we call this evaluation the \emph{pair sieve}. The algorithm and its
asymptotic cost are theirs. Our implementation evaluates every photon number of the next mode in one
sieve, without a cutoff, runs it in Triton kernels, and returns a condition number with every call,
which tells us when double-precision rounding is too large and the sieve must be redone in
double-double arithmetic. Double precision meets our error criterion up to about $N = 62$ detected
photons; double-double, at $58$ to $67$ times the cost of a double-precision term, extends it to
$N \approx 141$ (Sec.~\ref{sec:precision}).

\paragraph{Non-linear sites.} A non-linear site applies the phase $e^{-i\varphi\hat n^2}$ to one mode
between two interferometers. Because the mode holds at most $n$ photons, this phase equals a finite sum
$\sum_{b=1}^{B} w_b\, e^{-i\theta_b \hat n}$ of $B$ ordinary phase shifts with complex weights $w_b$,
so the circuit is a coherent sum of $B$ linear interferometers, which we call \emph{branches}. We sample
it exactly by rejection: draw from the incoherent mixture of the branches and accept with probability
$1/\|w\|_1^2$, following the rejection principle of Dias and K\"onig~\cite{DiasKoenig2025} in its
$\ell^1$ form, with no ancilla. A sample then costs $\|w\|_1^2$ times one single-photon sample, plus
$\kappa B$ branch permanents, with $\kappa = 0.55 \pm 0.08$ measured. The factor $\|w\|_1^2$ varies
smoothly with $\varphi$, but the branch count $B$ is set by the denominator of $\varphi/\pi$: it jumps
from $1$ at $\varphi = 0$ to about $n$ for a generic small phase.

\paragraph{Feed-forward.}\label{sec:adaptivealg} $K$ output modes are measured mid-circuit, and their
photon counts choose the interferometer applied to the remaining modes~\cite{CMS2021}. The sampler
draws the mid-circuit outcome step by step from the marginals of a sampler without feed-forward. It
then draws the final counts with a chain that starts from the $r$ photons already read, and is
abandoned and redrawn whenever a later photon lands on a measured mode. For a fixed interferometer the
expected number of chains is $\E[2^r] = \per(1 + V^\dagger V)$, where $V = U_0[a, t]$ is the
$K \times n$ block of the first interferometer that connects the occupied input modes $t$ to the
measured modes $a$. The cost therefore grows with the number of photons read mid-circuit,
$\bar r = nK/m$ on average, and not with the number of measurement rounds or with the rule that chooses
the interferometer. Oh and Lim give an exact sampler for Gaussian inputs with Gaussian final
measurement~\cite{OhLim2025adaptive}; ours takes single photons with photon-number-resolving detection.

\paragraph{Devices used tick by tick.}\label{sec:trainsalg} For an emitter train through a fixed
interferometer, each tick leaves the labels of the emitters' spin states unchanged, so a sample can be
drawn tick after tick: draw the spin labels carried by the tick's photons, draw where those photons are
detected with the single-photon sampler, then update the $2^k$ amplitudes of the spin register. We
compute that update as one large matrix product on the GPU. It takes more operations than a direct
expansion, $4^k/2$ against $k3^k$, but runs at full GPU speed, which makes it faster in practice
(Supplementary Sec.~\ref{supp:trains}). In a loop
device, the state between time bins is confined to the photons inside the loop. For an input of
definite photon number, the sampler keeps only the $\binom{p+L-1}{L-1}$ amplitudes with the current
number $p$ of photons in the loop, so its cost grows linearly with the length of the train at fixed
delay $L$. An input of indefinite photon number fills every photon number at once.

All these samplers spend most of their time on one operation: computing, for many samples at once,
the permanents (or loop hafnians) of all the submatrices of one shared matrix that a step needs,
reusing work across them and across repeated output modes, in double precision. No GPU library
provides this operation, so \texttt{photonic-engines} implements it once and every sampler uses it;
this shared kernel puts the samplers at the same engineering level
(Sec.~\ref{sec:parity}).\label{sec:primitive}

\section{The frontier for demultiplexed inputs}\label{sec:measured}\label{sec:demux}

When the photon train is demultiplexed into space, all four inputs reach one sample per second on one
H100 at a few tens of photonic inputs: $31.0$ for emitter photons, $38.3$ for single photons, $41.5$ for
squeezed light and $50.0$ for superpositions, at $m = 2n$. The differences come from how many photons
each input carries and from the overhead of its sampler. Changes to a single-photon device move the
frontier by very different amounts: replacing lossy single photons by coherent superpositions of the
same brightness lowers it by $14.2$, one non-linear site by $2.2$, and feed-forward on up to three
modes by at most $0.2$. Section~\ref{sec:fourinputs} compares the four inputs at nine mode ratios, and
Secs.~\ref{sec:coherence} to~\ref{sec:adaptive} treat coherence, non-linear sites and feed-forward.

\subsection{Four inputs at nine mode ratios}\label{sec:fourinputs}

\begin{table}[t]\centering\small
\caption{\textbf{The frontier at nine mode ratios.} $n^*$ in photonic inputs at one exact sample per
second on one H100, for universal full-depth interferometers with $m/n$ modes per photonic input,
$t_S$ with $S = 10^4$, and no post-selection. The emitter row is one quantum emitter producing a linear
cluster state, a chain of photons each entangled with its neighbours. Bold: values quoted in the text.}\label{tab:ratios}
\setlength{\tabcolsep}{4.5pt}
\begin{tabular}{@{}lccccccccc@{}}
\toprule
$m/n$ & 1 & 1.5 & 2 & 2.5 & 3 & 4 & 6 & 8 & 12\\
\midrule
single photons & \textbf{42.6} & 40.0 & \textbf{38.3} & 37.5 & 36.8 & 36.2 & 35.3 & 34.9 & 34.4\\
one emitter (linear cluster) & --- & --- & \textbf{31.0} & 30.5 & 30.2 & 29.8 & 29.5 & 29.2 & 29.0\\
superpositions, $\alpha = \pi/4$ & 53.1 & 51.1 & \textbf{50.0}$^{a}$ & 49.5 & 49.1 & 48.6 & 48.0 & 47.5 & 47.0\\
squeezed light, $0.5$ photons per squeezer$^{c}$ & 42.9 & 41.8 & \textbf{41.5} & 41.1 & 40.6 & 39.9 & 39.1 & 38.2 & 37.4\\
\midrule
single photons $+$ one non-linear site$^{b}$ & 40.1 & 37.4 & \textbf{36.1} & --- & 34.5 & 33.8 & 33.0 & 32.5 & 32.1\\
\bottomrule
\end{tabular}\\[2pt]
\parbox{0.95\textwidth}{\footnotesize $^{a}$ From a dedicated run; the nine-ratio scan of this row
gives $50.2$ at $m = 2n$. $^{b}$ $\varphi = \pi/2$. The $m/n = 2$ values of this row and of the
single-photon row come from one run, the other values of this row from an earlier scan. $^{c}$ From
the timing model calibrated on measurements, which reproduces $0.91$ to $1.00$ of the measured cost on
shots not used for the calibration, for the truncated statistic $t_{10^4}$; not measured directly.}
\end{table}

Table~\ref{tab:ratios} gives the frontier of the four inputs at nine mode ratios. For single photons
the measured $\beta_{\rm eff}$ follows Clifford and Clifford's $\beta(c)$ closely: $1.82$ against $1.81$
at $c = 2$. Emitter photons reach $7.3$ photonic inputs fewer than single photons at $m = 2n$, and
$5.4$ fewer at $m = 12n$. The gap comes from the extra terms the emitter sampler handles, a number that
grows like $n/2$ (Sec.~\ref{sec:emitteralg}); fitted over the measured range, it raises the effective
base to $\beta_{\rm eff} = 1.94$, against $1.82$ for single photons.

Squeezed light and superpositions reach more photonic inputs than single photons because their cost
grows with the number of photons detected, and many of their inputs carry no photon. For squeezed
light the pair sieve costs $2^{N/2}$ for $N$ detected photons; at $0.5$ photons per squeezer, a sample
at the frontier holds about $21$ detected photons on average, and only one sample in $10^4$ holds more
than $60$, against $38.3$ for single photons. Brighter squeezers lower the frontier quickly:
$n^* = 223.2$, $89.4$, $41.5$, $25.5$, $20.2$, $13.3$ and $9.7$ at $0.1$, $0.25$, $0.5$, $0.75$,
$0.9$, $1.2$ and $1.5$ photons per squeezer. Spectral impurity raises it: squeezers that emit into two
spectral modes, with a weight $\lambda_1 = 0.95$, $0.9$ or $0.8$ in the main one, reach $44.0$, $47.3$
and $54.3$.
A superposition source prepares $\cos\alpha\,|0\rangle + \sin\alpha\,|1\rangle$ and emits a photon with
probability $\sin^2\alpha$; at $\alpha = \pi/4$ this is one half, so a sample holds about $n/2$
photons. At $\alpha = \pi/2$, where every source emits a single photon, the same sampler reaches
$38.4$, as the single-photon sampler does on the same interferometers.

\subsection{Coherence}\label{sec:coherence}

Superposition sources and lossy single photons can have the same brightness, the probability
$p = \eta\sin^2\alpha$ that a source yields a detected photon, and differ only in coherence: a lossy
single photon is an incoherent mixture of vacuum and one photon. At fixed $p = 1/2$ we scan from lossy
single photons ($\eta = 0.5$, $\alpha = \pi/2$, no coherence) to coherent superpositions ($\eta = 1$,
$\alpha = \pi/4$). The frontier falls steadily, from $64.2$ to $56.8$, $54.1$, $52.4$, $51.1$ and
$50.0$ at $\eta = 0.5$, $0.6$, $0.7$, $0.8$, $0.9$ and $1$. For this sampler, coherence therefore
lowers the frontier by $14.2 \pm 0.2$ photonic inputs; the uncertainty is statistical and comes mostly
from the rare, expensive samples at the incoherent end. More than half of the drop occurs at the first
step away from zero coherence. Both ends run in the same sampler on the same kernel, and the frontier
changes smoothly and in one direction, so the result is conservative in the sense of
Sec.~\ref{sec:provisional}. Six points that fall steadily do not show that the sampler is optimal, so
they do not prove that the $14.2$ is intrinsic to coherence.

\subsection{Non-linear sites}\label{sec:nonlinear}

One site with phase $\varphi = \pi/2$ lowers the single-photon frontier by $2.2$ to $2.6$ photonic
inputs at every mode ratio (Table~\ref{tab:ratios}), and by $2.2$ at $m = 2n$, where both rows come
from the same run. This is a constant factor in cost, as the cost law $\|w\|_1^2(1+\kappa B)$ of
Sec.~\ref{sec:algorithms} predicts, not a change of exponent. Each further site lowers the frontier by
about $2.1$ more, the slope of a least-squares fit over one scan: $n^* = 36.0$, $34.3$, $30.3$ and
$21.4$ for one, two, four and eight sites, whether the sites sit in the same layer or one after
another (within $0.1$).

\begin{figure}[t]\centering
\includegraphics[width=0.62\textwidth]{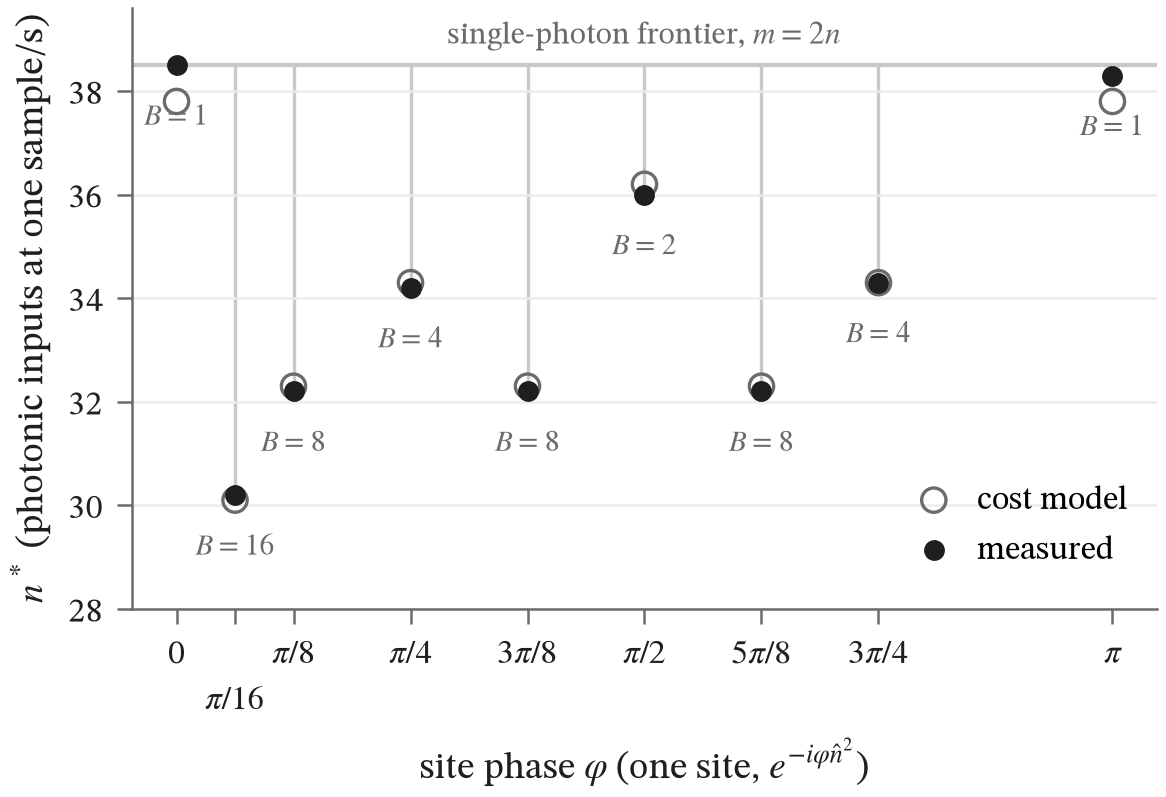}
\caption{\textbf{The frontier of one non-linear site against its phase.} $n^*$ at one exact sample
per second, $m = 2n$, one H100, measured (filled) and from the cost law $\|w\|_1^2(1+\kappa B)$
(open), with the branch count $B$ set by the denominator of $\varphi/\pi$. The horizontal line is
the scan's own $\varphi = 0$ point, a single-photon sampler in the same run.}\label{fig:nlbsphi}
\end{figure}

Against the phase, the frontier jumps and does not change in one direction (Fig.~\ref{fig:nlbsphi}).
Within one scan, $\varphi = \pi/2$ lies $2.5$ below the scan's own $\varphi = 0$ point, $\varphi =
\pi/16$ lies $8.3$ below, and $\varphi = \pi$ comes back to the $\varphi = 0$ point within timing
noise. The cost is set by the number of branches, not by the strength of the non-linearity, so this
result is provisional: the jumps come from the branch expansion the sampler uses, not from the physics
of the state.

\subsection{Feed-forward}\label{sec:adaptive}

\begin{figure}[t]\centering
\includegraphics[width=0.49\textwidth]{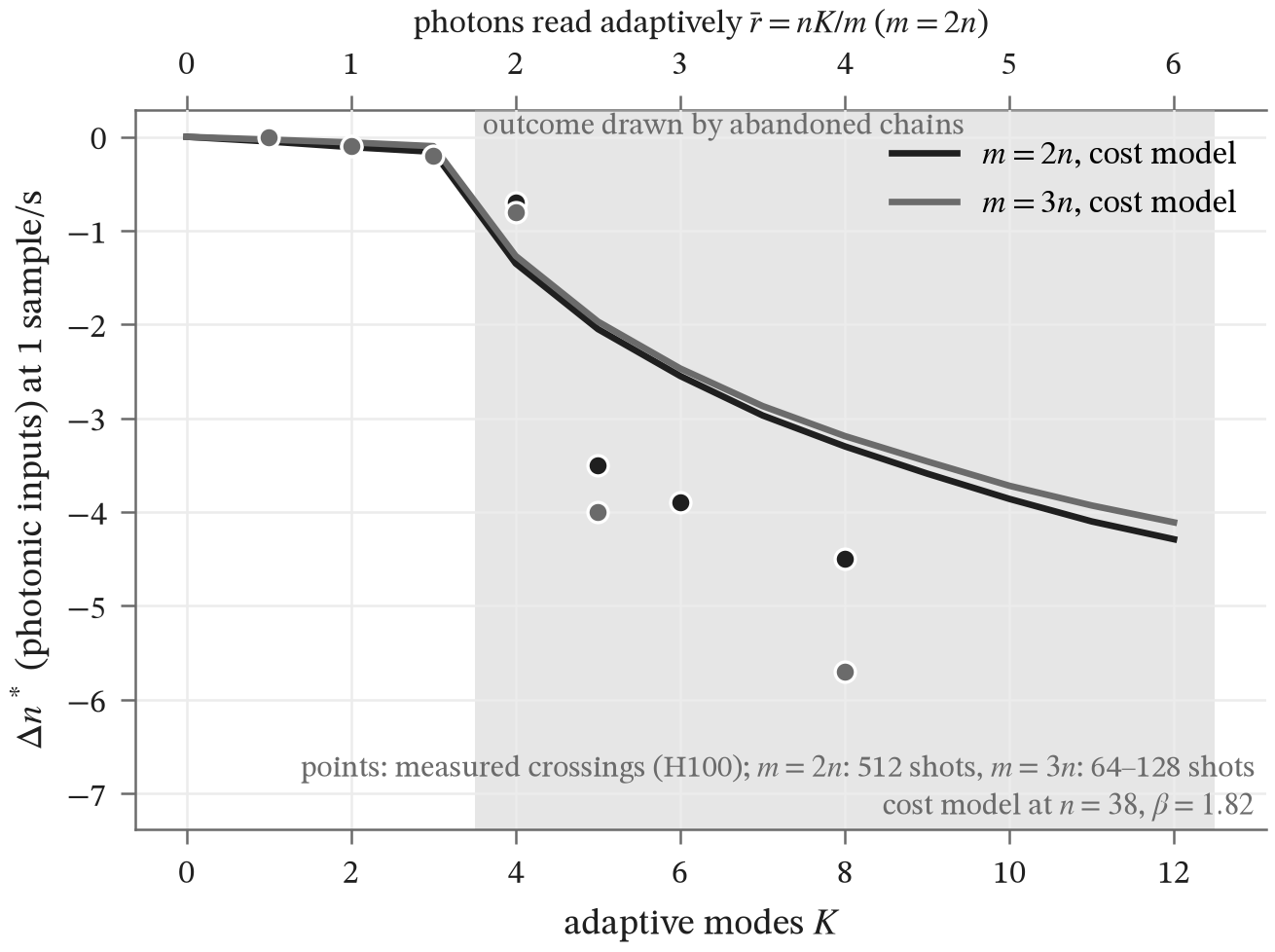}\hfill
\includegraphics[width=0.49\textwidth]{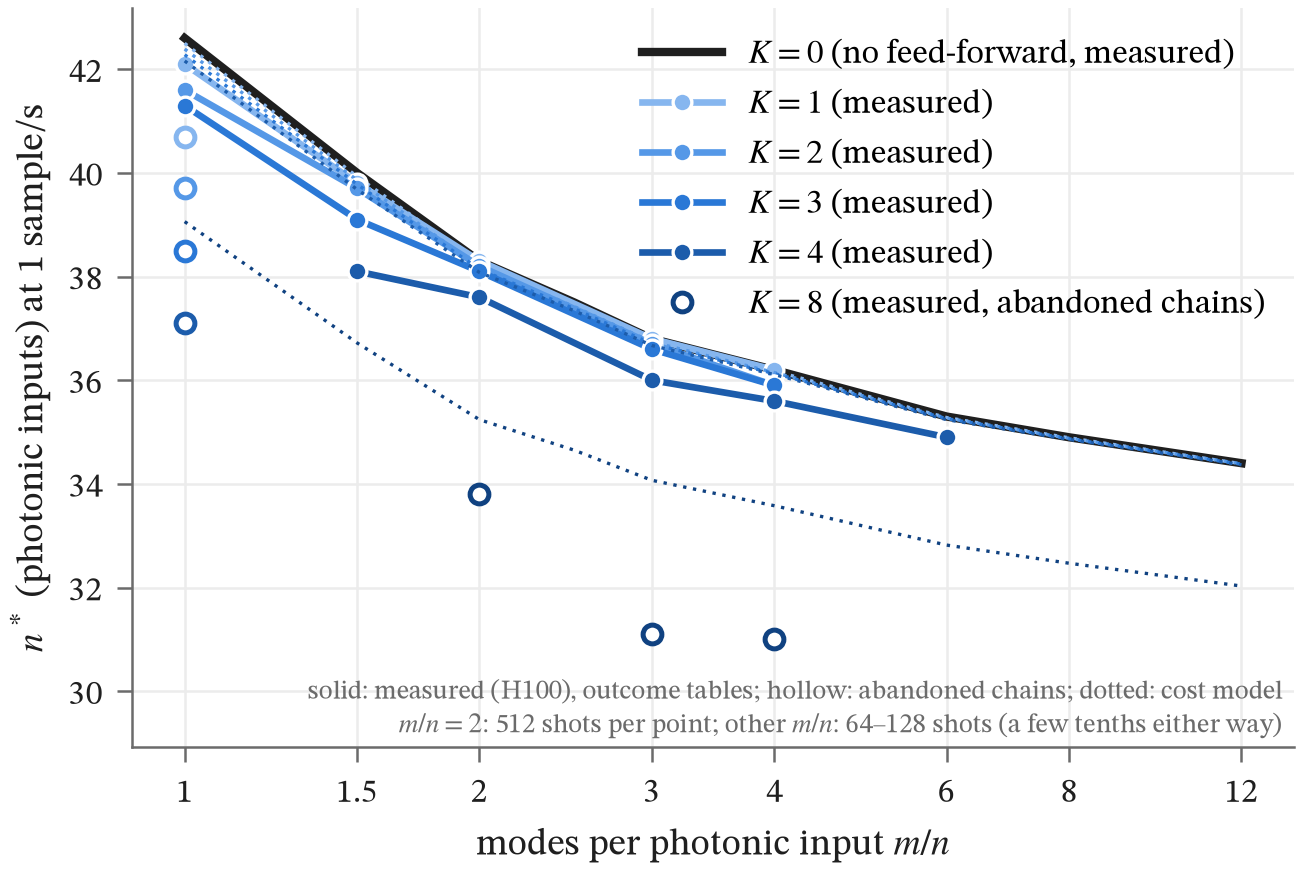}
\caption{\textbf{The cost of feed-forward.} Left: the drop of the frontier $\Delta n^*$ against the
number $K$ of modes measured mid-circuit (adaptive modes), with the mean number of photons read
mid-circuit, $\bar r = nK/m$, on the top axis; measured values and the cost model at $m = 2n$ and $3n$;
shaded, where the outcome is drawn by abandoned chains. Right: $n^*$ against the mode ratio for each
$K$. Points at $m/n \neq 2$ come from $64$ to $128$ shots and can be off by a few tenths of a photonic
input; the $m/n = 2$ values in the text come from $512$ shots.}\label{fig:adaptive}
\end{figure}

With $K$ modes measured mid-circuit and fed forward, the single-photon frontier at $m = 2n$ stays at
$38.3$, $38.2$ and $38.1$ for $K = 1$, $2$ and $3$, against $38.3$ without feed-forward. It drops to
$37.6$ at $K = 4$, and to $34.8$, $34.4$ and $33.8$ at $K = 5$, $6$ and $8$ (Fig.~\ref{fig:adaptive}).
Feed-forward on up to three modes therefore lowers the frontier by at most $0.2$, and on eight modes
by $4.5$. For $K \le 3$ the measured frontiers lie within $0.8$ of the cost model $\E[2^r]$ of
Sec.~\ref{sec:adaptivealg} at $m/n = 1$ to $4$ (fifteen measurements; thirty for all $K$ up to
$m/n = 6$). From $K = 4$ the table of mid-circuit outcomes no longer fits in memory, so the outcome is
drawn by chains that are abandoned and redrawn, as for the final counts; the frontier there is the
better of the two methods, and from $K = 5$ it lies below the cost model (Fig.~\ref{fig:adaptive},
left). When the interferometer depends on the measured counts, the number of chains has no closed
form. For interferometers drawn at random for each outcome, the exact outcome tables give about $2.0$
and $3.6$ chains per sample at $K = 1$ and $2$; measured on the GPU at $n = 30$ to $38$, the means are
$2.1$ and $4.1$ (ranges $1.8$ to $2.3$ and $3.5$ to $4.7$).

Because the cost grows as $\E[2^r]$, with $\bar r = nK/m$ photons read mid-circuit on average,
$K = \log n$ measured modes add only a polynomial overhead, and only $K \propto m$ changes the
exponent. The hardness parameters of Chabaud \emph{et al.} and of Oh and Lim, in contrast, bound
estimation, not sampling, as functions of $K$~\cite{CMS2021,OhLim2025adaptive}.

\section{The frontier for machines used tick by tick}\label{sec:pertick}

When a device processes its photons tick by tick, the simulator carries a register from one tick to
the next, and the size of that register, not the length of the photon train, sets the cost. For
emitter trains the register is the emitters' spins: about $21$ emitters reach one tick per second,
however many photons each emits. For time-bin loops it is the photon content of the loop, so the
frontier becomes a loop delay rather than a photon number. Section~\ref{sec:trains} treats emitter
trains, Sec.~\ref{sec:loops} loops, and Sec.~\ref{sec:borealis} replays a published squeezed-light
experiment.

\subsection{Emitter trains}\label{sec:trains}

Consider $k$ emitters, each sending one photon per tick into a fixed universal interferometer of $2k$
modes, with every tick detected. The cost of a sample then grows linearly with the length of the
train and exponentially with $k$: the time per tick stays constant within one percent from $4$ to
$400$ time bins at $k \le 7$, and does not change over $40$ ticks at $k = 16$ to $18$. From the second
tick on, a random interferometer leaves the emitters' spins almost maximally entangled: the rank of
the register across its middle reaches its maximum $2^{k/2}$, and the entanglement entropy comes within
about one bit of its maximum $k/2$ by the third tick, so no matrix-product compression, exact or
approximate, makes the register smaller. The frontier is therefore set by the register. The largest
register we measured, $k = 18$, takes $28\,$ms per tick and per sample; extrapolating the growth from
$k = 16$ to $18$ gives $21$ to $22$ emitters at one tick per second with the matrix-product update of
Sec.~\ref{sec:trainsalg}, and $19$ to $20$ with a direct expansion. A fixed universal machine with
about $21$ spin emitters is thus at the frontier from its second tick on, whatever the length of its
train.

How quickly the register becomes entangled depends on how far the interferometer mixes the modes. A
local interferometer of four layers of neighbouring beam splitters keeps the spins almost
unentangled: over sixteen ticks at $k = 8$ to $16$ the entanglement entropy stays near $0.1$ bit, and a
matrix-product description of bond dimension $1.6$ to $1.9$ captures $99.9\,\%$ of the state.
Feed-forward, which changes the spin gates or the routing according to earlier detections, changes
neither the entanglement of the register nor the low-order click statistics under a fixed global
interferometer, and once a tick takes longer to compute than to launch on the GPU ($k \ge 11$) it adds
no time per tick (Supplementary Fig.~\ref{fig:ff}).\label{sec:ff} Feed-forward makes the register
harder only when measurements entangle the spins directly: local optics with programmed parity
measurements, repeated until they succeed, reach maximal entanglement after eight to twelve layers.
The machine is then a spin quantum computer rather than a sampler, and a qubit simulator with the same
measurement operators is the better classical tool.

\subsection{Loops}\label{sec:loops}

\begin{figure}[t]\centering
\includegraphics[width=0.85\textwidth]{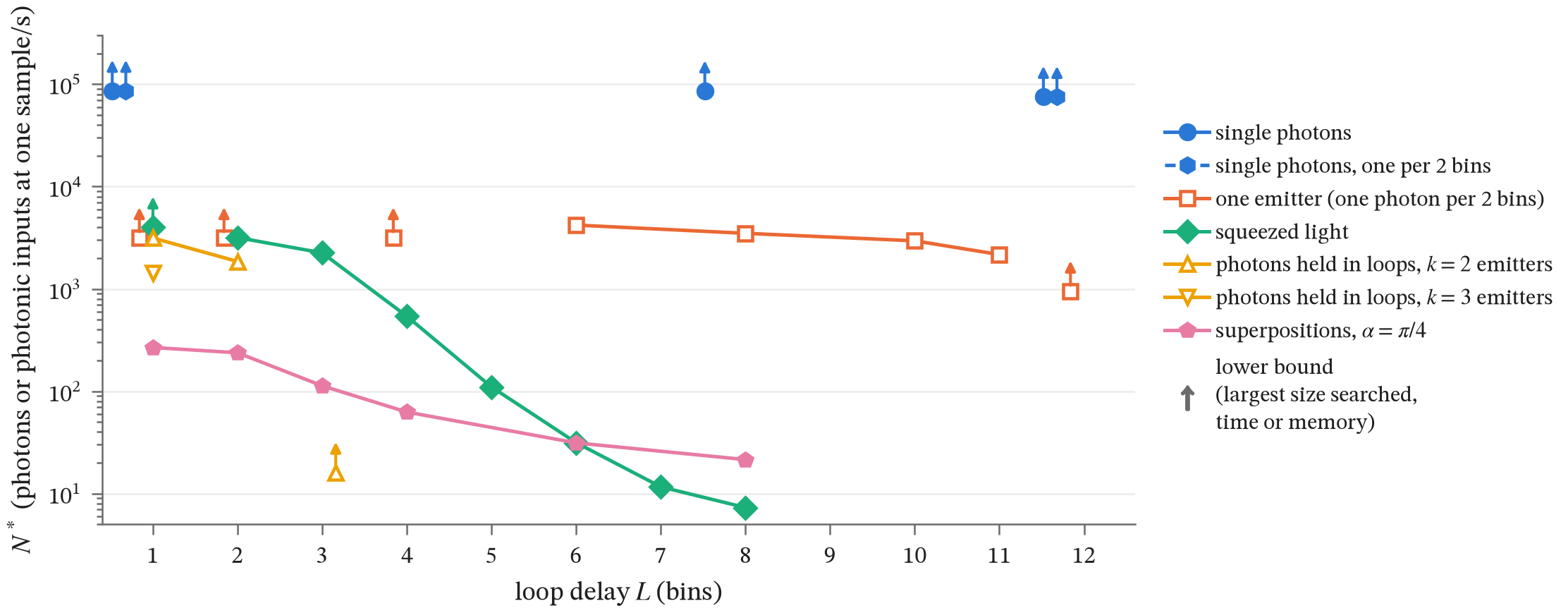}
\caption{\textbf{The loop reach per input.} $N^*$ at one exact sample per second of our sampler
against the delay $L$ of a single loop, one H100. With single photons a single loop is efficiently
simulable at any delay, so those results are reaches, not hardness boundaries. Open markers are
results outside the parity audit (\hollow). Filled markers are the single-photon, superposition and
squeezed results; superpositions fill every photon number in the loop, while the squeezed register is
truncated at a maximum photon number. In its single-photon limit ($\alpha = \pi/2$) the superposition
sampler reproduces the single-photon results within $3\,\%$ at every $L$. Arrows are lower bounds,
set by the largest size searched, the time limit of the table construction, or one GPU's memory: with
one emitter at $L \le 4$ the search stopped at $3142$ photons, still above one sample per second.}\label{fig:loops}
\end{figure}

Each loop result below is for a single loop of delay $L$ time bins, the elementary block of time-bin
loop architectures, which need several loops to be universal~\cite{Motes2014}. At fixed delay our
sampler's cost grows linearly with the length of the train, so its frontier is a loop delay rather
than a photon number (Fig.~\ref{fig:loops}). With single photons, a single loop of any delay is
efficiently simulable~\cite{Novak2025,Lubasch2018}: a sampler that draws each time bin as it leaves
the loop holds a single photon-number state between beam splitters, with $p + 1$ terms for $p$
photons at a splitter. Our single-photon loop results use this route, at a cost linear in the train
length and independent of $L$. They show how far our search went, not a hardness boundary; for single
photons, hardness requires several loops, where this shortcut fails. The shortcut does not apply to
the other inputs, whose results are the measured reach of our sampler.

With random beam-splitter settings the loop holds only a handful of photons at a time, nine to
thirteen on average at the emitter frontier. For an input of definite photon number the sampler keeps
only the amplitudes with the current number $p$ of photons in the loop, $\binom{p+L-1}{L-1}$ for $L$
slots. Single photons are sampled exactly at more than $7.5\cdot10^4$ photons per sample at every delay
measured, the largest size we searched ($10^5$ at $L = 12$). One emitter enlarges this register from
$\binom{p+L-1}{L-1}$ to $2\binom{p+L}{L}$ amplitudes, a factor $2(p+L)/L$ that depends on the loop
content but not on the length of the train; it
reaches $4204$ photons at $L = 6$ and $2166$ at $L = 11$, and from $L = 12$ building the register's
tables on the CPU, not sampling, stopped our runs, so those values are lower bounds. An input of
indefinite photon number fills every photon number at once: superpositions reach $267.4$ photons at
$L = 1$ and $21.4$ at $L = 8$, and squeezed light falls fastest with $L$, dropping below the $41.5$
photonic inputs of the demultiplexed sampler, which ignores the loop structure, at $L \approx 6$. With
a loop on each of the $2k$ rails of $k$ emitters, so that photons stay in the loops, the reach is
$1440$ photons at $k = 3$ ($L = 1$, a $12$-mode register); $k = 4$, a $16$-mode register, was not run
to its frontier, and beyond that one GPU's memory, not its speed, stops the sampler. Hardness for
devices used tick by tick therefore needs a large register \emph{and} many photons in it at
once.\label{sec:carried}\label{sec:conjunction} Full tables are in Supplementary
Sec.~\ref{supp:loops}.

\subsection{Replaying a squeezed-light experiment}\label{sec:borealis}

The $216$-mode, three-loop processor of Madsen \emph{et al.}~\cite{Madsen2022} published its
samples, so its classical cost can be measured on the run itself. For the low-brightness run
(squeezing parameter $0.53$, $1.044$ million samples, at most $53$ detected photons), drawing the same
number of exact samples from its published model takes $39$ to $44.5$ hours on one H100: $0.133\,$s
per sample when replaying the run's own patterns, and $0.154\,$s per sample for $10^4$ fresh exact
samples, in double precision. With a fitted uniform loss of $2.1\,\%$, that model reproduces the
device's photon-number distribution within the statistical noise of $10^4$ samples. This checks one
marginal, not the full output distribution, so the result concerns the published model rather than
an exact reproduction of the device. For the high-brightness run (squeezing parameter $1.10$, $124$ detected
photons on average), the median sample costs $2.4\cdot10^{16}$ H100-seconds, or $7.5\cdot10^8$
H100-years, in the double-double arithmetic its exactness requires. The authors estimated $9000$
years per sample on the Fugaku supercomputer~\cite{Madsen2022}, about $3\cdot10^{15}$ H100-seconds at
the two machines' peak double-precision rates; our figure is eight times higher, and that estimate
does not include the cost of the higher precision.
No other squeezed-light device is within reach of exact sampling; the classical simulations that have
matched squeezed-light experiments are approximate (Sec.~\ref{sec:discussion}).

\section{Loss and distinguishability}\label{sec:imperfections}

For inputs of fixed photon number, uniform loss and partial distinguishability act through one law: a
photon adds to the cost only if it is detected and indistinguishable from the others. The measured
single-photon and emitter frontiers follow this law, squeezed light does not, and imperfections make
the added elements of Sec.~\ref{sec:demux} more expensive. We first state how each imperfection is
modelled, then the law and its tests.

Uniform transmission $\eta$ commutes with the interferometer $U$, so a lossy sample is a lossless sample
on the photons that survive, a random subset whose size follows the binomial law $\mathrm{Bin}(n,\eta)$.
For partial distinguishability we use the common-overlap model: every pair of photons has the same
overlap $x$ between their internal states, and a Hong--Ou--Mandel visibility $V$ corresponds to
$x = \sqrt V$. Every sampler implements it as a mixture over which photons are distinguishable. For
single photons and superpositions the distinguishable photons are drawn as classical particles; for
emitter, feed-forward and non-linear circuits they are drawn jointly with the rest of the sample. For
squeezed light, loss changes the covariance matrix, and spectral impurity makes a shot the sum of
independent shots of each spectral mode. In feed-forward and non-linear circuits loss is applied at the
sources, before the first interferometer: it commutes with a passive interferometer but not with a
mid-circuit measurement or a non-linear site, so loss placed after those elements is a different
model, which the non-linear sampler also supports.

Take an input of fixed photon number $n$ before loss, and a cost that grows as $\beta^j$ in the number
$j$ of detected photons. Averaging over $j \sim \mathrm{Bin}(n,\eta)$ gives the expected cost of a
lossy sample,
\begin{align}
  \E[\beta^j] = (1-\eta+\eta\beta)^n.\label{eq:lossycost}
\end{align}
To leading exponential order, the frontier counted in \emph{emitted} photons is then
\begin{align}
  n^*(\eta) \approx n^*(1)\,\frac{\ln\beta}{\ln(1-\eta+\eta\beta)},\label{eq:losslaw}
\end{align}
and distinguishability enters the same formula through $\eta \to \eta x$. In words, a photon adds to
the cost only if it is detected and indistinguishable from the others.\label{sec:law}
Table~\ref{tab:loss} gives the measured frontiers. For the emitter, Eq.~\eqref{eq:losslaw} with
$\beta = 1.94$ matches every measured value within $0.3$. For single photons, with $\beta = 1.82$, it
overestimates the frontier by $0.3$ at $\eta = 0.9$ and by $2.6$ at $\eta = 0.5$; the law neglects the
polynomial prefactor of Eq.~\eqref{eq:beta}.
Squeezed light falls outside the law: its photon number is not fixed, and loss acts on its covariance,
thinning a photon-number distribution broader than Poissonian. Its frontier rises faster with loss than
Eq.~\eqref{eq:losslaw} allows, by a factor $96.7/41.5 = 2.33$ at $\eta = 0.5$, where the law allows at
most a factor of two. It is also the only input whose detected photon number at the frontier grows with
loss, from $20.7$ at $\eta = 1$ to $24.2$ at $\eta = 0.5$, because loss shortens the tail of its
photon-number distribution.

\begin{table}[t]\centering\small
\caption{\textbf{The frontier under loss.} $n^*$ in photonic inputs at $m = 2n$, one sample per second,
one H100, against the transmission $\eta$. The rows marked ``detected'' give the mean number of detected
photons at the frontier, $\eta n^*$.}\label{tab:loss}
\begin{tabular}{@{}lccccc@{}}
\toprule
 & $\eta = 1$ & $0.9$ & $0.8$ & $0.7$ & $0.5$\\
\midrule
single photons & 38.3 & 41.2 & 45.0 & 49.2 & 64.2\\
\quad detected & 38.3 & 37.1 & 36.0 & 34.4 & 32.1\\
one emitter (linear cluster) & \textbf{31.0} & 33.6 & 36.8 & 40.9 & 53.5\\
\quad detected & 31.0 & 30.2 & 29.4 & 28.6 & 26.8\\
superpositions, $\alpha = \pi/4$ & 50.0 & 54.0 & 58.3 & 64.8 & ---$^{a}$\\
squeezed light, $0.5$ photons per squeezer, $S = 10^4$ & 41.5 & 47.8 & 55.5 & 65.2 & 96.7\\
\bottomrule
\end{tabular}\\[2pt]
\parbox{0.9\linewidth}{\footnotesize $^{a}$\,Not measured: the cost per sample is too heavy-tailed to converge within the run budget.}
\end{table}

For distinguishability at $\eta = 1$, visibilities $V = 1$, $0.95$, $0.9$ and $0.8$ give single-photon
frontiers of $38.3$, $39.0$, $40.0$ and $41.8$, against $38.3$, $39.1$, $39.9$ and $41.7$ from the law,
and emitter frontiers of $31.0$, $31.6$, $32.2$ and $33.4$, against $31.0$, $31.6$, $32.2$ and $33.6$.
At $\eta = 0.7$ the same visibilities give $49.2$, $50.3$, $51.6$ and $53.7$ for single photons and
$40.9$, $41.7$, $42.6$ and $44.7$ for the emitter. Counted in detected photons, the emitter frontier lies
$7.3$ below the single-photon one at $\eta = 1$ and $5.3$ below at $\eta = 0.5$.

With imperfections, the added elements lower the frontier by more. One non-linear site lowers the
single-photon frontier by $2.2$, $2.6$ and $4.4$ at $\eta = 1$, $0.9$ and $0.7$ with $V = 1$, and by
$2.9$ and $3.7$ at $V = 0.9$ with $\eta = 1$ and $0.7$. The feed-forward sampler handles both
imperfections with no new algorithm, because loss and distinguishability only partition the input and
the feed-forward reads only photon counts. Measured in the same run as the sampler without
feed-forward, one to three measured modes lower the frontier by $0.6$ to $1.1$ at $\eta = 0.9$ and by
$0.9$ to $1.0$ at $V = 0.9$, against $0.0$ to $0.2$ without imperfections.

A transmission threshold means different things depending on how the device is
used.\label{sec:modes} If every train counts as a sample, the classical cost still grows as
$(1-\eta+\eta\beta)^n$, so the device eventually crosses the frontier at any $\eta > 0$, with $n^*$
growing as about $1/(\eta(\beta-1))$ at small $\eta$. If instead the device keeps only the trains in
which all $n$ photons are detected, its rate falls as $\eta^n$ while the classical cost grows as
$\beta^n$, so for $\eta < 1/\beta$ it never crosses: below $0.55$ for single photons and $0.51$ for the
emitter at $m = 2n$. These statements concern exact sampling, whose cost stays exponential at any
$\eta > 0$. The transmission thresholds of
Refs.~\cite{GarciaPatron2019,OszmaniecBrod2018,Qi2020} concern approximate simulation instead, whose cost
becomes polynomial once too few photons survive.

\section{Arithmetic and memory frontiers}\label{sec:companions}

Speed is not the only limit on exact simulation. Floating-point precision limits how far a sampler
stays exact, and memory limits strong simulation, which stores the whole output distribution. In
double precision our samplers meet their error criterion up to about $55$ single photons and $62$
detected photons of squeezed light, so near these sizes the number format, not the number of GPUs,
sets the frontier. Memory limits strong simulation to $13$ single photons at $m = 2n$, far below the
rate frontier.

\paragraph{Arithmetic precision.}\label{sec:precision} An exact sampler running in floating-point
arithmetic needs an error criterion. Ours is that the relative error of the probabilities computed at
each step of the chain stays below $10^{-3}$; the error of a whole sample of $n$ steps, measured as a
total-variation distance, is then at most $n$ times that of one step. We call \emph{fp64 accuracy
range} the largest photon number at which this criterion holds in double precision (fp64). For squeezed
light the error of a sieve is set by its condition number, which grows exponentially with the number
$N$ of detected photons; the criterion holds up to $N \approx 62$ in double precision and
$N \approx 141$ in double-double arithmetic, and the sampler checks it after every sieve and redoes the
sieve in double-double when it fails. For single photons we measured the true error of the production
kernel against an extended-precision version up to $38$ photons; it grows by a factor $2.35$ per
photon, which puts the limit near $55$ photons. The emitter kernel grows by $2.3$ per photon, and the
superposition rows stay at least $10^5$ below the criterion. At the frontier the per-step error is
therefore at least four orders of magnitude inside the criterion for single photons, emitters and
superpositions, about $3\cdot10^{-10}$ for single photons at $n^* = 38.3$, so a whole sample is exact to
better than $10^{-5}$. The squeezed result is closer: its largest samples, near $N = 60$, reach a
per-step error of about $2\cdot10^{-4}$ and a whole-sample bound of about $10^{-2}$; for these the timing
model uses double-double arithmetic. The guard, the measurements and the bounds are detailed in
Supplementary Sec.~\ref{supp:precision}.

Beyond this range the criterion needs double-double or quad-double arithmetic, which current GPUs do
not provide natively. For squeezed light a double-double sieve exists and costs $58$ to $67$ times a
double-precision term; for the other inputs no double-double kernel exists yet. Near this range the
frontier is therefore set by the \emph{number format}, not by the number of GPUs. For single photons,
the frontiers extrapolated to $72$ and $576$ GPUs, $n^* \approx 45.4$ and $48.9$ at $m = 2n$, still keep
the per-step error below $10^{-4}$, but from about $53$ photons the arithmetic needs error compensation,
and the criterion fails near $55$. For squeezed light the format binds earlier when $S$ is large: at
$S = 10^6$, $39\,\%$ of $t_S$ comes from $N > 64$, so the values at $S = 10^6$ and $10^8$ are
double-precision costs of calls that need double-double.

\paragraph{Memory.}\label{sec:memory} Strong simulation stores the full output distribution, after
which drawing a sample is a lookup. Its frontier is set by memory, not rate: the largest $n$ whose
distribution fits in the $80\,$GiB of one H100, stored as complex64 amplitudes and float32
probabilities, is $18$, $13$, $10$, $8$ and $6$ single photons at $m/n = 1$, $2$, $4$, $8$ and $16$.
Emitter and superposition inputs lie within $0.5$ of these values, and squeezed light reaches $7.2$
photonic inputs at $m = 2n$, with the distribution truncated so that the photon numbers left out carry
at most a probability $1/S$. Each doubling of memory adds only $0.36$ photon at $m = 2n$, so eight H100
add one photon. Between the two frontiers, from $n = 14$ to $38$ at $m = 2n$, only a chain-rule sampler
reaches the device. Supplementary Sec.~\ref{supp:slos} gives the strong simulators for every input, with
imperfections, and their validation.

\section{Devices and energy}\label{sec:hardware}

A fast, lossy device can outpace exact
classical reproduction of its output stream long before its typical sample is hard to simulate; loss
lengthens the emitter clusters needed to reach the frontier; and, for one H100, measured
GPU energy moves the threshold for an energetic advantage below twelve photons.

\paragraph{Reproducing a device's output stream.}\label{sec:competitive} The typical sample of every
current single-photon and emitter-based machine lies inside the one-hertz frontier. The high-brightness
squeezed-light run of Sec.~\ref{sec:borealis} lies far beyond it for exact sampling, though not for the
approximate samplers that have matched it. For a device, the relevant question is whether a classical
sampler can reproduce its exact output stream at the rate the device produces it. Consider a device
with $q$ sources at clock rate $f$, which emits $f/\lceil n/q\rceil$ trains of $n$ photons per second;
with transmission $\eta$, each train has $j \sim \mathrm{Bin}(n,\eta)$ surviving photons. A classical
sampler draws which photons survive and runs the $j$-photon sampler, so its cost per train is the
average of the measured time per sample over this binomial law, dominated by the rare trains in which
many photons survive. For a three-source entangled device at $1\,$GHz with $48$ photons in $96$ modes,
reproducing the whole stream exactly in real time needs about thirty H100 at $20\,\%$ end-to-end
transmission ($29$ at $20\,\%$, $580$ at $30\,\%$), while its typical train, with about ten
detected photons, lies far inside the one-hertz frontier. A two-source device at $80\,$MHz with $24$ photons in $48$ modes needs
one H100 at about $32\,\%$ transmission and eight at about $54\,\%$ (Supplementary
Sec.~\ref{supp:devices}). Such a device outpaces one GPU in the sense that the GPU cannot reproduce its
exact output distribution, including the rare large samples, at the device's emission rate; its
typical sample is not hard to simulate.

\paragraph{Emitter clusters.} Loss lengthens the cluster an emitter must produce to reach the
frontier: for one quantum-dot emitter the frontier in emitted photons is $31$, $34$, $37$, $41$ and
$54$ at $\eta = 1$, $0.9$, $0.8$, $0.7$ and $0.5$ (Table~\ref{tab:loss}), while the number detected
falls only from $31$ to $27$. Today's clusters of ten to fourteen photons are therefore seventeen to
forty-four emitted photons short of the frontier at any transmission. With $k$ emitters fed into separate modes,
each cluster can be $k$ times shorter, and the $2^k$ spin register adds no classical cost until
about $21$ emitters (Sec.~\ref{sec:trains}).

\paragraph{The energetic frontier.}\label{sec:energy} Energy per sample defines a frontier in the same
way as time per sample: the device parameters at which an exact sample costs a given energy on a fixed
machine. While the GPU is fully loaded, energy per sample is power times time, so this frontier also
follows $\beta^n$, and a budget changed by a factor $X$ moves it by $\ln X/\ln\beta$ photonic inputs,
as for rate. We measured the H100's board power while the production single-photon sampler runs at
each size, $270$ to $344\,$W, and multiplied it by the measured time per sample. At $n = 38$ and
$m = 76$ ($339\,$W, $1.29$ samples per second), an exact sample costs $7.3\cdot10^{-2}\,$Wh. This is
$7\,\%$ below the $7.8\cdot10^{-2}\,$Wh implied by the operation count of Clifford and Clifford's 2018
algorithm at $72.7\,$GFLOPS/W; our sampler runs their 2020 algorithm, which does less work. At $n = 12$
to $30$ the GPU is limited by how fast it issues samples rather than by arithmetic, and spends $2.4$ to
$58$ times that bound.

Soret \emph{et al.}~\cite{Soret2026} place the energetic advantage at $M_e^* = 15$ on their metric $M$,
an effective number of detected photons that accounts for loss, distinguishability and an approximation
tolerance. They take the classical cost as the operation count of Clifford and Clifford's 2018
algorithm at $72.7\,$GFLOPS/W, and a device energy of about $10^{-8}\,$Wh per sample. Averaging over
loss as they do, their equations give a crossing at $14.5$, consistent with their $15$, and at $16.7$
for a fixed number of lost photons, consistent with their $17$. We then change only the classical
side. At their device's $3.4$ to $3.8$ modes per detected
photon, the 2020 algorithm does two to two and a half times less work than the 2018 count, which raises
the threshold by about one photon, to $15.8$. For one H100, measured board energy lowers it below twelve: at
twelve photons the GPU already spends $2.0\cdot10^{-8}\,$Wh per sample, more than the device model at
any size in its range. At that size a sample takes $10^5$ to $10^6$ operations and the GPU is limited by
how fast it issues samples, so the threshold reflects the GPU's sampling rate at small sizes, not the
algorithm. The comparison keeps their task, modes, loss and device energy. Their device energy already
includes the fixed and cryogenic power of the device, whereas our side counts one H100 board only;
adding host, cooling and facility overhead to our side would lower the crossing further.

\section{Discussion}\label{sec:discussion}

\paragraph{What the frontier measures.} The frontier is conservative where our algorithms are mature
and provisional where they are not, and the shape of each measured curve says which
(Sec.~\ref{sec:provisional}). For a non-linear site, the jumps with the phase come from the branch
expansion our sampler uses; a chain rule that does not go through the branches would make the cost
independent of the exact value of $\varphi$. Coherence moves the frontier smoothly and in one
direction, so for this sampler the classical cost of these devices lies both in the number of photons
and in the coherence of their emission, and we have now measured the second; whether a better
sampler would reduce it is open. Approximate algorithms move the frontier further but introduce a sampling
error, and faster hardware moves it only logarithmically.\label{sec:designer}

\paragraph{Exact and approximate sampling.}\label{sec:gbslimits} The frontier can be defined for any
input state and detector, and the same code computes permanents or loop hafnians. For squeezed light
at experimental scale, however, the classical methods that matter are approximate: the classical
simulations that have matched squeezed-light experiments are approximate~\cite{Oh2024MPS,Dodd2025},
and no exact method has reached the published model of any device except the low-brightness run of
Sec.~\ref{sec:borealis}. An exact squeezed-light frontier must therefore not be read as ``safe up to
$n^*$ squeezers''; it measures the gap between exact and approximate classical sampling at each device.

\paragraph{Algorithms that use these regimes.}\label{sec:whichalgos} For the inputs considered here,
statistics in the classes covered by additive-error estimators can be estimated classically
(Supplementary Sec.~\ref{sec:estimation}). A useful algorithm beyond the frontier must therefore rest
on quantities outside those classes; which algorithms do is beyond the scope of this paper.

\paragraph{Limitations.}\label{sec:limitations}\label{sec:libraries} The frontier is measured on one
machine, for exact sampling only, with uniform imperfections and photon-number-resolving detection.
For squeezed-light devices, whose relevant classical competitors are approximate, our frontiers
therefore overstate the classical difficulty. Low-depth circuits are bounded, not measured. The
squeezed frontier comes from a timing model calibrated on measurements, which reproduces $0.91$ to
$1.00$ of the measured cost on shots not used for the calibration. Loop results beyond a delay of
eleven to thirteen bins, and multi-emitter results with photons held in loops beyond $k = 3$, are
limited by table construction and memory, and are lower bounds. The multi-GPU curves rest on three to
eight samples per point beyond $n \approx 42$. Strong simulation with feed-forward is not implemented.
Finally, much of the speed of our samplers comes from their shared GPU kernel
(Sec.~\ref{sec:primitive}), which no accelerator library provides; exploiting photon-number
conservation alone gives a factor of about $4800$ against a dense simulator for $3$ photons in a $10$-mode mesh,
before any parallelism (Supplementary Sec.~\ref{sec:nogpu}).

\section{Conclusion}\label{sec:conclusion}

The simulability frontier gives one scale on which to compare photonic devices: the size at which the
fastest exact sampler we have produces one sample per second on a given classical machine. Because the
cost of exact sampling grows exponentially with the number of detected photons, faster hardware moves
this frontier only logarithmically, whereas a better algorithm changes its growth rate.

On one NVIDIA H100 GPU, with twice as many modes as photonic inputs, the frontier lies at $38.3$ single
photons, $31.0$ photons from one quantum emitter, $50.0$ superposition sources and, from a calibrated
timing model, $41.5$ squeezed-light sources. For our samplers, coherence alone makes simulation
markedly harder: at equal brightness, coherent superpositions reach $14.2$ photonic inputs fewer than
lossy single photons; whether a better algorithm would close this gap is open. One non-linear site lowers the single-photon frontier by $2.2$, whereas
feed-forward on up to three modes barely changes it. When a device is used tick by tick, the state
carried between ticks sets the cost: about $21$ quantum emitters reach one tick per second, however
long their photon train. For inputs of fixed photon number, loss and partial distinguishability act
through one law, and near the accuracy range of double precision the number format, not the number of
GPUs, limits exact sampling.

The frontier also shows that a fast, lossy device can outpace a GPU without any of its samples being
hard to simulate, because reproducing its full output stream is dominated by rare large samples. Every
result is timed under one protocol in one open-source library, with an audit of the algorithm and code
behind each number, so these comparisons can be checked and repeated. It remains open which useful algorithms can
exploit the regimes beyond the frontier.

\section*{Acknowledgements}
The \texttt{photonic-engines} code and the manuscript were written with the assistance of generative
models from Anthropic (Claude). The algorithms, the code architecture, the benchmarking protocol, the
definitions and the analysis of the results were designed and written or reviewed by the authors, who
take responsibility for the whole.
We thank Rawad Mezher for his insightful feedback and inspiration.
This work has been co-funded by the European Commission by the Horizon-CL4 program under the grant agreement 101135288 for the EPIQUE project, by the PROQCIMA and TUF-TOPIQC programs within the French National Quantum Strategy (France 2030).

\section*{Code and data availability}
The \texttt{photonic-engines} library is open source under the MIT licence, at
\url{https://github.com/Quandela/photonic-engines}; version 0.2.1 is available from the Python Package
Index (\url{https://pypi.org/project/photonic-engines/}). The result files behind every number are in
the \texttt{results/keep/} directory of that GitHub repository, and the runs that produced them, by run stamp (Supplementary
Sec.~\ref{supp:tables}), at \url{https://github.com/Quandela/photonic-engines-runs}.

\clearpage
\makeatletter
\let\maketitle\pe@maketitle
\let\@maketitle\pe@@maketitle
\let\and\pe@and
\let\@author\pe@author
\def\@title{Supplementary Information for\\ ``The simulability frontiers of photonic quantum computing''}
\def\@date{\small \today}
\makeatother
\setcounter{section}{0}\setcounter{figure}{0}\setcounter{table}{0}\setcounter{equation}{0}
\setcounter{footnote}{0}
\renewcommand{\thesection}{S\arabic{section}}
\renewcommand{\thefigure}{S\arabic{figure}}
\renewcommand{\thetable}{S\arabic{table}}
\renewcommand{\theequation}{S\arabic{equation}}
\renewcommand{\theHsection}{S\arabic{section}}
\renewcommand{\theHfigure}{S\arabic{figure}}
\renewcommand{\theHtable}{S\arabic{table}}
\renewcommand{\theHequation}{S\arabic{equation}}
\maketitle

\section{Related work}\label{sec:sota}\label{supp:related}

\subsection{Estimation and sampling}\label{sec:estimation}

A single output amplitude of $n$ photons in a unitary $U$ can be estimated to additive error
$\varepsilon$ with $O(1/\varepsilon^2)$ samples of polynomial cost~\cite{Gurvits2005}, with
collisions~\cite{AaronsonHance2014}, and, in Lim and Oh's coherent-state form, for arbitrary input
and output states~\cite{LimOh2025}; Perceval ships a Gurvits backend~\cite{Perceval}. Pashayan,
Bartlett and Gross show that such estimation is strictly weaker than $\varepsilon$-simulation unless
the distribution is poly-sparse~\cite{Pashayan2020}. Lim and Oh's identity is general; what was
restricted to product inputs was the efficient sampling of the estimator's importance distribution.
The companion paper~\cite{Estimator} gives an exact sequential sampler for it with matrix-product-state
inputs, hence for every train emitted by $k$ spins, with a state-independent second moment, at
$O(nL\chi^2 + mn)$ per shot --- $7.5$--$80\,\mu$s on an H100 at $n = 10$--$40$.
Seron \emph{et al.}~\cite{Seron2026} show that linear statistics of boson sampling --- binned,
total-photon-number and threshold summaries --- are estimable in $O(m^2/\varepsilon^2)$ for Fock and
squeezed inputs, with loss and distinguishability; with the companion paper~\cite{Estimator} the same holds for
emitter-generated inputs. The remaining hardness is in sampling and in non-linear statistics.

\subsection{Exact samplers}\label{sec:inventory}

\paragraph{Fock inputs.} All published exact Fock samplers but one implement the same chain rule,
Clifford and Clifford's algorithm B: draw a random order of the $n$ input columns, grow the output
one photon at a time, and at step $k$ weight each candidate output mode by $|\per|^2$ of a
$k \times k$ submatrix, expanded by Laplace along the new row into $k$ cofactors. They differ in how
the $k$ cofactors are summed. C\&C 2018~\cite{CC2018} shares one row-signed Glynn sweep among them,
$2^{k-2}$ addends per step and $2^n$ per sample (Perceval/exqalibur, the CRAN
\texttt{BosonSampling} package, \texttt{BosonSampling.jl}). C\&C 2020~\cite{CC2020} collapses that
sweep on repeated \emph{output} rows, $\prod_\nu(s_\nu+1)/2$ addends, $\beta(c)^n$ per sample (their $\rho(c)$; we write $\beta$, Eq.~\eqref{eq:beta}) with
$\beta(1) = 1.69$ and $\beta(2) = 1.81$; it is our production algorithm. The generalised form of Brod
and Oszmaniec~\cite{BrodOszmaniec} signs and collapses the \emph{input} columns, $2^{k-1}$ per step
on an unbinned input (Piquasso, \texttt{theboss}); PiquassoBoost evaluates each cofactor by its own
output-collapsed Gray-code sweep; and DeepQuantum computes the full distribution,
$\binom{m+n-1}{n}$ amplitudes, before an inverse-CDF lookup. No public implementation reaches the
2020 exponent at the published work. \texttt{theboss} carries the 2020 submatrix speed-up and
multiplicity-aware Gray codes, but over the input multiplicity vector, which is unbinned for a Fock
input, so it walks exactly $2^{n+1}-4$ addends per sample (fitted base $2.024$ against ours $1.821$
at $m = 2n$ and $1.728$ at $m = n$);
PiquassoBoost reaches $\beta(c)^n$ at $k$ times the work per step, one full permanent per input
column, counted at $1.887^n$ at $m = 2n$. No wall-clock time has been published for the 2020
algorithm. These algorithms are timed against each other in Sec.~\ref{supp:envelope}. Our samplers and \texttt{theboss} agree with each other and with the enumerated law at the
shot-noise floor ($2\cdot10^5$ shots, $n \le 8$), and at $n = 10$--$14$ both reproduce the exact one-
and two-mode statistics and share the collision-type law ($10^4$ and $10^5$ samples).

\paragraph{Entangled and superposed inputs.} Two generic exact routes exist. Renema's
arbitrary-input marginal chain~\cite{Renema2021marginals} extends the Clifford--Clifford idea to any
pure input but states no runtime in the input's own parameters; Dias and K\"onig's rejection
reduction~\cite{DiasKoenig2025} is exact given component samplers and density oracles, at
$\chi\|c\|_2^2$ expected proposals, with $\chi = 2^n$ for a product of two-term superpositions.
Neither has been implemented. Dedicated samplers with a runtime in the natural parameter are given
in the companion paper~\cite{Methods}: a reduced-state chain rule with matrix Glynn cofactors for
emitter-generated inputs, and, for sources in a photon-number superposition~\cite{Renema2020SBS}, a
chain rule whose active source set starts at $n-k$ for $k$ detected photons, which makes a random
source order consistent (main text, Sec.~\ref{sec:algorithms}).

\paragraph{Squeezed inputs.} The chain rule over conditional Gaussian
states~\cite{QuesadaArrazola2020,Quesada2022,Bulmer2022} needs one loop hafnian per step; The Walrus,
Piquasso and DeepQuantum provide CPU oracles, and a CUDA chain-rule package released in 2026 reaches
$N \le 20$ detected photons. We evaluate the loop hafnians by the power-trace
inclusion--exclusion of Bj\"orklund, Gupt and Quesada~\cite{BGQ2019} with the collision grouping and
batched next-mode evaluation of Bulmer \emph{et al.}~\cite{Bulmer2022}, adding a per-call condition
number with a tolerance-driven choice of pairing and a double-double fallback, an exact all-counts
next-mode sieve, and GPU kernels (Sec.~\ref{sec:gbsalg}).

\paragraph{Per-tick machines.} Loops are simulated by a register march over the causal
cone~\cite{Novak2025}, and with single photons a single loop of any delay is efficiently
simulable~\cite{Novak2025,Lubasch2018}. Low-depth local circuits admit a treewidth sampler~\cite{OhTreewidth2022}
whose cost does not grow exponentially in $n$ at fixed depth (Sec.~\ref{sec:classes}).

\paragraph{Approximate samplers.} Metropolis independence sampling, matrix-product samplers and
cumulant emulators for Gaussian boson sampling~\cite{Oh2024MPS,Dodd2025} are excluded from the
frontier by its definition; they appear in Fig.~\ref{fig:money} as literature reference levels.

\subsection{Accelerated implementations}\label{sec:nogpu}

There is no published, benchmarked, open-source GPU exact Fock sampler.
GPU code exists for building blocks (standalone permanent libraries; Piquasso's CUDA permanent,
which its sampler does not call) and for approximate Gaussian boson sampling. The GBSKernels
library~\cite{GBSKernels} provides batched permanents, hafnians and loop hafnians on CPU and on one
GPU, with a conditional chain-rule sampler for Gaussian boson sampling that runs its chain on the host
and its hafnians on the GPU, validated in total variation against exact distributions; it publishes no
timings or sizes, and reaches Fock sampling only as a reduction. It is the only other public GPU
sampler for this task, and we ran it against ours on the instances of our squeezed row ($m = 2n$,
$0.5$ photons per squeezer, one H100, statistic $t_S$ at $S = 10^4$). Run as shipped, with its default
settings (v0.2.2, fp64, per-mode cutoff $6$, one shot per call, its fastest mode on these instances),
it draws the same distribution as ours (total variation at the noise floor for $n \le 4$) and
reaches one sample per second at about $5$ squeezed photonic inputs ($n^* \approx 4.7$--$5.0$); ours does at
$41.5$, from the timing model calibrated on measurements up to $n = 12$ ($2.9\,$ms per sample). At $6$
photonic inputs it takes $12.5\,$s per sample, against $1\,$ms for ours (Table~\ref{tab:gbskernels}). The
comparison is code against code: it does not separate the algorithm (a host-driven chain of batched
loop hafnians for GBSKernels, the pair sieve for ours) from its implementation.
\begin{table}[h]\centering\small
\begin{tabular}{lccccc}
\toprule
$n$ (squeezed photonic inputs, $m = 2n$) & $2$ & $3$ & $4$ & $5$ & $6$ \\
\midrule
GBSKernels, as shipped & $4.6\,$ms & $75\,$ms & $0.25\,$s & $1.7\,$s & $12.5\,$s \\
\texttt{photonic-engines} & $0.24\,$ms & $0.39\,$ms & $0.59\,$ms & $0.74\,$ms & $0.98\,$ms \\
ratio & $19$ & $190$ & $420$ & $2{,}300$ & $13{,}000$ \\
\bottomrule
\end{tabular}
\caption{Seconds per sample ($t_S$, $S = 10^4$) of the two public GPU samplers for Gaussian boson
sampling on the same instances and H100. With $26$--$256$ shots per size, most of the GBSKernels
statistic at $n \ge 3$ rests on the fitted tail of the cost per photon number; the plain mean over the
shots drawn gives the same crossing to within $0.3$ photonic inputs.}\label{tab:gbskernels}
\end{table}
On the CPU side we ran The Walrus (v0.22.0) the same way, its public sampler with every argument at
its default, one shot per call, on the $24$ cores of the host of our H100: it reaches one sample per
second at $27.7$ squeezed photonic inputs against our $41.5$. As shipped it is not exact, and both departures favour
its timing: its default per-mode cutoff of $5$ biases the per-mode photon numbers (by up to
$6.8\,\sigma$ at $n = 3$ over $20{,}000$ shots, removed with a cutoff of $12$), and its default
maximum of $30$ photons per shot redraws every larger shot, truncating a law whose $t_S$ strata reach
$46$--$50$ photons at $n \approx 28$. Set to sample exactly --- a cutoff of $12$ and a maximum at
the largest stratum $N_S$, with the law checked against ours at $n = 3$ and $12$ --- it reaches one
sample per second at $23.8$ photonic inputs (to about $\pm 1$, the upper bracket point resting on $44$
shots). Piquasso's Gaussian sampler (v8.0.1, in its own environment with the numba release it
supports), run the same way on the same cores, is biased at its default per-mode cutoff of $5$ as
well; at a cutoff of $12$, with the law checked against ours at $n = 3$ and $12$, it reaches one
sample per second at $35.3$ photonic inputs (to about $\pm 1.5$, the upper bracket point resting on $68$
shots).
The one accelerated exact point is the four-chip engine at $n = 40$, $m = 60$, at $\approx 80\,$s per
sample~\cite{Morse2024}, against $1.0\,$s for our sampler at the same $(n,m)$: a factor
$\approx 80$, or $+7.6$ photons at $\beta_{\rm eff} = 1.78$.
That implementation is not public
and its arithmetic is fixed point, widening from 64 to 128 bits, so it is a cited point rather than a
reproduction. Permanent libraries optimise one large permanent, whereas the chain rule needs
$\approx nm$ small, correlated sub-permanents per sample: the fastest published GPU permanent kernel
computes one $40 \times 40$ permanent in $14.2\,$s on a Quadro GV100, in fp64 with a relative error
of $2.7\cdot10^{-7}$ at $n = 40$ and $1.5\cdot10^{-4}$ at $45$ (a double-quad variant costs $10\,\%$
more and reaches $10^{-10}$)~\cite{SUperman}, while the chain rule draws a $40$-photon exact sample
in $1.0$--$1.3\,$s on an H100; the GPUs and tasks differ, so this illustrates the shape of the
workload rather than a head-to-head comparison. For time-bin loops we benchmarked CUDA-Q's photonics
simulation target (\texttt{orca-photonics}, CUDA-Q release 0.16.0), a CPU-based double-precision
simulator for state-vector simulations of photonic qudit circuits. On mesh and loop circuits of eight
to ten modes our GPU sampler is two to four orders of magnitude ($176$ to $10{,}939\times$) faster
than this CPU target. The difference comes from what each code represents. A dense qudit state
vector holds $\mathrm{level}^m$ amplitudes, and its build cost follows that (fitted exponent $1.009$,
$R^2 = 0.970$). Photon-number conservation confines the physics to $\binom{m+n-1}{n}$ states, about
$4800$ times fewer at a $3{:}10$ mesh, and our sampler never forms a state: it draws each sample
photon by photon from sub-permanents of the interferometer matrix. Once the state is built, the
target's cost per sample, $0.1$--$0.3\,\mu$s, is not lower than the chain rule's, so no shot count
makes the dense-state path cheaper.
ORCA's tensor-network photonic simulator sits on an unmerged development branch of CUDA-Q in ORCA's
fork on GitHub (\texttt{photonics\_tn\_simulator\_lm}, commit \texttt{3d66214}). We built it with one change, a build
fix that quotes one CMake generator expression so that CUDA 13's two include directories stay in one
list; no simulator code is touched. Its photonic kernels run only from C\texttt{++} in library mode,
and we drove it that way on the H100 with its defaults as shipped: qudit level $n + 1$, exact
contraction (\texttt{tensornet}) and matrix-product states (\texttt{tensornet-mps}) with bond
dimension at most $64$ and cutoffs $10^{-5}$. Its samples follow the brute-force law in all $21$
cases checked. We timed one call drawing $10^4$ samples after a warm-up, for it and for our generic
single-photon sampler on the same loop and mesh circuits. It takes $0.05$ to $0.09\,$s at four
photons and $4.3$ to $6.3\,$s at six (exact contraction does not finish the six-photon mesh in
$900\,$s); at seven photons on a loop it takes $306\,$s, and at eight the
call does not finish in $900\,$s on either target. Ours takes under $1\,$ms at four, six and eight
photons, so it is faster at every size tested, by at least $77\times$ at four photons and $4800\times$
at six. Other bond caps and contraction options were not tested.
We also ran the progressive simulator of Nov\'ak \emph{et al.}~\cite{Novak2025} as shipped (commit
\texttt{941400e}) on $24$ CPU cores, on one loop of delay $L$ with $N$ single photons in $m = 2N$ time
bins. Its samples were checked against permanents up to $N = 4$; one case ($L = 3$, $N = 2$) sits
$2.0$ to $2.7$ standard deviations off in two runs and is unresolved. On one loop its state between
beam splitters is a single basis state, so its algorithm is polynomial there; after a one-time
compilation, about ten minutes at $N = 256$, its cost per sample grows as about $N^{1.8}$, from its
implementation, and barely depends on $L$, while ours grows about linearly in $N$ and rises with $L$.
It draws one sample at a time; drawn as one batch of $10^4$ samples, ours is faster at every size and
delay tested, by $10$ to $290\times$ at $N = 96$--$128$ (the least at $L = 8$, the most at $L = 1$),
and at $N = 256$, in batches of $64$, by $5.1$, $1.9$ and $1.46\times$ at $L = 1$, $6$ and $8$. In
batches of $64$ at $N = 96$--$128$ it would be faster for $L = 6$--$8$, by up to $1.6\times$. As shipped it
cannot run $1000$ modes or more, a limit on the length of the numba tuples in its implementation rather
than of its algorithm, so it stops at $N = 499$, while its cost is still under one second per sample.
On two loops of delays $1$ and $2$, its multi-loop network against ours on the same delays, ours is
faster at every size tested. On three loops of delays $1$, $5$ and $25$ the two codes simulate
different instances below $13$ photons, and neither returned a sample at $13$, so there is no
comparison there. The two codes run
the same architecture at the same sizes but with different settings and boundary conventions, so this
is a reference, not a parity row.

\paragraph{Coverage of public engines.} Table~\ref{tab:coverage} lists every public engine we know
that samples one of our modalities, with the evidence behind its row: \emph{measured} (run as
shipped on our instances, the law checked first, the same statistic) or \emph{analysed} (its
algorithm read from its source or documentation and its cost taken from our implementation of the
same algorithm on our substrate, Supplementary Table~\ref{tab:T1}). An analysed row bounds the
algorithm, not the code. In summary: for the emitter, superposed, adaptive and non-linear inputs we
know of no public sampler other than full strong simulation of the output distribution; the public
single-photon algorithms, reimplemented on our substrate, fall below the production sampler; the
public squeezed-light samplers reach one sample per second at about $5$ squeezed photonic inputs on the GPU as
shipped and, set to sample exactly on $24$ CPU cores, at $24$ and $35$, against our $41.5$; and for
loops the released CUDA-Q target is slower at every size tested, ORCA's tensor-network simulator on
an unmerged CUDA-Q development branch is slower at every size tested, and the progressive simulator of
Nov\'ak \emph{et al.} on $24$ CPU cores is slower than ours at every size and delay tested on one
loop, on runs of $10^4$ samples, and at every size tested on two loops of delays $1$ and $2$; on
three loops of delays $1$, $5$ and $25$ the two codes simulate different instances below $13$
photons and neither returned a sample at $13$, so there is no comparison.
Low-depth circuits, where tensor-network and treewidth methods are the
right tools~\cite{OhTreewidth2022}, are outside the comparison (Sec.~\ref{sec:classes}).

\begin{table}[H]\centering\scriptsize
\caption{\textbf{Public engines by modality.} One H100 for ours; the hardware of each external
measurement is stated. $n^*$ at one sample per second, $m = 2n$; for an analysed row, $n^*$ is its
algorithm in our implementation (Supplementary Table~\ref{tab:T1}). Each modality opens with our
reference value.}\label{tab:coverage}
\setlength{\tabcolsep}{3pt}
\begin{tabular}{@{}L{2.0cm}L{3.0cm}L{5.0cm}L{5.2cm}@{}}
\toprule
modality & engine & evidence & result\\
\midrule
single photons & \textbf{ours} & measured, C\&C 2020 on the GPU (run 20261009\_141132) & $\mathbf{n^* = 38.3}$\\
 & Perceval & analysed: C\&C 2018 (documentation) & $n^* = 32.5$\\
 & theboss (CPU) & analysed and measured: input-side collapse (source); same law as ours and as brute force; addends per sample counted, $3.6$--$8.8\times$ ours at $n = 4$--$12$ & $n^* = 31.7$; as shipped $0.03$--$0.56\,$s per sample at $n = 10$--$14$\\
 & Piquasso & analysed: input-side collapse, contributed from theboss (source) & $n^* = 31.7$\\
 & PiquassoBoost & analysed: one sweep per cofactor (source) & $n^* = 30.9$\\
 & BosonSampling.jl, CRAN BosonSampling & analysed: C\&C 2018 (source; the optimised entry points of the Julia package are disabled) & $n^* = 32.5$\\
 & Lightworks (CPU) & analysed, v2.3.5: enumerates the full distribution, one permanent per output pattern (The Walrus), then samples & $n^* \approx 6$--$7$\\
 & DeepQuantum (GPU-capable) & analysed, v4.5.0: enumerates the full distribution, batched Ryser permanents, then samples & $n^* \approx 8$; at most $10$ in $80\,$GB\\
superposed, feed-forward & \textbf{ours} & measured (runs 20261007\_090041; 20261007\_022012, \_061124) & $\mathbf{n^* = 50.0}$ superposed; $\mathbf{38.3}$--$\mathbf{33.8}$ with $K = 1$--$8$ adaptive modes\\
 & Perceval & analysed: strong simulation of the output distribution, then sampling & memory-bound at $n = 13$ in $80\,$GiB\\
squeezed & \textbf{ours} & timing model calibrated on measured terms (Sec.~\ref{sec:statistic}) & $\mathbf{n^* = 41.5}$ (modelled crossing of $t_{10^4}$)\\
 & GBSKernels (GPU, H100) & measured, as shipped (run 20261011\_105434) & $n^* \approx 4.7$--$5.0$\\
 & The Walrus (24 CPU cores) & measured, exact setting and as shipped (runs 20261011\_131100, \_123902) & $n^* = 23.8$ exact; $27.7$ as shipped, not exact\\
 & Piquasso (24 CPU cores) & measured, per-mode cutoff $12$ (run 20261011\_150903) & $n^* = 35.3$; the default cutoff of $5$ is biased\\
loops & \textbf{ours} & measured, one loop of delay $L$, $T = 2N$ (Sec.~\ref{supp:loops}) & single loop, single photons: efficiently simulable~\cite{Novak2025,Lubasch2018}; our reach $\mathbf{N^* > 7.5\cdot10^4}$ at every delay measured (the search cap)\\
 & CUDA-Q photonics target \texttt{orca-photonics} (CPU, double-precision state vector of photonic qudit circuits) & measured, release 0.16.0 (run 20260923\_224734) & $176$--$10{,}939\times$ slower than ours, at $8$--$10$ modes\\
 & progressive simulator of Nov\'ak \emph{et al.}~\cite{Novak2025} (24 CPU cores) & measured, as shipped (commit \texttt{941400e}, runs 20261011\_163425, \_202121, \_230230; ours 20261012\_072530); one loop, $m = 2N$; steady cost after its one-time compilation & $10$--$290\times$ slower than ours at $N = 96$--$128$ (batches of $10^4$), $1.46$--$5.1\times$ at $N = 256$ (batches of $64$); runs to $N = 499$ (implementation limit)\\
 & tensor-network simulator, unmerged development branch\tablefootnote{Branch \texttt{photonics\_tn\_simulator\_lm} of \url{https://github.com/orcacomputing/cuda-quantum}, commit \texttt{3d66214} (2026-02-18).} & measured, H100, C\texttt{++} library mode, defaults as shipped, one build fix (runs 20261012\_022037, \_022335, \_041437, \_043025, \_052852); law checked, $21$ of $21$ cases & at least $77\times$ slower than ours at $n = 4$ and $4800\times$ at $n = 6$ (runs of $10^4$ samples); $10^4$ samples take over $900\,$s at $n = 8$\\
 & DeepQuantum delay loops & analysed, v4.5.0: Gaussian states with homodyne detection only; photon counting with delay loops is refused in its code & ---\\
emitter & \textbf{ours} & measured, parity build (run 20261006\_004008) & $\mathbf{n^* = 31.0}$\\
 & GraphiQ & analysed: generates emitter states, has no sampler & ---\\
\bottomrule
\end{tabular}
\end{table}

\subsection{Prior normalisations}

Table~\ref{tab:T2} lists what each prior statement of classical cost normalises by and what it
leaves free. They differ in machine, task (Gaussian against Fock, photon-number-resolving against
threshold), accuracy semantics (exact draws, bond-truncated matrix-product states, cumulant
truncation, correlated Metropolis chains) and metric (seconds per sample, batch totals, a single
worst-case event), and there is no shared instance set. Wakaura and Tanimae~\cite{Wakaura2026} use
the neighbouring phrase ``energy-advantage frontier'' for random circuit sampling on qubits.

\paragraph{Published photon thresholds.}\label{sec:thresholds}
Quoted thresholds for classical simulability refer to different tasks and resources. For standard
boson sampling, Clifford and Clifford proposed about \(50\) photons as the threshold for quantum
supremacy, given the cost of their exact sampler and the feasibility of large
permanents~\cite{CC2018}. Neville et al.\ demonstrated an empirically validated Markov-chain sampler at \(30\) photons and
concluded that an experimental demonstration would require more than \(50\) photons. Under their
specified source-rate, transmission and mode-scaling assumptions, their supplementary analysis gives
requirements of more than \(60\) or more than \(70\) photons for their chosen quantum-supremacy
criteria~\cite{Neville2017}.
A subsequent fine-grained complexity analysis gave \(98\) photons and \(500\) optical elements under
specified conjectures and a runtime linear in the number of elements~\cite{Dalzell2020}; Clifford and
Clifford's later paper cites an earlier version of that analysis as \(90\) input
photons~\cite{CC2020}. Other modalities introduce different measures: Bulmer et al.\ simulated
Gaussian boson sampling in \(100\) modes with up to \(60\) threshold-detector clicks on a
\(\sim 100{,}000\)-core supercomputer, and with number-resolving detection generated a single
\(92\)-photon event in \(82\) minutes~\cite{Bulmer2022}, while Nov\'ak et al.\ found that the memory
of their progressive simulator exceeds a supercomputer estimate for a three-loop architecture
\((1,5,25)\) from \(80\) time bins and \(40\) photons~\cite{Novak2025}. These figures
cannot be converted into a common photon threshold: they concern different sampling guarantees,
detectors, hardware assumptions, or resource limits. The main text instead reports the sample rate of evaluated exact algorithms on specified instances
and a fixed classical machine.

\begin{table}[H]\centering\scriptsize
\setlength{\tabcolsep}{4pt}
\caption{\textbf{Prior normalisations.} What each prior statement normalises by, and what it leaves
unnormalised.}\label{tab:T2}
\begin{tabular}{@{}L{4.0cm}L{4.4cm}L{4.8cm}@{}}
\toprule
statement & normalised by & left unnormalised\\
\midrule
C\&C 2018 / 2020~\cite{CC2018,CC2020} & photons, asymptotically & machine, rate, wall-clock\\
``the boundary''~\cite{Bulmer2022} & core-seconds on a named cluster & rate, machine class, input family\\
loop memory bounds~\cite{Novak2025} & memory bits of the causal cut & time per sample, photon number\\
lossy-simulability thresholds~\cite{GarciaPatron2019,OszmaniecBrod2018,Qi2020} & transmission, asymptotically & rate, machine, accuracy semantics\\
advantage runtimes~\cite{Madsen2022} & years on a named supercomputer & one configuration only; approximate competitors\\
approximate spoofing~\cite{Oh2024MPS,Dodd2025} & samples per wall-clock hour at a bond dimension or cumulant order & exactness; the benchmark that validates it\\
energy per sample~\cite{Soret2026} & joules, one architecture, one algorithm & rate, machine, algorithm generation\\
hardware requirements~\cite{Sund2024} & component counts & rate, classical cost\\
accelerated exact point~\cite{Morse2024} & seconds per sample & one $(n,m)$ only; arithmetic; reproducibility\\
state-vector scale~\cite{DeRaedt2026} & qubits on an exascale partition & photonic structure\\
treewidth-limited exact~\cite{OhTreewidth2022} & circuit structure & rate, machine\\
``advantage frontier''~\cite{Wakaura2026} & qubit random circuits & photonics; exact sampling\\
\midrule
\textbf{this paper} & \textbf{rate, machine, instances, algorithm and code path, statistic} & one machine; exact only (Sec.~\ref{sec:limitations})\\
\bottomrule
\end{tabular}
\end{table}

\section{Samplers and protocol in detail}\label{supp:samplers}\label{supp:chainrules}

The chain rules are stated here, not proved; the derivations are the companion methods
paper's~\cite{Methods}, except for the emitter trains (Sec.~\ref{supp:trains}).

\subsection{The modality grid}\label{sec:grid}\label{sec:engines}

\begin{figure}[H]\centering
\includegraphics[width=\textwidth]{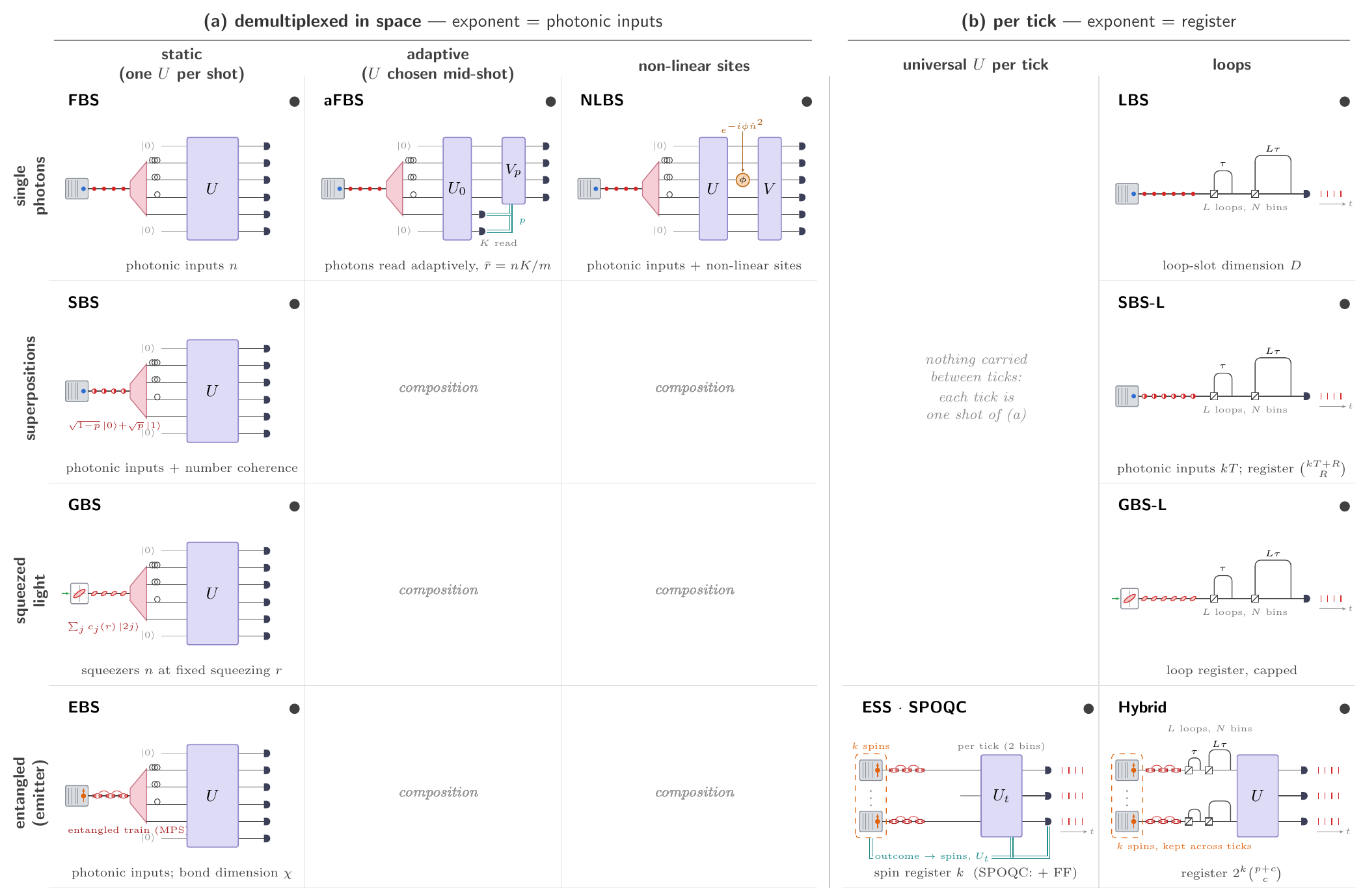}
\caption{\textbf{Modalities by usage of the photon train.} (a) Demultiplexed in space: one
interferometer per shot, the exponent is the number of photonic inputs. (b) Per tick: a register is carried
from tick to tick, the exponent is the register; the interferometer class (universal or loops) is
the second axis. Adaptive and non-linear samplers accept any input the demultiplexed chain rule
handles. Emitter photons are polarisation qubits handled as two time bins. Labels: FBS, aFBS, NLBS
--- Fock inputs, static, adaptive, with non-linear sites; SBS --- superposed sources; GBS ---
squeezed inputs; EBS --- emitter-generated entangled inputs, demultiplexed; ESS --- emitter trains
sampled tick by tick on the spin register, and with feed-forward the spin-optical
architecture~\cite{SPOQC2024}; LBS, SBS-L, GBS-L --- the three inputs through loops; Hybrid ---
emitters through loops, with carried photons.}\label{fig:modalities}
\end{figure}

A modality is an input crossed with what the circuit does between the sources and the detectors
(Fig.~\ref{fig:modalities}). Every cell either has a sampler, is a composition of other cells, or is
out of scope. Adaptive and non-linear samplers accept any input the demultiplexed chain rule handles.
The register that sets the exponent is the number of photonic inputs for Fock, superposed and emitter inputs;
detected photons for squeezed inputs; adaptively read photons $\bar r = nK/m$ for adaptivity, with
$K = \log n$ a polynomial overhead and $K \propto m$ a change of exponent; the loop delay for loops;
the emitter count $k$ for trains; and $2^k\binom{p+c}{c}$ for $c$ carried modes holding $p$ photons.
The number of stages never enters the exponent. State injection~\cite{Monbroussou2024,Monbroussou2025}
is a sub-case of adaptivity with the same parameters. Out of scope: threshold and click detection
(Gaussian device rows are labelled upper bounds), Schmidt-mode distinguishability of a squeezer, a full
treewidth sampler for low-depth circuits, fusion- and measurement-based optics, and non-linear
resources beyond single-mode phases.

\subsection{Exact Fock samplers on one instance family}\label{supp:envelope}

\begin{figure}[H]\centering
\includegraphics[width=\textwidth]{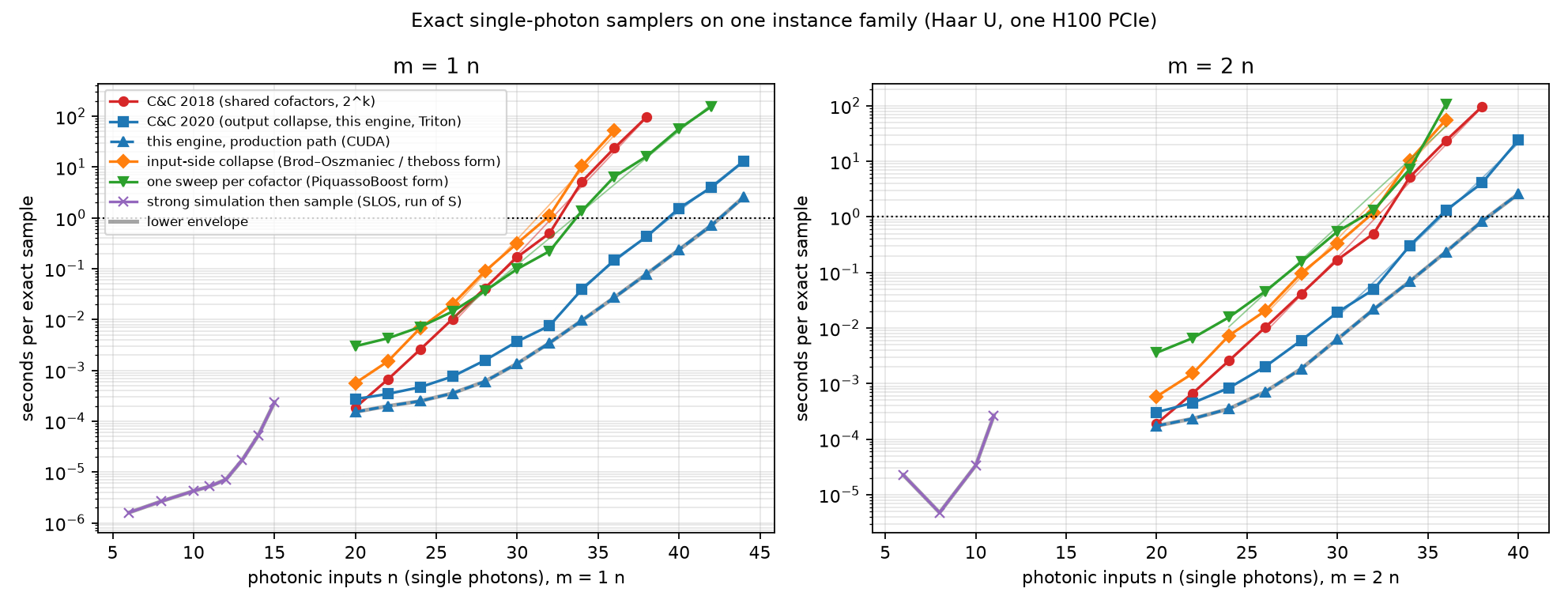}
\caption{\textbf{The algorithm envelope on one instance family.} Seconds per exact sample against
$n$ for C\&C 2018, the input-side collapse, one sweep per cofactor, C\&C 2020 on the Triton
substrate and on the production CUDA kernel, and strong-then-sample, at $m = n$ (left) and
$m = 2n$ (right); run stamps in Supplementary Sec.~\ref{supp:tables}. Each curve names its substrate.}\label{fig:envelope}
\end{figure}

Table~\ref{tab:T1} lists the algorithms and Fig.~\ref{fig:envelope} their measured cost. Every
algorithm is reimplemented in one vocabulary on the same Triton substrate; the per-addend throughput
of the four Triton kernels is the same to a factor $1.6$, and the two reimplemented forms run their
addends faster than our own Triton 2020 kernel, so no library speed-up is claimed. The Gray-code walk
is implemented as PiquassoBoost's structure ($k$ sweeps per step), the Gray-code order having the
same addend count as Glynn. Addend counts along 200 GPU-drawn chains give $1.814$ against
$\beta(2) = 1.808$ and $1.680$ against $\beta(1) = 1.6875$. The lower envelope is the production curve
above $n \approx 12$; below it strong-then-sample (a lookup after one build) wins, and the crossover
is set by memory, not rate. The production kernel is one thread per Gray lane in CUDA with tiers keyed
by $k$; the $m$ candidate weights are $m$ inner products of length $k$ followed by one draw.

\paragraph{Production sampler, in detail.} The production sampler is Clifford and Clifford's
algorithm~B with the 2020 output-side collapse~\cite{CC2020}: at step $k$ the $k$ cofactors are the
coefficients of one row-signed Glynn sum walked in Gray order and collapsed over repeated output rows,
in a CUDA kernel with one thread per Gray lane. At one sample per second ($m = 2n / m = n$), $n^*$ is
$32.5 / 32.5$ for C\&C 2018, $31.7 / 31.8$ for the input-side collapse, $30.9 / 33.7$ for one sweep per
cofactor, $35.7 / 40.0$ for C\&C 2020 on Triton and $38.3 / 42.6$ on the production CUDA kernel;
strong-then-sample is limited by memory at $n \le 11$--$13 / 15$--$18$.
The two C\&C 2020 rows run the same algorithm on the same instances and the same card: their gap of
$2.6$ photonic inputs is a constant factor of about $5$ in time per sample between the Triton and the
CUDA kernel, and since the squeezed-light sampler runs on Triton, it favours the single-photon row.
Over five Haar-random interferometers the single-photon frontier at $m = 2n$ is $38.37$ with a
standard deviation of $0.02$ (range $0.05$), so the quoted $38.3$ does not depend on the instance.

\begin{table}[H]\centering\scriptsize
\setlength{\tabcolsep}{3.5pt}
\caption{\textbf{Algorithm inventory and measured envelope.} $n^*$ at one exact sample per second,
one H100 PCIe, $m = 2n / m = n$, Haar instances. Upper block: the exact Fock samplers on one instance
family, each reimplemented in this code base; middle block: inputs beyond single photons; lower
block: prior exact routes we cite but do not implement. The companion methods paper~\cite{Methods}
gives the chain rules of the middle block; every $n^*$ in the last column is measured here. The
non-linear row's two values come from two runs (the $m = 2n$ value is the re-read of
Table~\ref{tab:ratios}).}\label{tab:T1}
\begin{tabular}{@{}L{3.3cm}L{2.9cm}L{3.0cm}L{2.3cm}l@{}}
\toprule
algorithm & cost per sample & public implementation & our treatment & $n^*$ ($2n/n$)\\
\midrule
\multicolumn{5}{@{}l}{\emph{Exact Fock samplers, one instance family}}\\
C\&C 2018~\cite{CC2018} & $O(n2^n)$, $2^{k-2}$ addends/step & Perceval/exqalibur, CRAN, \texttt{BosonSampling.jl} & reimplemented, Triton & $32.5 / 32.5$\\
generalised C\&C~\cite{BrodOszmaniec} & $2^{k-1}$/step (input-side collapse) & Piquasso, \texttt{theboss} & reimplemented, Triton & $31.7 / 31.8$\\
one sweep per cofactor & $k$ collapsed sweeps/step & PiquassoBoost & reimplemented, Triton & $30.9 / 33.7$\\
C\&C 2020~\cite{CC2020} & $O(n\beta(c)^n)$; $\beta(2){=}1.81$, $\beta(1){=}1.69$ & \texttt{theboss} (input radix, base $2.024$); PiquassoBoost (output collapse at $k\times$ work) & reimplemented, Triton & $35.7 / 40.0$\\
C\&C 2020, production kernel & same, one thread per Gray lane & --- & \textbf{ours (CUDA)} & $\mathbf{38.3 / 42.6}$\\
strong-then-sample & $\binom{m+n-1}{n}$ amplitudes, then a lookup & DeepQuantum & ours & $\le 11$--$13 / 15$--$18$\\
\midrule
\multicolumn{5}{@{}l}{\emph{Inputs beyond single photons}}\\
emitter reduced-state chain & Fock cost $\times$ a term count $\sim n/2$ & none & ours~\cite{Methods} & $31.0 / {-}$\\
offset-start source chain (superposed), with the 2020 collapse & $\approx 2^k n(m+2k)$ & none & ours~\cite{Methods} & $50.0 / 53.1$\\
conditional-Gaussian chain, loop-hafnian pair sieve & $2^{N/2}N^3$ in \emph{detected} photons & The Walrus, Piquasso, DeepQuantum (oracles); a CUDA package to $N \le 20$ & ours (Triton sieve) & $41.5 / 42.9$\\
branch-mixture rejection (non-linear site) & $\|c\|_1^2(1+\kappa B)$, acceptance $1/\|c\|_1^2$ & none & ours~\cite{Methods} & $36.1 / 40.1$\\
adaptive exchangeability chain & $\E[2^r]$ (cost model), $\bar r = nK/m$ & none & ours~\cite{Methods} & $38.3$ ($K{=}1$) \dots $33.8$ ($K{=}8$)\\
photon-number-sector register march (emitters through loops) & linear in train, $\binom{n+L-1}{L-1}$ & none & ours & $N^* = 4204$ ($L = 6$)\\
one-loop Fock chain (single photons through a loop) & linear in train, independent of $L$ & progressive simulator of Nov\'ak \emph{et al.}~\cite{Novak2025} & ours & $N^* > 7.5\cdot10^4$\\
label-diagonal pass, Glynn as GEMM & $4^k/2$ flops at GEMM speed & none & ours & $k \approx 21$--$22$\\
\midrule
\multicolumn{5}{@{}l}{\emph{Prior exact routes, cited and not implemented}}\\
arbitrary-input marginal chain~\cite{Renema2021marginals} & no runtime in the input's own parameters & none & cited as prior art & ---\\
superposition rejection~\cite{DiasKoenig2025} & $\chi\|c\|_2^2$ proposals; $\chi = 2^n$ here & none & cited as prior art & ---\\
prefix-amplitude sampler~\cite{BGL2022} & $O(g)$ amplitudes after a Fock cutoff & none & cited as prior art & ---\\
auxiliary-mode gadget (non-linear)~\cite{Spagnolo2021} & a larger linear instance plus post-selection & none public & cited as prior art & ---\\
treewidth sampler~\cite{OhTreewidth2022} & $O(Mn^42^n)$ at Haar depth; $2^{O(D/c)}\mathrm{poly}(n)$ at depth $D$ & none & bounded, not implemented & no frontier in $n$\\
\bottomrule
\end{tabular}
\end{table}

\subsection{Emitter-generated and superposed inputs}

The emitter sampler marches the reduced state of the photons already placed, with Kraus-purified
marginals and matrix Glynn cofactors ($2^k \times 2^k$ blocks). The superposed sampler, conditioned
on $k$ clicks, walks a random source order with the active set offset to $n-k$, which makes that order
consistent, and keeps the 2020 collision collapse.
The emitter sampler handles loss by the label unravelling of the trace, one term per gap; the traced
form would cost $4^{\rm gaps}$ and flattens the lossy curve. For superposed sources the marginal of the
first $j$ placed rows, conditioned on $k$ clicks, is a sum over residual sets of excluded sources of a
$\lambda$-filtered Glynn sum; loss is a Bernoulli mask (total-variation distance $2\cdot10^{-16}$). The
graded basis $\binom{m+n}{n}$ is the strong side, at $1.8\times$ the time and $1.08\times$ the memory
of the Fock kernel at $n = 11$. Against the best existing kernel the superposed row's residual is at
most $0.07$ photonic input (Sec.~\ref{supp:kernels}).

\subsection{Squeezed inputs}

Every loop-hafnian call returns its condition number, with the pairing and grouping chosen under a
round-off tolerance and a double-double fallback; one sieve covers all photon numbers of the next mode,
without a cutoff and normalised exactly.
The next mode's photon-number law is one loop hafnian evaluated as a polynomial in that mode's
Bargmann variable, so a single evaluation gives every photon number of that mode; the normaliser is a
closed-form Gaussian integral and the draw an exact inverse CDF. The pair sieve runs in Triton kernels
with gathered subset matrices and collision-aware enumeration; repeated output modes are grouped as in
Bulmer \emph{et al.}, so a pattern with collisions costs $\prod_i(r_i+1)\,2^{N_f/2}$ terms rather than
$2^{N/2}$. At $n = 40$ a shot costs $0.086\,$s. Loss is exact as a covariance map: a lossy squeezed
vacuum is a pure one at smaller squeezing plus a classical displacement, and a general lossy transfer
matrix is handled by a Williamson split of the output covariance. The sieve's condition number
$\kappa$ is set by the pairing, searched over candidates; against a double-double reference on 275
sieves the fp64 error is at most $0.74\,\varepsilon\kappa$ (median $0.031\,\varepsilon\kappa$) with
$\log_{10}\kappa \approx 0.40 + 0.197N$; double-double uses Ozaki products on the tensor cores.

\subsection{Adaptive linear optics}

$n$ photons enter $U_0$; $K$ output modes are read with photon-number resolution in $\Lambda$ levels;
after each level a classical function of the counts so far picks the unitary applied to the modes
still unmeasured~\cite{CMS2021}. Each of the $\binom{n+K}{K}$ adaptive outcomes $p$ has its own
cumulative $U^p$, and $P(p,s) = |\per U^p[\omega,t]|^2/\omega!$; $K$ enters the cost, $\Lambda$ and the
policy do not. Oh and Lim make mean values estimable for $K = O(1)$ and BQP-complete for
$K = \mathrm{poly}(n)$~\cite{OhLim2025adaptive}.

The adaptive outcome is drawn level by level: the conditionals are $K_\lambda$-mode marginals of a
non-adaptive sampler, computed from the counts' generating function $\per(1 + V^\dagger(Z-1)V)$ on a
table graded by total degree and inverted binomially. This is exact and rejection-free while the table
fits; it is $\Theta(n^K)$ ($10^8$ entries at $K = 4$, $n = 38$; $10^{10}$ at $K = 5$), and beyond that $p$
is drawn by aborted chains. The final counts given $p$ use exchangeability: the law given ``exactly
$p$'' equals the law given ``the first $r = |p|$ photons of the ordered chain land on the adaptive modes
as $p$ and no later photon does''. The chain starts with the $r$ adaptive photons in place --- their
input columns drawn polynomially from suffix tables --- and the remaining $n - r$ photons run with the
adaptive modes forbidden, aborting at the first photon that lands there, on average
$(n-r+1)/(h+1)$ steps in for $h$ offending photons. A rejection-free route is closed: conditioning on
the mode constraint turns the Fock input into a superposition of input subsets whose posterior is not
diagonal (off-diagonal mass $28$ to $56\,\%$ of the diagonal at $n = 5$), and the exact conditional
marginal then costs $4^k$ per step. With feed-forward the chain count is $\sum_p T_{U^p}[p]$ over the
outcomes that occur, $T_{U^p}[p]$ being the chains started for $p$ under the interferometer $U^p$ the
policy chose. In the regime $K \propto m$ every route here fails and only the projected state vector
remains, which is the memory frontier (Sec.~\ref{supp:slos}). Lossy and partially distinguishable
adaptive circuits are the lossless one on the surviving subset, with the policy seeing the lossy
counts, and each interfering species runs the same feed-forward as its own instance, singletons routed
classically level by level; this is exact against an enumerated mixture.

\subsection{Non-linear sites}

The random-photon-order chain does not carry over to a non-linear site (total-variation distance
$0.115$ against the exact law), and the true-marginal chain costs $4^k$ against $2^k$; the
branch-mixture rejection of the main text is therefore the sampler used.

\subsection{Emitter trains and feed-forward}\label{supp:trains}

For label-preserving emitters the Kraus operator of a tick is diagonal in the spin-label basis after
the gate. Sampling is therefore a trajectory: draw the label word, draw the pattern from the
$k$-photon Fock sampler, and update the register. The $2^k$ register amplitudes are one Glynn sum turned into a batched GEMM: $0.17 / 0.71 /
3.6 / 28\,$ms per tick-shot at $k = 12 / 14 / 16 / 18$ against $0.41 / 3.2 / 29 / 273$ with the
Laplace recurrence; multiplicity collapse inside that GEMM adds $1.4\times$ at $k = 16$ and $1.9\times$
at $k = 18$. With precomputed Kraus sets the reach is $k = 7$ ($19\,$GB; $418$ bins, about $2{,}900$
photons, at one sequence per second, measured at $n = 400$). Per-stream spin gates and unitaries chosen
from earlier outcomes are implemented and verified against branch enumeration. Under a static global
$U$ the labels enter only through $P(x)$, and $\sum_x P(x)\langle n_i\rangle_x$ is blind to $P(x)$ when
$\langle n_i\rangle_x \approx k/m$; a repeat-until-success parity brickwork with routing feed-forward
shows the same one-bit area-law plateau as the static one ($0.76$ to $1.39$ bits at $k = 4$--$16$
against a maximum of $k/2$). Three regimes follow: the Fock-sampler regime (tick one, or local optics
with no program), the emitter-sampler regime (global optics, the register is the exponent) and the
spin-computer regime. The strong side under feed-forward (conditional-state marginals, a Markov
feed-forward at $S_{\rm bin}^2\chi^3$ per stage) is not built.

\begin{figure}[H]\centering
\includegraphics[width=0.7\textwidth]{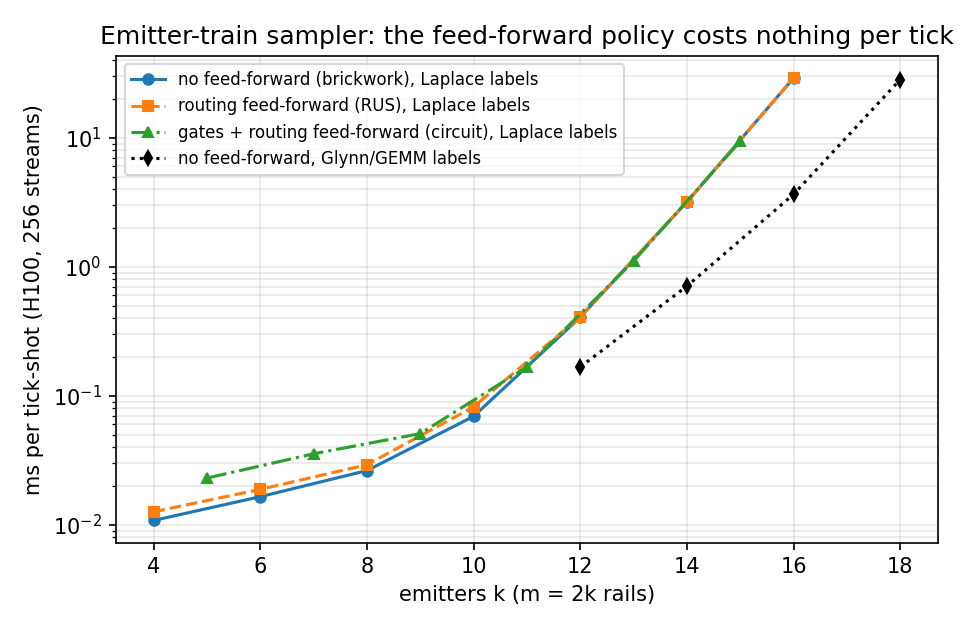}
\caption{\textbf{Feed-forward on emitter gates.} Milliseconds per tick-shot of the emitter-train
sampler against the number of emitters $k$ ($m = 2k$ rails; one H100, 256 streams), for a static
brickwork, routing fed forward, and gates and routing fed forward, all with the same label pass;
dotted, the static brickwork with the Glynn/GEMM label pass. Once a tick is compute-bound ($k \ge 11$)
the policy costs nothing per tick, ratio $1.00 \pm 0.01$ under both policies; below that, a tick is
launch-bound and the per-stream gather of the policy adds $1.1$ to $1.7\times$.}\label{fig:ff}
\end{figure}

\subsection{Loop register march}

Because emission adds one photon, the beam splitters conserve the number and detection removes what it
counts, a trajectory has a definite photon number in the loop whenever the input does. The exact
sampler carries only the register states with the current count, $\binom{n+L-1}{L-1}$ amplitudes (times
two for an emitter's spin). For an input of indefinite photon number every count up to $N$ is carried
at once and the register is the dense $\binom{N+L+1}{L+1}$: polynomial in $N$ at fixed $L$, but of
degree $L+1$.

\subsection{Low-depth circuits}\label{supp:depth}

On a depth-$D$ local mesh at $m = cn$ the treewidth is $\approx 2D/c + 1$~\cite{OhTreewidth2022}. At
Haar depth the treewidth sampler costs $O(Mn^42^n)$ against our $1.80^n$ for Fock inputs and
$O(MN^32^N)$ against our $2^{N/2}$ for squeezed inputs, so no measured $n^*$ moves. Calibrated to our
measured rate, it overtakes the chain rule below $D^* \approx 12$--$17$ layers at $n = 38$, $m = 76$, and
below $\approx 17$--$20$ at the squeezed frontier point. Demultiplexing through a loop unrolls to a
structured low-depth $U$, which is this class. Every device placed in this paper is universal or
loop-based.

\subsection{Device replay, reference machine}\label{sec:replay}

For a published experiment we set $S = S_{\rm exp}$ and stratify over that run's published samples,
since the cost of an exact sample depends on its photon pattern alone. Each call is tagged
\emph{measured} (whole evaluation timed), \emph{estimated} (term count times a measured per-term rate) or
\emph{beyond-kernel} (extrapolated), with its precision tier. The reference machine is one NVIDIA H100
PCIe: 114 streaming multiprocessors, $80\,$GB HBM2e at $2\,$TB/s, $26\,$TFLOPS fp64 ($51$ with the fp64
tensor cores, used only by the double-double Gaussian path), $350\,$W TDP and $\approx 280\,$W measured
during sampling. For $G$ GPUs, $n^*(G) = n^*(1) + \ln G/\ln\beta_{\rm eff}$ for independent samples,
verified by direct timing up to $\approx 4$ photons past the one-GPU frontier.

\section{Per-tick machines in detail}\label{supp:loops}

\subsection{Loops}

The measured reach of our sampler on a single loop of delay $L$, $T = 2N$ bins, one sample per
second, one H100. With single photons a single loop is efficiently simulable at any
delay~\cite{Novak2025,Lubasch2018}, so the single-photon rows are reaches, not hardness boundaries; the
other rows are the measured reach of our sampler:

\begin{center}\small
\begin{tabular}{@{}L{6.0cm}L{9.4cm}@{}}
\toprule
row (\hollow: outside the parity audit) & $N^*$ at one sample per second \\
\midrule
$N$ single photons, bins $0\dots N-1$ (one basis state between splitters) & $> 7.5\cdot10^4$ at $L = 1 / 8 / 12$ (the search cap; $10^5$ at $L = 12$); cost independent of $L$\\
single photons at one per two bins & $> 7.5\cdot10^4$ at $L = 1 / 12$ (the search cap)\\
\hollow one emitter, linear-cluster train & $> 3142$ for $L \le 4$ (the search cap); $4204 / 3493 / 2950 / 2166$ at $L = 6 / 8 / 10 / 11$; $\ge 943$ at $L = 12$\\
superposed sources, $\alpha = \pi/4$ (dense register) & $267.4 / 237.4 / 112.9 / 62.5 / 31.3 / 21.4$ at $L = 1 / 2 / 3 / 4 / 6 / 8$\\ %
squeezed, one per bin, $0.5$ photons each (capped register) & $\ge 4000 / 3171 / 2250 / 540 / 110 / 31.4 / 11.6 / 7.3$ at $L = 1 \dots 8$\\
\hollow $k$ emitters, a loop on each of $2k$ rails (provisional) & $k = 2$: $3143 / 1853$ at $L = 1 / 2$; $k = 3$: $1440$ at $L = 1$; beyond, one GPU's memory\\ %
\bottomrule
\end{tabular}
\end{center}

A loop is easy to simulate when its photon number is definite, because $p$ photons in $L$ slots form a register
of $\binom{p+L-1}{L-1}$ states, set by the occupancy and not by the train length, whereas an input whose
loop content is a superposition of photon numbers fills every occupancy at once, dense up to the total
photon number or capped at the source's tail. On the universal reference, a published loop processor
detecting sixteen photons in thirty-two bins sits far inside a frontier of $38.3$ photonic inputs --- one H100
reproduces its $10^6$-sample dataset about $2500$ times faster --- and for a $128$-bin successor one H100
sustains $10^4$ exact samples per second up to about twenty-one detected photons, $10^3$ per second to
twenty-five and one per second to thirty-six, against a Haar reference at $m = 128$; a banded loop
network is easier than Haar.

\subsection{Carried photons}

For $k$ emitters, each dual-rail on two rails, a loop of delay $L$ on every one of the $2k$ rails and a
static Haar mixer after the loops, the register has $2k(1+L)$ modes; it is validated against the
unrolled oracle and measured with the same photon-number sectors and the same statistic as the loop
rows, and like the emitter loop rows it is \hollow. The rows are provisional: $k = 2$ reaches
$N^* = 3143$ at $L = 1$ and $1853$ at $L = 2$; $k = 3$ reaches $1440$ at $L = 1$. Beyond these points
one GPU's memory stops the sampler: at $k = 2$, $L = 3$ the reach is $N^* \ge 16$ (out of memory from
$N = 56$), at $k = 3$, $L = 2$ a single shot runs at $24$ photons, and $k = 4$, $L = 1$ was not run
to its frontier: two shots at $36$ photons still ran above one sample per second. The largest sectors
at these points hold $1.5\cdot10^8$ ($k = 2$, $L = 3$), $6.7\cdot10^7$ ($k = 3$, $L = 2$) and
$1.2\cdot10^8$ amplitudes ($k = 4$, $L = 1$), and those cells carry no frontier value.

\subsection{Squeezed light in loops}

Our chain rule is structure-blind, yet a register on the loop slots is worse than it at a $(1,6,36)$
ladder --- $\binom{N+44}{44}$ against $2^{N/2}N^3$, $2\cdot10^{16}$ against $8\cdot10^{6}$ at $N = 20$ and
$8\cdot10^{40}$ against $10^{25}$ at $N = 125$ --- the same conclusion as the measured crossing at
$L \approx 6$, reached by counting. For the high-brightness run of the three-loop processor (maximum
$N = 219$) the per-term rates are measured to $N = 128$ and $\kappa$ to $N = 58$ (on grouped pairings to
$N = 62$), extrapolated beyond;
cross-mode grouping would bring the median to $2.4\cdot10^{13}$--$2.1\cdot10^{14}$ H100-seconds, but only
in quad-double arithmetic, for which no kernel exists, and the run's mean is carried by its brightest
few samples. The timing model reproduces the measured low-brightness time to $0.977$.
Devices on the squeezed axis, at the published mode ratio, squeezing and loss, $S = 10^4$, with
threshold-detector rows as photon-number-resolving upper bounds: the three-loop processor's settings
($m/n = 1$, $r = 1.1$, $\eta = 0.33$) give $n^* = 39.6$ against $216$ squeezers on the device; an
$m/n = 2.88$, $\eta = 0.43$ machine gives $19.1$ at $r = 1.2$ and $9.6$ at $r = 1.4$ against $50$; an
$m/n = 8$, $r = 0.9$, $\eta = 0.51$ machine gives $38.3$ against $1024$. $G$ GPUs add $\ln G/\ln\beta$
squeezed photonic inputs, about seven per decade of GPUs at the headline point.

\section{The frontier in one table}\label{sec:T3}

\begin{table}[H]\centering\scriptsize
\caption{The one-hertz frontier. One H100 PCIe, $m = 2n$ unless stated, $t_S$ at $S = 10^4$. Class:
(i) universal full-depth, (ii) loops, (iii) low-depth local. A hollow row (\hollow) is outside the
parity audit and enters no comparison.}\label{tab:T3}
\setlength{\tabcolsep}{2pt}
\begin{tabular}{@{}L{1.0cm}L{3.0cm}L{3.6cm}L{0.7cm}L{1.6cm}L{1.9cm}L{2.7cm}@{}}
\toprule
usage & row & $n^*$ (or register) & class & precision & parity & run\\
\midrule
demux & Fock, C\&C 2020 CUDA & \textbf{38.3} ($42.6$ at $m=n$) & (i) & fp64 & yes & 20261009\_141132\\
demux & Fock, the five other algorithms & $32.5 / 31.7 / 30.9 / 35.7$; strong $\le 11$--$13$ & (i) & fp64 & substrate named & 20260930\_082501 (strong: \_141657)\\
demux & Fock $+$ adaptive $K = 1 \dots 6, 8$ & $38.3$ / $38.2$ / $38.1$ / $37.6$ / $34.8$ / $34.4$ / $33.8$ & (i) & fp64 & route named from $K = 4$ & 20261007\_022012 / \_030408 / \_061124\\
demux & Fock $+$ one non-linear site & $36.1$ ($2.1$ per site) & (i) & fp64 & chain yes; amplitudes PyTorch & 20261009\_141132 (per-site slope 20260923\_080913)\\
demux & emitter ($\chi = 2$) & \textbf{31.0} & (i) & fp64 & yes (Fock limit) & 20261006\_004008\\
demux & superposed, $\alpha = \pi/4$ & \textbf{50.0} ($53.1$ at $m=n$) & (i) & fp64 & yes (Fock limit) & 20261007\_090041 / 20261010\_045711\\
demux & squeezed, $0.5$ ph/squeezer & \textbf{41.5} ($37.6 / 35.7$ at $10^6 / 10^8$) & (i) & fp64 at $S = 10^4$; dd beyond & own kernel & 20260925 model\\
per tick & emitter trains, static universal $U$ & $k \approx 21$--$22$ at one tick/s (extrapolated from $k = 18$) & (i) & fp64 & yes & 20260919\_084900\\
per tick & single photons, loop delay $L$ & $N^* > 7.5\cdot10^4$ (every $L$ measured; search cap) & (ii) & fp64 & polynomial chain & 20261016\_023739\\
per tick & one emitter, loop delay $L$ & \hollow $N^* > 3142$ ($L \le 4$), $4204$ ($L = 6$), $2166$ ($L = 11$) & (ii) & fp64 & \textbf{no} (host-side tables) & 20261016\_023739\\
per tick & superposed, loop delay $L$ & $267.4 \dots 21.4$ ($L = 1 \dots 8$) & (ii) & fp64 & yes (Fock limit) & 20261001\_101803\\
per tick & squeezed, loop delay $L$ & $\ge 4000 \dots 7.3$ ($L = 1 \dots 8$) & (ii) & fp64 & capped register & 20261002\_203546\\
per tick & squeezed, device replay, low brightness & $0.154\,$s/sample; whole run $39$--$44.5\,$h & (ii) & fp64 & measured & 20260927\_190700\\
per tick & squeezed, device replay, high brightness & median $2.4\cdot10^{16}$ H100-s & (ii) & dd & estimated & tiers 20260930\\
per tick & $k$ emitters through loops & \hollow $N^* = 3143 / 1853$ ($k = 2$, $L = 1 / 2$), $1440$ ($k = 3$, $L = 1$); memory-limited beyond & (ii) & fp64 & \textbf{no} (host-side tables) & 20261016\_023739\\
--- & low-depth local $U$ & no frontier in $n$; $D^* \approx 12$--$17$ at $n = 38$ & (iii) & --- & bounded & ---\\
\bottomrule
\end{tabular}
\end{table}

\section{Devices against a real-time classical sampler}\label{supp:devices}

Two readings share one rate model. A device with $S$ sources at clock $f$ emits $f/\lceil n/S\rceil$
trains per second. (A) Post-selected on exactly $k$ detected photons, the device delivers
$(f/\lceil n/S\rceil)\binom nk \eta^k(1-\eta)^{n-k}$ $k$-photon samples per second, each costing $t(k,m)$
classically. (B) For the whole lossy stream, a classical sampler's cost per train is the binomial
average of $t(k,m)$ under the run-of-$S$ statistic ($S$ = the trains of a $3600\,$s run). Measured cost
table: $t(2,96) = 52\,$ns, $t(16,96) = 3.2\,\mu$s, $t(24,96) = 0.42\,$ms, $t(30,96) = 14\,$ms,
$t(36,96) = 0.45\,$s, $t(38,96) = 1.3\,$s, $t(45,96) = 94\,$s.

\begin{table}[H]\centering\scriptsize
\caption{Devices, described by (sources, clock, photons / modes), against a classical sampler
reproducing their exact stream in real time. Ascella~\cite{Maring2024} is a published
device.}\label{tab:devices}
\setlength{\tabcolsep}{4pt}
\begin{tabular}{@{}L{3.0cm}lL{4.2cm}L{2.4cm}L{1.9cm}@{}}
\toprule
device (sources, clock, photons / modes) & trains/s & (B) $\eta$ to tie $1/8/64$ H100 ($\langle k\rangle$) & (A) post-selected optimum & 1\,Hz frontier\\
\midrule
1, 80\,MHz, 12 / 24 & $6.7\cdot10^6$ & $73\,\%$ (8.7) / never / never & $k=12$ at $92\,\%$ & every $k$ inside\\
2, 80\,MHz, 24 / 48 & $6.7\cdot10^6$ & \textbf{32\,\% (7.7) / 54\,\% (12.9) / 72\,\% (17.3)} & $k=15$ at $52\,\%$ & every $k$ inside\\
3 entangled, 1\,GHz, 48 / 96 & $6.2\cdot10^7$ & throughput-bound / \textbf{13.5\,\% (6.5) / 22.8\,\% (11.0)} & $k=17$ at $23\,\%$ (8 H100) & \textbf{$k=38$ detected}\\
Ascella (1, 80\,MHz, 6 / 12) & $1.3\cdot10^7$ & $75\,\%$ (4.5) / never / never & $k=6$ at $96.5\,\%$ & every $k$ inside\\
\bottomrule
\end{tabular}
\end{table}

H100-equivalents needed to reproduce the whole stream in real time: the three-source device needs
$4.4$ at $\eta = 8\,\%$, $10$ at $15\,\%$, $29$ at $20\,\%$ and $580$ at $30\,\%$; the two-source device
between $0.26$ and $0.52$ up to $20\,\%$, and $5.3$ at $50\,\%$; the one-source devices stay below one
H100 up to $\eta \approx 70\,\%$. For the three-source device the Fock cost table is the conservative
reading: entangled emission is costed by the emitter sampler, which needs about eight fewer detected
photons at equal cost, so every transmission in its row is a lower bound. Its one-H100 tie at
$1.5\,\%$ is a throughput statement, not a hardness one --- $6\cdot10^7$ trains per second of zero- to
three-photon samples exceed what this sampler emits in batch --- so it is quoted from eight H100
upwards. Below $\eta = 1/\beta(c)$ ($0.54$ at $m = 2n$) post-selected hardware loses a factor per added
photon and never crosses the classical curve.

\section{Strong simulation on one layered basis}\label{supp:slos}

Strong simulation returns the probability $p_{\mathbf t}$ of every output occupation
$\mathbf t \in \mathbb N^m$ (of every one below a cutoff when the photon number is unbounded). Every
strong simulator of the code base is one algorithm on one object: the layered occupation basis of the
SLOS method~\cite{Heurtel2023SLOS}, graded by total photon number. This section states how each input
of the paper runs on it, how imperfections act on it, and the memory frontier that follows.

\paragraph{The object.} The layer $S_k$ is the set of $k$-photon occupations,
$|S_k| = M_k = \binom{m+k-1}{k}$, and the raising edges into it are the pairs
$(\mathbf t, \mathbf t + e_j)$ with $\mathbf t \in S_{k-1}$, $mM_{k-1}$ of them. For $v \in \C^m$ the
raising pass $\mathcal R(v)$ maps a coefficient vector over $S_{k-1}$ to one over $S_k$,
\begin{equation}
(\mathcal R(v)\,c)_{\mathbf t + e_j} \mathrel{+}= v_j\, c_{\mathbf t}
\quad\text{for every edge}, \label{eq:raise}
\end{equation}
which is multiplication of the generating polynomial $\sum_{\mathbf t} c_{\mathbf t} w^{\mathbf t}$ by
the linear form $v^\top w$. Coefficients are raw (of the monomial $\prod_j (a_j^\dagger)^{t_j}\ket 0$), so
$p_{\mathbf t} = |c_{\mathbf t}|^2\, \mathbf t!/\mathbf s!$ for an input occupation $\mathbf s$. Per-mode
photon caps prune every layer without changing an admissible amplitude, since they are closed under
photon removal. Each engine below is a pass structure on this object: which vectors are raised, how
many coefficient vectors are carried, and how the result is closed.

\paragraph{Fock inputs and coherent superpositions.} With $u_i = U[:, i]$ and photons in modes
$i_1, \dots, i_n$, the output coefficients are $c^{(n)} = \mathcal R(u_{i_n}) \cdots \mathcal R(u_{i_1})\,1$,
one pass per photon~\cite{Heurtel2023SLOS}: the layer-$k$ coefficients are sub-permanents shared by
every output. A superposition of $K$ Fock components is the same sweep with a component axis of length
$K$, summed coherently before the output factor $\mathbf t!$ is applied. On one H100 the reach is
$n = 11$ at $m = 22$ ($1.29\cdot10^{8}$ output states) in $98.9\,$ms and $15.4\,$GiB per unitary.

\paragraph{Products of superposed sources.} Source $i$ in $\sum_{q<d} a_{i,q}\ket q$ makes the photon
number indefinite, so the recurrence runs over sources and fills every layer at once:
\begin{equation}
\psi^{(i)}[k] = \sum_{q<d} \frac{a_{i,q}}{\sqrt{q!}}\, \mathcal R(u_i)^q\, \psi^{(i-1)}[k-q],
\qquad p_{\mathbf t} = |\psi^{(n)}[|\mathbf t|]_{\mathbf t}|^2\, \mathbf t!. \label{eq:graded}
\end{equation}
The mass of layer $k$ is the input's photon-number law, independent of $U$. For $d = 2$ the basis has
$\binom{m+n}{n} = (1 + 1/c)\,M_n$ states, $1.5$ times the Fock table at $c = 2$, and $a_i = (0, 1)$
reproduces the Fock recurrence bit for bit. At $n = 11$, $m = 22$ the graded kernel takes $180\,$ms
and $16.6\,$GiB, the same reach as Fock.

\paragraph{Emitter-generated inputs.} A train from one emitter of dimension $\chi$ is a matrix-product
state $\sum_x \bra f M^{(n)}_{x_n}\cdots M^{(1)}_{x_1}\ket{s_0}\ket x$~\cite{Schon2005}. Its $L^n$ Fock
components need not be expanded: the bond index rides the recurrence. With
$C^{(k)}_j = \sum_l U_{j,\mathrm{modes}[k][l]}\, M^{(k)}_l \in \C^{\chi\times\chi}$,
\begin{equation}
c^{(k)}_{\cdot,\, \mathbf t + e_j} \mathrel{+}= C^{(k)}_j\, c^{(k-1)}_{\cdot,\, \mathbf t}
\quad\text{for every edge into layer } k, \label{eq:slab}
\end{equation}
and the bond is closed at the end, by $\bra f$ for the projected amplitudes or by a trace for
$p_{\mathbf t} = \mathbf t! \sum_\sigma |c^{(n)}_{\sigma,\mathbf t}|^2$. Started from each basis vector of
the emitter, the same pass gives the Kraus operators of the stage, with $\sum K^\dagger K = 1$ when the
emission is isometric. The cost is $\chi$ coefficient vectors per layer, not $L^n$ sweeps: the
emitter's mixed reduced states cost the sampler a factor $\sim n/2$, the strong side only $\chi$. At
$n = 11$, $m = 22$, $\chi = 2$ the kernel takes $304\,$ms and $34.1\,$GiB.

\paragraph{Gaussian inputs.} Squeezed vacua (squeezing $r_s$ on mode $\mathrm{mode}_s$, optional
displacement) have no photon cap; the output's raw coefficients are the Taylor coefficients of
$F(w) = \mathcal N \exp(\tfrac12 w^\top B w + \zeta^\top w)$ with
$B = -\sum_s \tanh r_s\, u_s u_s^\top$, $u_s = U[:, \mathrm{mode}_s]$. Euler's identity for the
homogeneous part $F_N$ of degree $N$, $\sum_i w_i \partial_i F_N = N F_N$, applied to
$\sum_i w_i \partial_i F = (\zeta^\top w + w^\top B w)F$, gives the layer recurrence
\begin{equation}
N\, c_N = \mathcal R(\zeta)\, c_{N-1} - \sum_s \tanh r_s\, \mathcal R(u_s)^2\, c_{N-2}, \qquad
c_0 = \mathcal N, \label{eq:euler}
\end{equation}
raising passes only, on the same edges. Layer $N$ depends on the two below it and never on a layer
above, so every $p_{\mathbf t}$ with $|\mathbf t| \le N_{\rm cut}$ is exact; and since the layer masses
are the input's photon-number law, the mass left out by a global cutoff is known in closed form before
any pass runs, so $N_{\rm cut}$ is chosen for a prescribed tail $\epsilon$. The per-amplitude recursion of
Miatto and Quesada~\cite{MiattoQuesada} also advances by photon number but reaches lower weights through
lowering edges, which this object does not store; Euler's identity sums over all modes at once. The
recursion agrees with per-pattern loop hafnians~\cite{Bulmer2022} to $10^{-15}$. The unoptimised GPU
kernel reaches six squeezed photonic inputs at $m = 12$ ($N_{\rm cut} = 18$) in $417\,$s and $12.7\,$GiB, short of the
counted reach below because it stores its edges. Under loss a truncated, thinned table is a lower bound
on each detected block, short by at most the tail; the exact lossy table needs the density-matrix
recurrence on $2m$ modes and is not built.

\paragraph{Imperfections on the same layers.} Uniform loss commutes with $U$ and is applied at the
output by thinning down the layers, $q_{k-1}(\mathbf t) = \sum_j q_k(\mathbf t + e_j)(t_j+1)/k$, with the
detected distribution $\binom nk \eta^k (1-\eta)^{n-k} q_k$ on layer $k$: one probability pass per layer,
against $(4/3)^n$ lossless sweeps for enumerating surviving subsets at $m = 2n$. Partial
distinguishability is the mixture over the set $G$ of photons in the common internal state, weight
$x^{|G|}(1-x)^{n-|G|}$, with the lossless engine on $G$ and each other photon added classically on the
same edges, $q_{k+1}(\mathbf t + e_j) \mathrel{+}= q_k(\mathbf t)|U_{ji}|^2$. Keeping at most $b$ such
photons costs $\sum_{b' \le b}\binom n{b'}$ sweeps, with a total-variation error at most
$\Pr[\mathrm{Bin}(n, 1-x) > b]$ returned with the result: at $n = 11$ and $x = 0.949$, $67$ sweeps for
$b = 2$ (bound $0.016$) against $2^{11}$ for the exact mixture. For an emitter input a distinguishable
photon stays coupled to the bond through its label, so the pass is resolved by species before the trace.

\paragraph{Non-linear sites and adaptive measurements.} On the $n$-photon space a phase
$\exp(-if(\hat n_x))$ is a sum of at most $n+1$ linear phase shifters, with the discrete Fourier
coefficients of $j \mapsto e^{-if(j)}$, so a circuit with sites is a coherent sum of linear circuits and
its strong side is that many sweeps summed, or one sweep followed by diagonal phases and later
interferometers pushed through the basis as two-mode gates. An adaptive circuit with $K$ measured modes
is $\binom{n+K}{K}$ sweeps, one per outcome under its own interferometer; it is built, and is the oracle
of the adaptive sampler. The projected state-vector engine that the regime $K \propto m$ would need is
not built.

\paragraph{The memory frontier.} Once the table exists sampling is a lookup, so the frontier of strong
simulation is the largest $n$ whose table fits in $80\,$GiB, a count rather than a benchmark. In
complex64 amplitudes and float32 probabilities the Fock builder that pulls its last layer straight into
$p$ peaks at $8M_{n-1} + 4M_n$ bytes, the emitter one at $8\chi M_{n-1} + 4M_n$, the superposed one at
$8\binom{m+n-1}{n-1} + 4\binom{m+n}{n}$. At $m = cn$ the table grows per photon by
$\beta_{\rm mem} = (1+c)^{1+c}/c^c$, $6.75$ at $c = 2$ against $\beta \approx 1.8$ for sampling, so a
doubling of memory adds $1/\log_2 \beta_{\rm mem}$ photons, $0.36$ at $c = 2$: eight H100 move the
frontier from $13.5$ to $14.6$. For inputs with a fixed photon number the frontier is the size of one
last layer and the builders differ only in the parent layer (Table~\ref{tab:memory}): emitter inputs
sit at most $0.3$ photon below Fock and superposed sources within $0.5$ photonic input, so strong
simulation has one memory frontier, a property of the output space, and all the modality dependence
lives in sampling. A Gaussian table holds a whole photon-number distribution per squeezer up to a cutoff
about seven times the mean photon number at the frontier, so with $\epsilon = 1/S$, $S = 10^4$, it
reaches $7.2$ squeezed photonic inputs at $c = 2$ against $13.5$ Fock photons ($8.3$ at $\epsilon = 10^{-3}$, $5.9$ at
$10^{-6}$), and lower again under loss.

\begin{table}[h]\centering\small
\caption{Memory frontier of strong simulation at $80\,$GiB, $m = cn$: integer $n$ / interpolated $n^*$.
Fixed-photon-number inputs in the pull form; squeezed inputs at $0.5$ photons per squeezer with the
cutoff at $\epsilon = 10^{-4}$, build peak.}\label{tab:memory}
\begin{tabular}{@{}rrccccc@{}}
\toprule
$c$ & $\beta_{\rm mem}$ & table alone & Fock & emitter ($\chi = 2$) & superposed & squeezed\\
\midrule
1  & 4.00  & 19 / 19.1 & 18 / 18.6 & 18 / 18.3 & 18 / 18.1 & 12.2\\
2  & 6.75  & 13 / 13.7 & 13 / 13.5 & 13 / 13.3 & 13 / 13.2 & 7.2\\
4  & 12.21 & 10 / 10.4 & 10 / 10.3 & 10 / 10.1 & 10 / 10.2 & 4.2\\
8  & 23.09 & 8 / 8.2   & 8 / 8.2   & 8 / 8.1   & 8 / 8.1   & 2.5\\
16 & 44.84 & 6 / 6.8   & 6 / 6.7   & 6 / 6.7   & 6 / 6.7   & ---\\
\bottomrule
\end{tabular}
\end{table}

\paragraph{Where the table is faster.} Between the two frontiers, $n = 14$ to $38$ at $c = 2$ for Fock
inputs, only the chain-rule samplers exist. Below the memory frontier the table is an alternative that
is built in seconds and then draws about $10^8$ samples per second, so it wins for more than
$10^6$--$10^7$ draws or whenever the probabilities themselves are wanted. A partial table does not move
the sampling frontier: drawing from a stored set of mass $M$ and otherwise from the chain rule gains at
most $1/(1-M)$, and for Fock inputs, whose probabilities are close to Porter--Thomas, $M$ is small
beyond the memory frontier (a speed-up of at most $1.5$ at ten times past it). For squeezed inputs the
cheap shots fit in memory and the cost is in the bright ones: at $38$ squeezed photonic inputs the patterns with at most
six photons are $3.5\,\%$ of the shots and $10^{-6}$ of the cost.

\paragraph{Validation.} Every strong engine is checked against an oracle that shares no code with it:
Fock against Ryser permanents and Perceval's SLOS backend; superposed sources against the Fock kernel at
$a = (0,1)$ and an independent amplitude filter ($2\cdot10^{-16}$--$4\cdot10^{-15}$ to $n = 40$ at $U = 1$); the
emitter against the $L^n$ expansion and an independent permanent sum at $n \le 4$, with the traced
distribution equal to the sum of projections and the Kraus set resolving the identity; Gaussian inputs
against loop hafnians and a truncated product oracle to $10^{-15}$; loss against input-side enumeration;
distinguishability against the $(n!)^2$ Gram-matrix oracle to $10^{-16}$; non-linear sites against dense
and branch-sum oracles to $10^{-16}$. Every pass is plain tensor arithmetic, so each engine is also
differentiable in $U$, the emitter operators, $\eta$, $x$ and the squeezing.

\section{Kernel engineering, and the parity audit}\label{supp:kernels}
The measured engineering gains, and then the audit Sec.~\ref{sec:parity} promises: one row per
quoted number with its algorithm, code path, measured or modelled status, arithmetic, statistic and
run stamp, and the reason a hollow row is hollow. Table~\ref{tab:audit} is that audit; the arithmetic
of each engine is in Table~\ref{tab:precision}. A row is \emph{at parity} when it runs the fastest
algorithm we have for its input on kernels at the engineering level of Sec.~\ref{sec:parity} under the common protocol, and, where
the input has a Fock limit, reproduces the Fock frontier there within $1.3\times$ per sample;
\emph{own engine} when no common substrate exists and the row runs the best known algorithm on a kernel
of comparable engineering; \emph{outside} (\hollow) otherwise.

{\scriptsize\setlength{\tabcolsep}{2pt}\setlength{\LTcapwidth}{\textwidth}
\begin{longtable}{@{}L{2.1cm}L{2.3cm}L{2.2cm}L{2.1cm}L{1.5cm}L{2.5cm}L{2.1cm}@{}}
\caption{The parity audit, one row per quoted number. Statistic: bracketed mean crossing of one
sample per second unless stated; ``modelled'', ``extrapolated'' and ``estimated'' mark the numbers not
read from a measured bracket. Runs are
the stamps of the kept result files (Supplementary Sec.~\ref{supp:tables}).}\label{tab:audit}\\
\toprule
row & quoted & algorithm & kernel & statistic & run & audit\\
\midrule
\endfirsthead
\toprule
row & quoted & algorithm & kernel & statistic & run & audit\\
\midrule
\endhead
\bottomrule
\endlastfoot
Fock, $m = 2n$ / $m = n$ & $38.3$ / $42.6$ & C\&C 2020 (output collapse) & CUDA cofactor kernel & $\ge 2048$ shots & 20261009\_141132 & reference\\
Fock, four algorithms (Table~\ref{tab:T1}) & $32.5$ / $31.7$ / $30.9$ / $35.7$ & 2018 / input / per-cofactor / 2020 & one common Triton kernel & bracket & 20260930\_082501 & algorithm comparison, not a frontier row\\
Fock, loss $\eta = 0.9 / 0.8 / 0.7 / 0.5$ & $41.2$ / $45.0$ / $49.2$ / $64.2$ & 2020 on the surviving photons & CUDA cofactor kernel & $\ge 2048$ shots; heavy tail at $\eta \le 0.7$ & 20261008\_071215 / \_085321 / \_104426 / \_192343 & at parity (same engine)\\
Fock, distinguishability, $\eta = 1$ / $0.7$ & $38.3$--$41.8$ / $49.2$--$53.7$ & 2020 per interfering species & CUDA cofactor kernel & bracket at $\eta = 1$ ($V < 1$ cells from a separate, earlier run); $\ge 2048$ shots at $\eta = 0.7$ & 20260921\_051530; 20261008\_104426, 20261009\_195643 & at parity (same engine)\\
Fock $+$ adaptive, $K = 1 \dots 6, 8$ & $38.3$ / $38.2$ / $38.1$; $37.6$ / $34.8$ / $34.4$ / $33.8$ & 2020 prefix chain $+$ exact level marginals; aborted chains from $K = 4$ & CUDA cofactor kernel; level tables on the host & $512$ shots & 20261007\_022012 / \_030408 / \_061124 & at parity for $K \le 3$; from $K = 4$ limited by the implementation, route named\\
Fock $+$ non-linear sites, $L = 1$ ($2 / 4 / 8$) & $36.1$ ($34.3 / 30.3 / 21.4$) & rejection from the branch mixture, 2020 proposals & proposals: CUDA cofactor kernel; branch amplitudes: batched Glynn in PyTorch & $\ge 2048$ shots ($L = 1$) & 20261009\_141132 ($L = 2, 4, 8$: 20260923\_080913) & chain at parity; part of the per-site cost is the amplitude pass\\
emitter, $\chi = 2$, $m/n = 2$ ($2.5 \dots 12$) & $31.0$ ($30.5 \dots 29.0$) & emitter chain rule with the 2020 collapse & CUDA emitter kernel ($2\times2$ chains) & bracket & 20261006\_004008 (20261002\_003041, 20261003\_103957) & at parity, residual $1.32$--$1.48\times$ per needed operation (the lane-split scan)\\
emitter, loss $\eta = 0.9 / 0.8 / 0.7 / 0.5$; distinguishability $\eta = 1$ / $0.7$ & $33.6$ / $36.8$ / $40.9$ / $53.5$; $31.0$--$33.4$ / $40.9$--$44.7$ & as above & as above & $\ge 2048$ shots ($\eta < 1$) & 20261008\_210721 / \_122031 / \_133143; 20261003\_121759; 20261009\_195643 & at parity\\
emitter trains & $k \approx 21$--$22$ per tick & Glynn label amplitudes, sign sum collapsed & cuBLAS GEMM & per-tick time, extrapolated from $k = 18$ & 20260919\_084900 & at parity\\
superposed, $\alpha = \pi/4$, $m = 2n$ (nine ratios) & $50.0$ ($53.1 \dots 47.0$) & random-source-order chain with the 2020 collapse & Fock step kernel ($s_0 = 0$), $\mu$- and $\lambda$-forms & $\ge 512$ counted shots & 20261007\_090041 (20261010\_045711) & at parity (Fock limit $38.4$ vs $38.4$); residual $\le 0.07$ photonic input\\
superposed, coherence and loss scans & $64.2 \dots 50.0$; $54.0 / 58.3 / 64.8$ & as above & as above & shot floor, $\ge 1024$ counted & 20261007\_122313; 20261007\_094729 / \_183855, 20261008\_233436 & at parity (twin limit $64.6$ vs $64.4$)\\
squeezed, $S = 10^4$ & $41.5$ ($42.9$ at $m = n$) & pair-sieve loop-hafnian chain & Triton sieve; double-double tier & $t_S$, modelled (validated $0.91$--$1.00$) & 20260925 model; tiers 20260930 & own engine\\
squeezed, device replay & $0.154\,$s/sample; high brightness: median $2.4\cdot10^{16}$ H100-s & as above & as above & measured; high brightness estimated (double-double) & 20260927\_190700; tiers 20260930 & own engine\\
single photons through a loop & $N^* > 7.5\cdot10^4$ at every $L$ measured (search cap) & Fock chain on the loop, one basis state between splitters & PyTorch & bracket & 20261016\_023739 & polynomial at any $L$; a reach, not a frontier\\
emitters through loops & one emitter: $N^* > 3142$ ($L \le 4$), $4204$ ($L = 6$), $2166$ ($L = 11$); $k = 2$: $3143$ / $1853$ ($L = 1$ / $2$); $k = 3$: $1440$ ($L = 1$) & spins $\otimes$ carried-Fock sectors & sector tables built on the host; PyTorch & $t_S$, $\ge 64$ shots & 20261016\_023739 & \hollow outside: host-side tables\\
superposed through a loop & $267.4 \dots 21.4$ ($L = 1 \dots 8$) & register propagation & PyTorch complex128 register & bracket & 20261001\_101803 (limit 20261005\_170728) & at parity (Fock limit within $1.15\times$ per sample, except at the $L = 1$ rate step)\\
squeezed through a loop & $\ge 4000 \dots 7.3$ ($L = 1 \dots 8$) & capped register & PyTorch & $t_S$ & 20261002\_203546 & own engine\\
\end{longtable}}

Against the best existing kernel the superposed-source row's residual is at most $0.07$ photonic input; a
higher-degree kernel that does not yet exist would add $0.14$--$0.18$ photonic input at $n = 46$ and
$0.35$--$0.43$ at $n = 60$, so the row is a conservative bound.

\section{Numerical precision}\label{supp:precision}

\subsection{The criterion, and what is measured}

Every sampler here is exact in the sense of the chain rule: the probability weights $w(r)$ of each
step are computed from a finite sum, and a shot is one inverse-CDF draw per step. Finite precision
enters in the weights and in the uniform variate of each draw. We say an implementation meets the
criterion at size $n$ when the relative rounding error of its step weights is below $10^{-3}$. Each
step's conditional law is then within $10^{-3}$ in total variation of the exact one; the whole sample,
a product of $n$ such steps, is within the sum of the per-step distances, at most $n \cdot 10^{-3}$.
The criterion itself is therefore not below the statistical resolution of the longest runs
($S = 10^8$, $10^{-4}$); what is, is the measured error of the emitter and superposed rows, at least
four orders inside the criterion, which bounds their whole sample below $10^{-5}$ (method 1 below).
The squeezed row's error follows a measured law (method 3) and comes within one to three orders of the
criterion in the upper strata of its crossing, $N \approx 50$--$60$; the Fock error is measured
against extended precision (method 1) up to $k = 38$, about six orders inside the criterion at the
frontier. The \textbf{fp64 accuracy range} of an engine is the largest photon
number at which its weights meet the criterion in double precision; beyond it the algorithm is
unchanged but the arithmetic needs a double-double factor or a compensated sum.

A rounding error can be measured in four ways, and which one applies is decided by what the
engine's sum allows.

\begin{enumerate}
\item \textbf{A true error against a higher-precision evaluation.} Where the kernel is plain
C\texttt{++} --- the thread-per-lane kernel of the emitter chain, and a CPU emulation of the
production Fock cofactor kernel --- the same code compiled in 80-bit extended precision
($2^{-64} \approx 5\cdot10^{-20}$ per operation) on the same instance gives the fp64 error, every
addend included: to about $10^{-18}$ relative for the emitter, and for the Fock cofactors at
$k = 38$ ($2^{37}$ addends, about forty minutes on $24$ host threads) to about $2^{-11}$ of the fp64
error, under $0.1\,\%$ of it. Each Fock size was measured on one instance.
\item \textbf{A lower bound from disagreement.} Where no higher-precision evaluation is affordable,
two fp64 evaluations of the same sum that
round differently --- two lane segmentations, two kernels, two algorithms --- differ by at most the
sum of their errors, so the larger error is at least half the disagreement. We state it as a lower
bound.
\item \textbf{A measured error law.} The sieves return $\kappa = \sum_S|\mathrm{term}_S|/\max|c|$
per step; the error is bounded by $\varepsilon\kappa$ and was measured against a double-double
reference on 275 sieves at $\le 0.74\,\varepsilon\kappa$.
\item \textbf{A conditioning guard.} The adaptive engine's graded-table inversion is an alternating
sum used only to size batches; its relative accuracy is poor and it never enters a weight.
\end{enumerate}

\subsection{Per engine}

\begin{table}[H]\centering\scriptsize
\setlength{\tabcolsep}{3pt}
\caption{Rounding error per engine, the method it was measured by, its growth law, and the fp64
frontier the $10^{-3}$ criterion puts it at. ``as Fock'' means the engine's step sum \emph{is} the
Fock sum, so the same law applies.}\label{tab:precision}
\begin{tabular}{@{}L{2.3cm}L{2.6cm}L{1.3cm}L{3.6cm}L{1.6cm}L{2.2cm}@{}}
\toprule
engine & the step sum & method & measured & growth & fp64 frontier\\
\midrule
Fock (2020, CUDA) & Glynn cofactors, $2^{k-1}$ addends, collapsed & 1 &
extended precision $3.6\cdot10^{-13}$ ($k{=}30$), $1.9\cdot10^{-11}$ ($34$), $3.4\cdot10^{-10}$ ($38$);
within a factor $2$ with one repeated row; one instance per $k$ %
& $\times 2.35$ /photon & $10^{-4}$ near $n \approx 53$;
$\mathbf{10^{-3}}$ near $\mathbf{n \approx 55}$ (extrapolated)\\
emitter & the same walk with $2\times2$ chains, $2^{K-2}$ collapsed addends per term & 1, 2 &
extended precision $1.1$--$1.3\cdot10^{-13}$ ($K{=}26$), $6.3$--$7.7\cdot10^{-13}$ ($28$),
$2.4$--$2.5\cdot10^{-12}$ ($30$); unchanged with one repeated row %
& $\times 2.3$ /photon & $10^{-6}$ at $K \approx 45$; $\mathbf{10^{-3}}$ at $\mathbf{K \approx 53}$\\ %
superposed & rectangular leave-one-out over $2^{q-1}$ signs; three forms by the silent-source count
$s_0$: the Fock kernel ($s_0{=}0$), the division-free $\mu$-form ($1 \le s_0 \le 18$), the $\lambda$-form above & 1, 2 &
$\mu$ ($s_0{=}18$): $3.2\cdot10^{-15}$ ($q{=}16$), $8.8\cdot10^{-15}$ ($20$), $1.4\cdot10^{-13}$ ($24$);
$\lambda$ ($s_0{=}20$): $6.7\cdot10^{-14}$, $2.7\cdot10^{-13}$, $3.6\cdot10^{-12}$
& $\mu$: $\times 1.2$ /photon; $\lambda$: $\times 2$--$2.8$ & $\mu$: none in reach; $\lambda$: $10^{-3}$ at $q \approx 46$,
run only at $q \le n - 20$\\
non-linear & the Fock sum per branch; acceptance a ratio of squared sums & as Fock, $\times 2$ on the acceptance & as Fock & as Fock & as Fock\\
adaptive & the Fock sum per round; the table inversion guards & as Fock, 4 & as Fock & as Fock & as Fock\\
emitter trains & $2^{k-1}$ signs $\times$ $2^k$ amplitudes per stream, renormalised each tick & 2 &
against a cancellation-free recurrence: $2.5\cdot10^{-14}$ ($k{=}16$), $4.6\cdot10^{-14}$ ($18$),
$1.1\cdot10^{-13}$ ($20$), the same on $|a|^2$ %
& $\times 1.4$ /emitter & $10^{-6}$ at $k \approx 50$; beyond the register ($k \le 24$)\\
squeezed (loop-hafnian sieves) & $\sum_S \pm$ terms, $\kappa$ returned & 3 & $3\cdot10^{-14}$ at
$n = 20$, and by the law at every $N$ & $\times 1.57$ /detected photon & $\mathbf{N \approx 62}$
fp64; $N \approx 141$ double-double\\
\bottomrule
\end{tabular}
\end{table}

Four notes. (i) The Fock and emitter true errors grow alike, by $2.35$ and $2.3$ per photon, and at
$30$ photons they are $3.6\cdot10^{-13}$ and $2.5\cdot10^{-12}$. (ii) That growth is the growth of the collapsed Glynn sum's cancellation:
the binomial weights of the output-side collapse against the alternating signs. It is the same law
in every engine built on that sum, which is why one policy covers them all. (iii) Where each
engine's quoted row sits: the Fock frontier at $n^* = 38.3$ carries about $3\cdot10^{-10}$ per step
and the multi-GPU extrapolation at $n^*(576) \approx 48.9$ stays below $10^{-4}$ by the extrapolated
law; the emitter row at
$n^* = 31.0$ carries $2.5\cdot10^{-12}$; the superposed rows at $n^* = 50.0$ and $53.1$ sit at
least $10^5$ below the criterion --- the $\lambda$-form, the only one whose error grows like Fock's,
runs at $q \le n - 20$ ($\le 10^{-8}$ there), the $\mu$-form carries every high-occupancy stratum
that occurs ($\le 10^{-12}$), and the shots that would place $q \approx n$ photons have probability
$\le 10^{-15}$ at $\alpha = \pi/4$, so fp64 does not limit that row although its $n^*$ is where Fock's
frontier would be; the trains at $k \approx 22.6$ emitters carry about
$2\cdot10^{-13}$, and their errors do not accumulate across ticks because the register is
renormalised at each one; the non-linear and adaptive rows sit where Fock does; and the squeezed
row is the one that evaluates part of its own tail in double-double. %
(iv) A disagreement bounds the error only between evaluations that round differently where the
error is made. Two kernels that walk the same collapsed order share the cancellation: on the
emitter at $K = 30$ they agree to $4\cdot10^{-14}$ while each is $2.5\cdot10^{-12}$ from the
extended-precision value, a fifty-fold under-estimate; two segmentations of the same kernel differ
by $8\cdot10^{-12}$ and do bound it. %
The Fock kernels show the same effect: the production CUDA kernel and the Triton kernel with the same
segmentation agree to $3$--$6\cdot10^{-14}$ while each is $3.6\cdot10^{-13}$ from the
extended-precision value at $k = 30$, a disagreement thirteen times below the error at $k = 30$ and
sixty-five times below it at $k = 32$. %

\subsection{The draw's own resolution}

A weight more precise than the uniform variate that draws from it is wasted, and on one engine that
is the binding limit rather than the arithmetic. The Fock device draw uses a counter-based float32
uniform cast to float64: a resolution of $2^{-24} \approx 6\cdot10^{-8}$ on the cumulative
distribution. So up to $n \approx 44$, by the extrapolated law of the cofactor error, the draw, rather than the cofactor sum, sets the precision floor of the
Fock engine. This floor is four orders of magnitude below the criterion; we state it because a mode of
probability below $10^{-7}$ is then drawn with a relative error of order one. The emitter,
train, non-linear and adaptive draws use 53-bit uniforms. A float64 draw for the
Fock kernel is a one-line change, but it alters every sample bit-for-bit, so it is deferred rather
than made mid-campaign.

\subsection{The policy}

Monitor where a cheap bound exists (the sieves' $\kappa$, recorded per shot; the adaptive table's
negativity guard). Measure once, at the frontier size, where no bound exists: a true error where
the kernel compiles in extended precision, a lower bound from two rounding orders where it does
not, and state the number with its growth law. Compensate only if the measured error on a weight
exceeds $10^{-4}$ at the quoted size. The emitter and superposed rows sit at least four orders below
the criterion; the squeezed strata from $N \approx 59$ exceed $10^{-4}$ by the measured law, and the
timing model evaluates every stratum beyond the fp64 range in double-double; the Fock row, measured
against extended precision, sits about six orders below the criterion at its frontier. Every sampler's statistics object carries the measured error law or bound of its own
row.

\subsection{Beyond fp64}

The bright squeezed-light runs of Sec.~\ref{sec:borealis} would need quad-double, for which no kernel
exists. For Fock, emitter and superposed inputs no double-double kernel exists yet, so its cost there is
not measured; it is not needed for Fock inputs below $n \approx 45$. The squeezed frontier's
$S$-dependence crosses the fp64 frontier: the $S = 10^4$ run stops at $N_S = 60$, inside fp64, but at
$S = 10^6$ some $39\,\%$ of $t_S$ comes from $N > 64$. Cross-mode grouping of the sieve costs about five
decades of $\kappa$, which brings the double-double reach from $N \approx 141$ to $117$.

\subsection{The double-double sensitivity for squeezed inputs}

Where the fp64 tier ends the sieve can switch to double-double, and that absorbs the extra decades
of condition number that cross-mode grouping costs, which fp64 cannot. At high collision density the
combination is cheaper than fp64 without grouping: measured on the H100 at $m = n = 30$ and two
photons per squeezer, the same calls ran $8.7$ to $18$ times faster in double-double with grouping
than in fp64. Recomputing the frontier with the sieve free to make that choice leaves the headline at
$41.5$, gains on the collision-dense points ($m = n$ at $44.7$) and loses on the collision-poor
tails. We quote the fp64 numbers as the frontier and this paragraph as the sensitivity, because the
sieve's own switching rule does not yet always take the cheaper arithmetic.

\subsection{Precision of the frontier samplers, as previously summarised in the main text}
The finite-precision implementation of an
algorithmically exact sampler needs an error criterion; ours is that the relative error of the
probability weights at each step of the chain is at most $10^{-3}$. This is a per-step criterion: the
total-variation distance of a whole $n$-step sample from the exact law is bounded only by the sum of the
per-step distances, at most $n$ times the per-step error. The draw has its own resolution: the single-photon
sampler draws from a float32 uniform, $2^{-24}$ on the cumulative weights, while the other samplers use
53-bit uniforms. The \textbf{fp64 accuracy range} is the largest photon number that meets the
criterion in fp64. For squeezed inputs the
sieve's condition number grows as $10^{0.40+0.197N}$, and the criterion holds to $N \approx 62$
detected photons in fp64 and to $N \approx 141$ in double-double. The sampler guards that range: it
checks $\varepsilon\kappa \le 10^{-3}$ after every sieve and redoes the sieve otherwise. The largest
measured value is $8.4\cdot10^{-4}$, at $N = 54$ on a grouped pairing; at $N = 60$--$62$ the guard
excludes grouped pairings, and ungrouped ones sit $1.3$ to $2.9$ decades under the fit at $N = 54$--$58$.
For single photons, the production
cofactor kernel measured against a CPU emulation of the same kernel in extended precision has a true
error of $3.6\cdot10^{-13}$ at $k = 30$ and $3.4\cdot10^{-10}$ at $k = 38$ (one instance per $k$), a
factor $2.35$ per photon, within a factor $2$ with a repeated row; extrapolating that growth, the criterion is reached near $n \approx 55$.
The emitter kernel, compiled in extended precision, gives a true error of $2.5\cdot10^{-12}$ at $30$
placed photons and a factor $2.3$ per photon; the superposed rows, measured against extended
precision as well, sit at least $10^5$ below the criterion because their shots place about $n/2$
photons. For these three rows the per-step error at the frontier is at least four orders of magnitude
inside the criterion --- about $3\cdot10^{-10}$ for single photons at $n^* = 38.3$ --- and the whole-sample
bound is below $10^{-5}$.
The squeezed row is not: by its measured law the per-step error is about $2\cdot10^{-6}$ at $N = 50$
and $2\cdot10^{-4}$ at $N = 60$, so the upper strata of its $t_{10^4}$ crossing are one to three
orders inside the criterion, with a whole-sample bound of up to about $10^{-2}$ at $N \approx 60$; the
timing model evaluates every stratum beyond the fp64 range in double-double
(Supplementary Sec.~\ref{supp:precision}). At $N = 50$--$62$ this error is bounded, not measured
directly, since one whole double-double sieve there takes hours to weeks: the law error
$\le 0.74\,\varepsilon\kappa$ is validated against double-double on 275 sieves up to $N = 48$ and
applied above it with $\kappa$ measured to $N = 58$ (to $62$ on grouped pairings), and the guard
bounds every sieve.

Beyond this range the criterion needs double-double or quad-double arithmetic, which current
GPUs do not provide natively. For squeezed inputs the double-double sieve exists and costs $58$ to
$67$ times an fp64 term; for single-photon, emitter and superposed inputs no double-double kernel exists yet.
Near this range the frontier is set by the \emph{number format}, not by the number of GPUs. For single
photons the multi-GPU extrapolations $n^*(72) \approx 45.4$ and $n^*(576) \approx 48.9$ at $m = 2n$
still stay below $10^{-4}$ by the extrapolated law; compensation is needed from about $53$ photons and
the criterion fails near $55$. The format binds earlier for the squeezed row's
dependence on $S$: at $S = 10^6$ some $39\,\%$ of $t_S$ comes from $N > 64$, so the $S = 10^6$ and
$10^8$ values are fp64 costs of calls that need double-double.

\section{Result files}\label{supp:tables}
Every number in the paper comes from one of the result files below. A run stamp names one GPU run
(date and time of its start, UTC); every run is one NVIDIA H100 PCIe unless the row says otherwise.
Each kept file is in the \texttt{results/keep/} directory of
\url{https://github.com/Quandela/photonic-engines}, and each run, by its stamp (log, parameters, raw output), at
\url{https://github.com/Quandela/photonic-engines-runs} (Supp.~\ref{supp:repro}).
A row whose model was calibrated on measured runs, rather than measured itself, says so.

{\small
\begin{longtable}{@{}L{4.3cm}L{3.0cm}L{7.6cm}@{}}
\toprule
quantity (where quoted) & run stamp & kept file(s) \\
\midrule
\endhead
\bottomrule
\endfoot
Fock production row at $m = n$ and $m = 2n$, and the non-linear site at $m = 2n$, on the canonical
256-shot batches (Tabs.~\ref{tab:T1}, \ref{tab:ratios}, \ref{tab:T3})
 & 20261009\_141132
 & \path|r30_fock_c{1,2}_20261009_h100.csv|, \path|r30_nlbs_c2_20261009_h100.csv| \\
Fock inputs, every other algorithm of the inventory; the envelope; cofactor precision
(Tab.~\ref{tab:T1}, Secs.~\ref{sec:inventory}, \ref{sec:fourinputs}, \ref{supp:precision})
 & 20260930\_082501; envelope at $m = 2n$: 20260930\_071754; strong-then-sample: 20260930\_141657
 & \path|threshold_{2018,2020,input,percof}_c{1,2}_20260930_h100.csv|,
   \path|algorithm_envelope_{c1,c2,strong}_20260930_h100.csv|,
   \path|algorithm_terms_gpu_20260930_h100.csv|,
   \path|cof_precision{,_collide}_20260930_h100.csv| \\
Fock inputs at the other mode ratios (Tab.~\ref{tab:ratios})
 & 20260920\_092711 & \path|frontier.csv| (Fock rows) \\
Emitter inputs at $m = 2n$, $\eta = 1$, on the parity build; its Fock-limit check (Tab.~\ref{tab:ratios},
Tab.~\ref{tab:T3})
 & 20261006\_004008; kernel throughput check 20261005\_234844
 & \path|frontier_loss_emitter_r18_53c8494_eta1_20261006_h100.csv|,
   \path|threshold_emitter_fock_c2_53c8494_20261006_h100.csv|,
   \path|emitter_fock_throughput_4x_53c8494_20261005_h100.csv| \\
Emitter inputs at the other mode ratios (Tab.~\ref{tab:ratios})
 & 20261002\_003041, 20261003\_103957 ($m/n = 12$)
 & \path|frontier.csv| (emitter rows), \path|frontier_auto_emitter_r*.csv|,
   \path|frontier_auto12_20261003_h100.csv| \\
Squeezed inputs: the frontier from the calibrated timing model, its validation on unseen shots,
the precision tiers (Tab.~\ref{tab:ratios}, Sec.~\ref{sec:gbsalg})
 & 20260924\_234127, 20260925\_075752, 20260925\_090352; validation 20260925\_095009; tiers: model,
   no new run
 & \path|gbs_timing_frontier_final_be_20260925_h100.csv|, \path|gbs_frontier_grid_20260925_h100.csv|,
   \path|frontier_gbs_20260925_h100.csv|, \path|gbs_timing_frontier_tiers_floor_20260930.csv|,
   \path|gbs_borealis_replay_tiers_floor_20260930.json|, \path|gbs_timing_curve_c2_tiers_floor_20261005.csv|
   (the curve of Fig.~\ref{fig:money}) \\
Squeezed inputs: the double-double sieve and its rate (Secs.~\ref{sec:gbsalg}, \ref{sec:precision})
 & 20260928\_204415, 20260929\_103413
 & \path|gbs_pairing_dd_20260928b_h100.csv|, \path|gbs_timing_rate_dd_20260928b_h100.json|,
   \path|gbs_borealis_replay_tiers_job30_20260929.json|, \path|gbs_timing_frontier_tiers_job30_20260929.csv| \\
Squeezed inputs at high photon count (Sec.~\ref{sec:precision}, Supp.~\ref{supp:precision})
 & 20260929\_210408, 20260929\_231151 & \path|gbs_highN_auto{,_b}_20260929_h100.json| \\
Exact samples of a published squeezed-light experiment (Sec.~\ref{sec:borealis})
 & 20260927\_190700 & \path|gbs_borealis_fig3a_exact_20260927_h100.csv| \\
Non-linear sites: phase, mode ratio, number of sites (Secs.~\ref{sec:fourinputs},
\ref{sec:frontier})
 & 20260922\_211901 & \path|nlbs_{phi,ratio,sites}_20260922_h100.csv| \\
Non-linear sites under loss and distinguishability (Sec.~\ref{sec:law})
 & 20260923\_111505 & \path|nlbs_imperfections_20260923_h100.csv| \\
Feed-forward: the adaptive row at $m = 2n$, $K = 1 \dots 8$, 512 shots (Tabs.~\ref{tab:T1},
\ref{tab:T3}, Sec.~\ref{sec:adaptive})
 & 20261007\_022012, 20261007\_030408, 20261007\_061124
 & \path|afbs_K{1,23,4,5,6,8,8b}_x4_20261007_h100.csv| \\
Feed-forward: crossings at other mode ratios (64--128 shots) and the costs under imperfections (128 shots,
Sec.~\ref{sec:adaptive})
 & 20260922\_074317 \dots\ 20260923\_181143
 & \path|adaptive_crossings.csv|, \path|adaptive_imperfect_crossings.csv|,
   \path|adaptive_imp3_{eta0.9,x0.95}_20260923.csv| \\
Emitter trains with fusion and feed-forward (Secs.~\ref{sec:trains}, \ref{sec:ff})
 & 20260918\_084620, 20260919\_084900
 & \path|esbs_fusion_*.csv|, \path|f7_feedforward_cost_20261002.csv| \\
Loss, Fock inputs, $\eta = 0.9 / 0.8 / 0.7 / 0.5$ (Tab.~\ref{tab:loss})
 & 20261008\_071215, 20261008\_085321, 20261008\_104426, 20261008\_192343
 & \path|r30_fock_eta{09,08,07,05}_20261008_h100.csv| \\
Loss, emitter inputs, $\eta = 0.9 / 0.8 / 0.7 / 0.5$ (Tab.~\ref{tab:loss})
 & 20261008\_210721 (0.9, 0.8), 20261008\_122031, 20261008\_133143
 & \path|r30_emitter_eta{09,08,07,05}_20261008_h100.csv| \\
Distinguishability at $\eta = 0.7$, Fock and emitter (Sec.~\ref{sec:law}, Sec.~\ref{sec:energy})
 & 20261009\_195643 & \path|r30_{fock,emitter}_eta07_v{095,09,08}_20261010_h100.csv| ($V = 1$:
   the $\eta = 0.7$ loss files) \\
Distinguishability at $\eta = 1$, $V < 1$ (Sec.~\ref{sec:law}, Sec.~\ref{sec:energy})
 & 20261003\_121759 (emitter); Fock: 20260921\_051530
 & \path|frontier_indist_run2.csv|, \path|frontier_indist_emitter_auto_emitter_eta1_V*.csv| \\
Loops: single photons, one emitter (Sec.~\ref{sec:loops})
 & 20261016\_023739; single photons at $N = 10^5$: 20261016\_021007
 & \path|hybrid_frontier_fock_chain{,_spaced}_L*_20261016_h100.csv|,
   \path|hybrid_frontier_emitter_v3_L*_20261016_h100.csv|,
   \path|hybrid_release_check_*_20261016_h100.csv| \\
Loops: superposed sources, and its single-photon control (Sec.~\ref{sec:loops})
 & 20261001\_101803; Fock-limit test 20261005\_170728
 & \path|hybrid_frontier_{fock,sbs}_20261001_h100.csv|, \path|hybrid_frontier_sbs_halfpi_20261005_h100.csv| \\
Loops: squeezed inputs (Sec.~\ref{sec:loops})
 & 20261002\_203546 & \path|gbs_loop_frontier_20261002_h100.csv|, \path|gbs_loop_count{,_borealis}_20261002.json| \\
Loops: $k$ emitters with carried photons (Sec.~\ref{sec:carried})
 & 20261016\_023739 & \path|hybrid_frontier_hybrid_v3_k{2,3,4}_L*_20261016_h100.csv| \\
Cost table at device mode counts; the device rows (Tab.~\ref{tab:devices}, Sec.~\ref{sec:competitive})
 & 20261011\_205906 & \path|cost_table_20261011_h100.csv|, \path|postselect_v2_20261011_h100.csv| \\
Precision of the emitter and train weights (Tab.~\ref{tab:precision}); CPU long-double references
 & 20261004\_223624 (GPU), CPU references
 & \path|emitter_precision{,_collide}_20261004_h100.csv|, \path|emitter_precision_k{26,28,30}{,_collide}_20261004_h100.npz|,
   \path|emitter_precision_reference_20261004_cpu.csv|, \path|sequence_precision_20261004_h100.csv| \\
Output distribution against an independent library (Sec.~\ref{sec:sota})
 & 20261005\_131153 & \path|theboss_tvd_exact_20261003.csv|, \path|theboss_tvd_ours_gpu_exact_20261005.csv|,
   \path|theboss_tvd_large_20261005.csv| \\
A dense-state photonics target: build cost and per-sample slope (Sec.~\ref{sec:sota})
 & 20260923\_224734, 20260924\_134210; slopes 20261005\_130135 (CPU target)
 & \path|tn_{mesh,loop}{,_ours}_20260923.csv|, \path|tn_{mesh,loop}_20260924.csv|,
   \path|tn_{mesh,loop}_shots_20261005.csv| \\
Superposed sources: the Fock-limit test (Sec.~\ref{sec:fourinputs})
 & 20261006\_012054
 & \path|limit_{fock_lossless,rso_pi2_lossless}_final_20261006_h100.csv| \\
Incoherent-twin end of the coherence scan, $\eta = 0.5$ (Sec.~\ref{sec:coherence}): the Fock
$\eta = 0.5$ loss point & 20261008\_192343 & \path|r30_fock_eta05_20261008_h100.csv| \\
Superposed sources: the coherence scan at $\eta = 0.6 / 0.7 / 0.8 / 0.9$ (Sec.~\ref{sec:coherence})
 & 20261007\_122313
 & \path|floor_coherence_eta{0.6,0.7,0.8,0.9}_N1024_r28_20261007_h100.csv| \\
Superposed sources at $\alpha = \pi/4$: the run-of-$S$ statistic $t_S$
 & 20261006\_132352
 & \path|sbs_rso_final_run_cost{,_fit}_r22_20261006_h100.csv| \\
Superposed sources at $\alpha = \pi/4$ with the shot floor: every mode ratio (Tabs.~\ref{tab:ratios}, \ref{tab:T3})
 & 20261010\_045711 & \path|floor_frontier_sbs_N512_20261010_045711_h100.csv| \\
Superposed sources at $m = 2n$ with the shot floor (headline, Tabs.~\ref{tab:ratios}, \ref{tab:T3})
 & 20261007\_090041 & \path|floor_sbs_eta1_N512_r28_20261007_h100.csv| \\
Superposed sources under loss, $\eta = 0.9 / 0.8 / 0.7$ (Tab.~\ref{tab:loss})
 & 20261007\_094729, 20261007\_183855, 20261008\_233436
 & \path|floor_sbs_eta0.9_N1024_r28_20261007_h100.csv|, \path|floor_sbs_eta0.8_N4096_r28_20261007_h100.csv|,
   \path|floor_sbs_eta0.7_N16384_20261008_233436_h100.csv| \\
Superposed sources under loss, $\eta = 0.5$ & \multicolumn{2}{l}{not measured: cost too heavy-tailed to converge within the run budget} \\
\end{longtable}}

\section{Validation and imperfection oracles}\label{supp:validation}

Every sampler is checked against a strong oracle that shares no code with it, and every kernel
against an independent evaluation of the same quantity. The table below is the ledger; the gates
are run in continuous integration on the CPU and by a device check script before every measurement
campaign.

\begin{table}[H]\centering\scriptsize
\setlength{\tabcolsep}{4pt}
\caption{Validation ledger. ``floor'' means the total-variation distance between the sampler's
histogram and the exact distribution is at the shot-noise floor of the number of shots
used.}\label{tab:validation}
\begin{tabular}{@{}L{3.0cm}L{5.2cm}L{4.6cm}@{}}
\toprule
object & oracle & agreement\\
\midrule
Fock cofactor kernel & an independent Ryser evaluation, blocks with repeated rows \emph{and}
columns, 1 / 4 / $65536$ lanes & within the $10^{-10}$ gate to $k-1 \le 11$; $8\cdot10^{-14}$ at
$k-1 = 14$, $1\cdot10^{-11}$ at $20$, $4\cdot10^{-11}$ at $23$ --- the complex128 cancellation of
Glynn's sum, on average a factor $4$ per two photons\\
the four Fock chain-rule algorithms (seven algorithm--backend pairs) & the closed-form addend count; each other, from the same seed (identical
chains by construction, so a wrong term moves a draw) & counts exact; identical draws at
$n = 12$, $m = 24$, 512 shots, on both the Triton and the CUDA backend\\
Fock sampler & the exact permanent distribution & total-variation distance at the floor,
$\chi^2$ $z \le 2.6$; second moments at $n = 16 / 20 / 24$ with $\max|z| \le 3.8$ over
$528 / 820 / 1176$ pairs, under the Bonferroni threshold\\
batched distinct unitaries & each circuit's own exact distribution, and the others' & at the floor
against its own, $z > 10$ against the others, both kernels and both backends\\
emitter sampler & the concatenated matrix-product expansion on the unrolled modes, through a
block-diagonal unitary & sequence probabilities to $10^{-9}$, bin marginals and the joint of two
bins to $10^{-12}$, sampling at the floor\\
emitter under loss & thinned marginals against thinned samples & at the floor\\
superposed-source sampler & the strong Dicke simulation at $n \le 12$; Perceval; the Fock twin's
photon law from the same seed & exact at $\eta = 1$; total-variation distance within shot noise;
lossy mask $2\cdot10^{-16}$\\
non-linear branch expansion & the reconstruction of $\exp(-i f(\hat n_x))$ on the $n$-photon space
& $10^{-12}$, with numerical zeros below $10^{-14}$ dropped\\
non-linear sampler & dense $n$-photon and branch-sum oracles & $10^{-16}$\\
adaptive sampler & enumerated branch tables and the strong joint & at $n = 16$ with $40{,}000$
shots, every acceptance within $1.4\sigma$ of its predicted $P(p)$\\
loop-hafnian pair sieve & brute-force matchings; The Walrus; the polynomial coefficients at four
values of the variable & $4\cdot10^{-15}$ to $N = 10$; $7\cdot10^{-13}$ at $N = 6, 11, 16$;
$10^{-14}$ on the coefficients\\
squeezed probabilities & The Walrus; Piquasso; a truncated product-basis strong table &
$10^{-12}$ displaced, $10^{-13}$ pure, $10^{-8}$ lossy (the truncation tail)\\
squeezed sampler & $\chi^2$ against the exact law, lossless, at $\eta = 0.6$, displaced, and with
two Schmidt modes & pass at every point; $\chi^2 = 35.8$ on $36$ degrees of freedom at $20{,}000$
shots\\
loop register march & the unrolled unitary through the generic engine & at the floor\\
carried-photon register & the emitter engine on the unrolled network & 42 tests, exact\\
imperfection oracles (loss, distinguishability) & an enumerated mixture over surviving subsets and
over interfering species; the Gram oracle & $10^{-13}$ (distinguishability, Gram); $10^{-9}$ to $10^{-12}$ (loss)\\
\bottomrule
\end{tabular}
\end{table}

Two remarks that belong with the ledger rather than in the main text. First, the Glynn sum's
cancellation is the binding numerical fact of the Fock side, and it is the reason
Sec.~\ref{sec:precision} gives arithmetic a frontier of its own: a hierarchical, many-lane
reduction is three to four times more accurate than a sequential one, and an independent
double-precision reference carries a comparable error, so beyond $k - 1 \approx 23$ the reference
has to be an extended-precision evaluation of the kernel code, which the quoted fp64 accuracy range
now rests on up to $k = 38$. Second, the strongest cheap
check in the whole suite is the same-seed identity between algorithms: the chains are identical by
construction, so any incorrect term in any of the seven algorithm--backend pairs would move a draw.

\section{The \texttt{photonic-engines} library}\label{supp:software}\label{supp:repro}

Every number in this paper was produced by one Python library, \texttt{photonic-engines}, built on
PyTorch with its kernels written in Triton and CUDA. It is open source under the MIT licence, at
\url{https://github.com/Quandela/photonic-engines}; version 0.2.1 is available from the Python Package
Index (\url{https://pypi.org/project/photonic-engines/}).
The numbers reported here come from runs made with pre-release revisions of the library; version 0.2.1
contains the same code paths, but the numbers have not been re-measured with it. The result files of
those runs are in the repository's \texttt{results/keep/} directory, and the runs themselves, by run
stamp, at \url{https://github.com/Quandela/photonic-engines-runs} (Sec.~\ref{supp:tables}).

\paragraph{Scope.} The library covers the four classical tasks of Sec.~\ref{sec:intro} on the inputs
of this paper. \emph{Probability computation}: single amplitudes and probabilities (permanents, loop
hafnians, and their sums for superposed inputs), which are also the inner step of the samplers.
\emph{Sampling}: exact samplers for Fock, emitter-generated, photon-number-superposed and squeezed
inputs through a universal interferometer, for adaptive circuits (mid-circuit photon counting with
feed-forward on the interferometer), for circuits with single-mode photon-number phases between
interferometers, for emitter trains, with the emitters' spin register carried bin by bin, and for time-bin loop
devices with and without carried photons (Sec.~\ref{sec:algorithms} and
Supplementary Sec.~\ref{supp:samplers}). \emph{Strong simulation}: the full output distribution
(amplitudes, probabilities, Kraus operators) for Fock, superposed, matrix-product and Gaussian inputs
on one layered basis (Supplementary Sec.~\ref{supp:slos}). \emph{Estimation}: the Gurvits-type
estimator of output probabilities for emitter-generated inputs~\cite{Estimator}. Adapters expose the
engines to Perceval and MerLin.

\paragraph{Architecture.} One shared vocabulary --- input states, output spaces, and the uniform-loss
and partial-distinguishability models --- is used by every engine, and the engines are used by the
adapters, never the reverse. Loss and distinguishability are therefore one model, applied identically
across inputs, which lets Sec.~\ref{sec:law} state one law for all of them; the $\eta = 1$
path of every engine reduces exactly to the lossless one, and this is tested. The detector model
belongs to the same vocabulary: photon-number-resolving or threshold detection, a detection efficiency
(part of the loss model), and dark counts --- Poisson-distributed extra counts per output mode and
detection window, independent of the light --- applied as one output-side layer after loss and before
the threshold clip; where a strong simulation truncates the dark counts, the dropped probability is
returned rather than renormalised away. The samplers spend
their time in a small set of shared kernels, chiefly the batched cofactor primitive of
Sec.~\ref{sec:primitive}; the parity audit rests on this sharing
(Sec.~\ref{sec:parity}, Supplementary Sec.~\ref{supp:kernels}). Optional dependencies (Triton, CuPy,
Perceval, MPI) are imported lazily, so the CPU reference paths run without them. The strong
simulators are written as PyTorch operations, so their probabilities, amplitudes and Kraus operators
are differentiable with respect to the unitary and, for emitter inputs, the emitter operators; the
MerLin adapter's layers are differentiable end to end on this basis.

\paragraph{Using the library.} A unitary is a complex \texttt{[m, m]} PyTorch tensor (a batch of
circuits is \texttt{[B, m, m]}), and every engine takes the same input objects. Strong simulation
and sampling of single photons, with \texttt{haar(m, seed)} a Haar-random unitary:
\begin{lstlisting}
import torch
from photonic_engines.slos import FockSLOSKernel
from photonic_engines.clifford import fock_sample
from photonic_engines.clifford.variants import strong_then_sample

U = haar(12)
# strong simulation: every output probability of |1,0,1,0,1,0,0,...>, differentiable in U
k = FockSLOSKernel(n=3, m=12, input_occupation=[1, 0] * 3 + [0] * 6, mask="fock", device="cpu")
p = k.probs(U.to(torch.complex64))                   # [C(14,3)] over k.layers.final_states
# exact sampling on the GPU: 10 000 output patterns of 6 photons in modes 0..5
gen = torch.Generator(device="cuda").manual_seed(0)
occ = fock_sample(U.cuda(), in_modes=list(range(6)), shots=10_000, generator=gen)   # [10000, 12]
# the same sampler with uniform loss and partial distinguishability (V_HOM = x^2)
occ = fock_sample(U.cuda(), list(range(6)), 10_000, loss=0.8, indist=0.95)
# strong-then-sample: one build of the full distribution, then a lookup per sample
occ = strong_then_sample(haar(8), [0, 1, 2, 3], 1000)
\end{lstlisting}
The other inputs use the same calls. An emitter is declared once as a matrix-product state; its
strong simulator carries the spin as a bond index, its sampler draws without the full distribution,
and the estimator of the companion paper~\cite{Estimator} returns single probabilities to additive
error:
\begin{lstlisting}
import math
from photonic_engines import EmitterMPS, spin_gate
from photonic_engines.slos import EmitterSLOSKernel
from photonic_engines.slos.estimate import EmitterGurvitsEstimator
from photonic_engines.clifford import emitter_sample

prep = spin_gate(math.pi / 2, 0)                     # |up> -> |+>
lc = EmitterMPS.caterpillar(4, [prep] + [spin_gate(math.pi / 2, 0)] * 3)   # linear cluster, m = 8
U8 = haar(8)
p = EmitterSLOSKernel(lc, device="cpu").probs(U8)   # strong simulation, spin traced
occ = emitter_sample(lc, U8, shots=1000, backend="torch")      # exact sampling ("triton" on a GPU)
est = EmitterGurvitsEstimator(lc, sampling="gaussian", seed=0)
e = est.probability(U8, (1, 0, 1, 0, 1, 0, 1, 0), shots=20_000)   # value and standard error
\end{lstlisting}
Superposed and squeezed sources, a non-linear site and an adaptive circuit:
\begin{lstlisting}
from photonic_engines import ProductSuperposition, GaussianInput, clifford, gaussian
from photonic_engines.nonlinear import NonlinearCircuit
from photonic_engines.adaptive import AdaptiveCircuit

sbs = ProductSuperposition.sbs(sources=4, m=8)       # cos(a)|0> + sin(a)|1> per source, a = pi/4
occ = clifford.sample(sbs, U8, 50)                   # chain rule on the indefinite photon number
st = GaussianInput.smsv(3, 6, r=GaussianInput.r_for_mean(0.5))   # 3 squeezers, 0.5 photons each
occ = gaussian.sample(st, haar(6, 1), 500, seed=0, device="cuda")  # no photon cutoff
prob = gaussian.pure_probability(st, haar(6, 1), (1, 1, 0, 0, 0, 0))   # one loop hafnian
C = NonlinearCircuit([haar(6, 1), haar(6, 2)], [0, 1, 2], [(1, 3, math.pi / 2)])  # Kerr site, mode 3
occ = C.sample(2000)                                 # branch mixture with rejection
policy = lambda lam, cnt: haar(5, 10 + cnt[0])       # feed-forward on the count of mode 1
A = AdaptiveCircuit(haar(6, 1), [0, 2, 4], [[1]], policy)
occ = A.sample(2000)
\end{lstlisting}
Loop devices and emitter trains have their own modules (\texttt{timeloop}, \texttt{spin\_register},
\texttt{sequence}), and adapters expose the engines to Perceval and MerLin. The documentation walks
through every engine with runnable examples, and a table maps each task and input to its engine.

\paragraph{Validation.} Each engine is checked against an oracle that shares no code with it, and
each kernel against an independent evaluation of the same quantity; the ledger is
Supplementary Sec.~\ref{supp:validation}. The CPU checks run in continuous integration, and a device
check runs before every measurement campaign.

\paragraph{Reproducing a frontier point.} One script brackets a frontier point:
\texttt{bench/find\_threshold.py} with the engine (\texttt{fock}, \texttt{emitter}, \texttt{sbs},
\texttt{nlbs}, \texttt{gbs}), the mode ratio $c$, the target rate and the imperfections as arguments;
the adaptive, loop and per-tick rows have their own scripts in \texttt{bench/}, one per question.
For example, the headline single-photon point and its $\eta = 0.5$ counterpart are
\begin{lstlisting}[language=bash]
python3 bench/find_threshold.py --engine fock --ratio 2 --target 1.0
python3 bench/find_threshold.py --engine fock --ratio 2 --target 1.0 --loss 0.5
\end{lstlisting} The
instance at $n$ is the Haar-random unitary drawn from the seed $s_0 + n$ (and $s_0 + n + 1000\,i$ for
the further interferometers of a non-linear or adaptive circuit), with the input in modes
$0 \dots n-1$; every timed run draws its shots from a fresh seed, and the first run, which compiles
the kernels, is not counted. At each $n$ the shots ramp $1, 2, 4, \dots$ until the time budget, the
sustained rate is the number of samples over the time of the runs above a minimum shot count, and the
bracketing and statistic are those of Sec.~\ref{sec:statistic}. Each measurement writes a directory
named by its run stamp (date and time, \texttt{YYYYMMDD\_HHMMSS}); the stamp is the source given for
every number in the paper, and the files kept are listed in Supplementary Sec.~\ref{supp:tables}.
The measurements ran on one NVIDIA H100 PCIe (80\,GB), driver 580.126.20, CUDA 13.0, Python 3.12.3,
PyTorch 2.14.0 and Triton 3.8.0; the full pinned environment ships with the code.

\section{Benchmark protocol as a proposed shared standard}\label{supp:protocol}

Every number in this paper is a measurement, and a measurement is only useful if someone else can
make it. What follows is the protocol we used, written as the thing a second group would need in
order to produce a number comparable with ours --- and, where our own choices are arbitrary, said
to be arbitrary.

\paragraph{1. The task.} Draw exact samples from the output distribution of a specified input
through a specified interferometer with specified detectors, with \emph{no} post-selection and no
photon cutoff. ``Exact'' is defined by the arithmetic criterion of
Sec.~\ref{sec:precision} --- relative error of a sample's probability weights at most $10^{-3}$ ---
and a benchmark that cannot state its arithmetic has not stated its task.

\paragraph{2. The instances.} Haar-random $U$ at $m = cn$, with the seed a published function of
$(n, c, \text{job})$ so that the instance family is reproducible; the input occupying modes
$0 \dots n-1$. The mode ratio $c$ is reported, never averaged over. A structured interferometer
(a loop transfer matrix, a local mesh) is a \emph{different} instance family and is reported as one,
with its class named (Sec.~\ref{sec:classes}).

\paragraph{3. The axis.} The number of \textbf{photonic inputs} (Sec.~\ref{sec:frontier}) --- input
modes that receive a single photon, one photon of an emitter's train, a vacuum--single-photon
superposition or a single-mode squeezed state --- and not the detected photon number, so that inputs with different
photon statistics sit on one axis. Where a register sets the exponent, the register parameter
replaces it. Both conventions have to be stated, because a squeezed row quoted in detected photons
and a Fock row quoted in photonic inputs differ by the whole point of the comparison.

\paragraph{4. The statistic.} $t_S$: the cost per sample of a run of $S$ samples, stratified over
the exact photon-number law and truncated at the $1 - 1/S$ quantile with the tail fitted inside
(Sec.~\ref{sec:statistic}). $S = 10^4$ is the headline and $10^6$, $10^8$ the sensitivity. A mean
over unconditioned shots is admissible only where it is shown to agree with $t_S$, which it does
wherever the detected photon number is bounded by the number of photonic inputs. Few-shot means are never
quoted. For a device row, $S$ is that experiment's own run length.

\paragraph{5. The budget and the bracket.} $n^*$ is not fitted from a cost curve; it is bracketed.
Two consecutive integers $n$ and $n+1$ are measured at the same budget until each has at least 64
samples --- a $\pm 5\,\%$ determination at the frontier needs 165 to 330 --- one of them above the
target rate and the other below, and $n^*$ is the log-interpolation between them. The bracket, the
two rates and the shot counts are published with the number. At low transmission the cost is
tail-dominated and the bracket is the noisy part, so the cure is shots and not a different
estimator. The Fock point at $\eta = 0.5$, $n^* = 64.2$, which is also the incoherent-twin end of
the coherence scan, is such a case: a residual tail remains in its bracket, and more shots could
move it below $64$, which would lower the coherence shift of $14.2$ by the same amount.

\paragraph{6. The machine, and how to convert off it.} One accelerator, named by model and
measured power, with the arithmetic precision of every kernel stated. A result on $G$ devices is
reported as $n^*(G) = n^*(1) + \ln G/\ln\beta_{\rm eff}$ only if that scaling has been verified by
direct timing, and with the range over which it was verified. The general conversion is
$\Delta n^* = \ln X/\ln\beta$ for any constant factor $X$ (Sec.~\ref{sec:deltan}), so a reader
with a different budget can shift the curve measured on one machine instead of discarding it.

\paragraph{7. Parity across rows.} The requirement of Sec.~\ref{sec:parity}, restated as a rule for
a table: two rows may be compared only if they run the fastest algorithm available for their input,
on kernels at the same engineering level, under the same bracketing protocol and the same statistic.
Publish the audit --- row against (algorithm, code path, measured or modelled, arithmetic, statistic,
run stamp) --- and mark the rows that fail it, so that a reader can tell which part of a difference
between rows is algorithmic.

\paragraph{8. Correctness.} Every sampler against a strong oracle sharing no code with it, and
every kernel against an independent evaluation of the same quantity, with the agreement published
(Supplement~\ref{supp:validation}). Three gates are cheap and we recommend all of them: a
total-variation distance at the shot-noise floor against an exactly enumerable instance; a
second-moment test on pairs of modes at a size where enumeration is impossible; and, where several
algorithms share a chain, the same-seed identity between them.

\paragraph{9. What must accompany a number.} The instance family and seed rule; the axis
convention; $S$; the bracket with its rates and shot counts; the machine and the arithmetic; the
algorithm and code path; the oracle and its agreement; and the result file, kept.
Everything in this paper carries a run stamp for exactly this reason.

\paragraph{10. What we do not claim to have settled.} The choice of $1\,$Hz, of $m = 2n$ as the
headline ratio, of $S = 10^4$, and of $10^{-3}$ as the arithmetic tolerance are conventions. They
are stated so that they can be changed: the rate and the budget move the curve by a known
logarithm, the mode ratio is scanned, $S$ matters only where the photon number is unbounded, and
the tolerance is the one parameter whose effect we have measured on both sides but not optimised.

\end{document}